\documentclass[12pt,review,times]{elsarticle}
\journal{Progress in Particle and Nuclear Physics}

\usepackage{amssymb,amsmath,comment}

\usepackage{tikz}
\usepackage{tikz-3dplot}
\usetikzlibrary{decorations.pathmorphing}
\tikzset{snake it/.style={decorate, decoration=snake}}

\usepackage{hyperref}

\newcommand{\be}{\begin{equation}} \newcommand{\ee}{\end{equation}}
\newcommand{\bea}{\begin{eqnarray}} \newcommand{\eea}{\end{eqnarray}}
\newcommand{\cO}{\mathcal{O}}
\newcommand{\cN}{\mathcal{N}}

\newcommand{\mt}[1]{\textrm{#1}}
\newcommand{\la}{\lambda}

\usepackage{epsfig}

\graphicspath{{figures2/}}
\def\d{\mathrm{d}}

\def\tens{{\cal T}_b}
\def\pot{V}
\def\Z{{\cal Z}}
\def\W{{\cal W}}
\def\k{\kappa}

\def\Nf{N_\mt{f}}
\def\Nc{N_\mt{c}}
\def\ls{\ell_s}
\def\alphap{\ell^2_s}
\def\laYM{\lambda_\mt{YM}}

\begin{document}

\begin{frontmatter}

\title{Holographic approach to compact stars and their binary mergers}

\author[1,2]{Carlos Hoyos}
\ead{hoyoscarlos@uniovi.es}

\author[3,4]{Niko Jokela}
\ead{niko.jokela@helsinki.fi}

\author[3,4]{Aleksi Vuorinen}
\ead{aleksi.vuorinen@helsinki.fi}

\affiliation[1]{organization={Departamento de F\'{\i}sica, Universidad de Oviedo},
            addressline={c/ Federico Garc\'{\i}a Lorca 18}, 
            city={ES-33007, Oviedo},
            country={Spain}}
\affiliation[2]{organization={Instituto de Ciencias y Tecnolog\'{\i}as Espaciales de Asturias (ICTEA)},
           addressline={c/ Independencia, 13}, 
            city={ES-33047, Oviedo},
            country={Spain}}
\affiliation[3]{organization={Deptartment of Physics},
            addressline={P.O.~Box 64}, 
            city={FI-00014 University of Helsinki},
            country={Finland}}
\affiliation[4]{organization={Helsinki Institute of Physics},
            addressline={P.O.~Box 64}, 
            city={FI-00014 University of Helsinki},
            country={Finland}}

\begin{abstract}

\noindent In this review article, we describe the role of holography in deciphering the physics of dense QCD matter, relevant for the description of compact stars and their binary mergers. We review the strengths and limitations of the holographic duality in describing strongly interacting matter at large baryon density, walk the reader through the most important results derived using the holographic approach so far, and highlight a number of outstanding open problems in the field. Finally, we discuss how we foresee holography contributing to compact-star physics in the coming years.

\end{abstract}

\begin{keyword}
Quantum Chromodynamics \sep AdS/CFT duality \sep Quark matter  \sep Nuclear matter \sep Neutron stars \\
\textit{Preprint numbers: } HIP-2021-49/TH

\end{keyword}

\end{frontmatter}

\newpage
\tableofcontents
\newpage

%%%%%%%%%%%%%%%%%%%%%%%%%%%%%%%%%%%%%%%
%%%%%%%%%%%%%%%%%%%%%%%%%%%%%%%%%%%%%%%
\section{Introduction}\label{sec:intro}
%%%%%%%%%%%%%%%%%%%%%%%%%%%%%%%%%%%%%%%
%%%%%%%%%%%%%%%%%%%%%%%%%%%%%%%%%%%%%%%

\begin{figure}[!t]
    \centering
        \includegraphics[height=9.3cm]{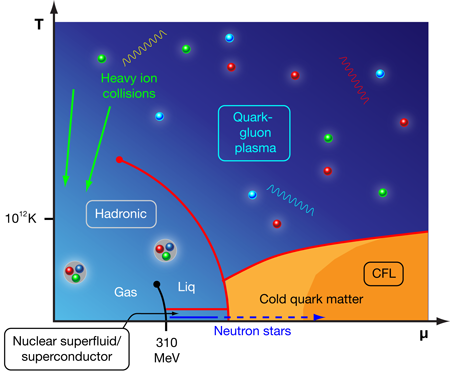} \\
    \caption{A schematic view of the QCD phase diagram spanned by the quark chemical potential $\mu$ and the temperature $T$, taken from \cite{PhaseDiag}. At high energy densities, one eventually enters the deconfined phase of the theory, which at high temperatures and moderately low densities is separated from the hadronic one by a crossover transition. Whether a critical point and a line of discontinuous first-order phase transitions leading all the way to the $T=0$ limit (as indicated here by the red line ending with a dot) exists is currently unclear, although arguments in favor of its existence have recently been made based on symmetry considerations \cite{Cherman:2017tey,Cherman:2018jir,Cherman:2020hbe}. At low temperatures and very high densities, quark pairing is expected to lead to various color superconducting phases, including the Color-Flavor-Locked (CFL) phase that would be dominant for asymptotically high values of $\mu$ should only three quark flavors exist \cite{Alford:1998mk,Son:1998uk}. Note that pairing can also occur in nuclear matter, and this is indeed expected to take place inside neutron stars; we, however, omit the treatment of this phenomenon  due to the lack of related holographic studies.}
    \label{fig:phase_diag}
\end{figure}

Although the very name of the strong nuclear force refers to the sizable nature of its interaction, the properties of the underlying quantum field theory, Quantum Chromodynamics (QCD), are often approached through an expansion around the limit of no interactions \cite{Brambilla:2014jmp,Ghiglieri:2020dpq}. As is well known, the reason for this apparent contradiction has its roots in the limitations of nonperturbative field theory tools, in particular lattice QCD, in describing a number of physically relevant limits, such as real-time phenomena or sizable baryon densities \cite{deForcrand:2010ys}. In the review article at hand, written from the perspective of another nonperturbative method, holography \cite{Maldacena:1997re,Gubser:1998bc}, we shall pay particular attention to the latter of these limits: dense QCD matter, responsible for the rich and very nontrivial phase structure visible in the lower right-hand corner of fig.~\ref{fig:phase_diag}.

\begin{figure}[!t]
    \centering
        \includegraphics[height=6.2cm]{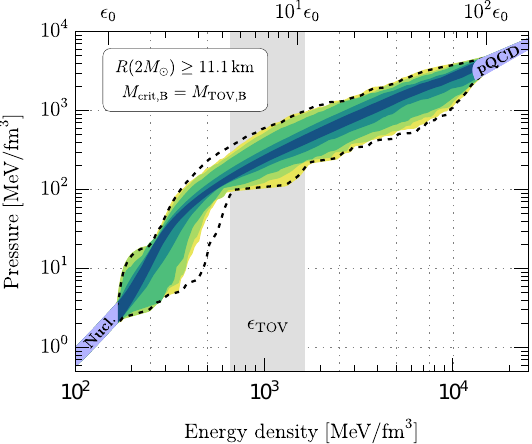}$\;\;\;\;$\includegraphics[height=5.8cm]{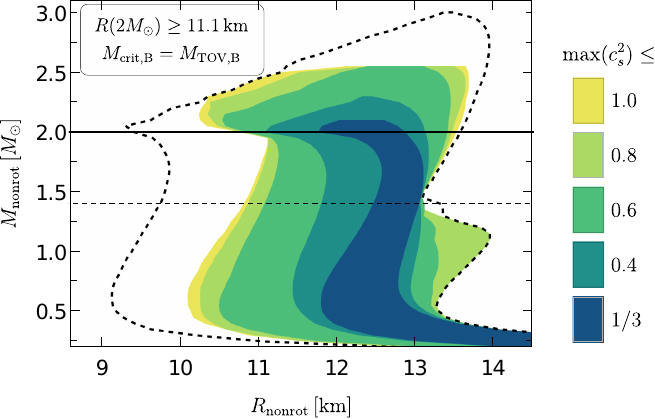} \\
    \caption{A state-of-the-art model-independent result for the EoS of neutron-star matter (left) and the corresponding mass-radius (MR) relation (right), taken from \cite{Annala:2021gom}. The color coding in the figure corresponds to the maximal value that the speed of sound squared attains at any density, with the conformal limit corresponding to $1/3$.}
    \label{fig:epsmr}
\end{figure}

At high enough baryon density, irrespective of the temperature, QCD matter is expected to enter a deconfined phase similar to the more familiar quark-gluon plasma, created and observed in ultrarelativistic heavy-ion collisions \cite{Shuryak:2014zxa}. At low temperatures, this phase is commonly referred to as quark matter, emphasizing the fact that the system behaves as a predominantly fermionic one at low temperatures, characterized by the presence of a Fermi sea of filled quark states at weak coupling \cite{Shuryak:1980tp,Rajagopal:2000wf}. At present, it is not clear whether such matter exists anywhere in our current Universe, but whether it can be found inside the cores of the most massive neutron stars is a topic of active ongoing research \cite{Annala:2019puf}.\footnote{Note that we use the term \textit{neutron star} to refer to stars composed of either nuclear matter or a combination of nuclear and quark matter, while the term \textit{compact star} is used for a broader class of stars, including more exotic objects such as quark stars \cite{Weber:2004kj}. A somewhat special class of stars is formed by so-called twin stars \cite{Alvarez-Castillo:2018pve}, for which the neutron-star mass-radius curve splits to two separate parts due to a strong first-order phase transition. While twin stars are often included under the umbrella of neutron stars, we note that some of the results presented here (e.g., those of \cite{Annala:2021gom}, shown in fig.~\ref{fig:epsmr}) do not apply for them.}

Neutron stars are the remnants of old massive stars that undergo a supernova explosion and a subsequent gravitational collapse \cite{Glendenning:1997wn,Lattimer:2004pg}. This requires that the mass of the progenitor, i.e.~the original hydrogen-burning star, is simultaneously sufficiently high ($M\gtrsim 10M_\odot$) for a supernova explosion to take place but also sufficiently low ($M\lesssim 25M_\odot$) so that its core does not directly collapse into a black hole. Should the Fermi and interaction repulsion of the QCD matter be able to resist the collapse, the end result of the process is an object of approx.~1-2$M_\odot$ in mass and only approx.~11-13km in radius, making its average density comparable to the nuclear saturation density $n_s\approx 0.16$/fm$^3$. This implies that the maximal densities expected to be reached in the centers of maximally massive neutron stars --- i.e.~ones just on the brink of collapse into a black hole --- are likely several times above $n_s$, thereby exceeding the density of any system that can be created in an Earth-based laboratory \cite{Senger:2020wvj}.

Even at the densities corresponding to neutron-star cores, QCD remains a strongly coupled theory, so approaches based on weak-coupling expansions either in quark or nuclear matter are eventually doomed for failure. Indeed, modern Chiral Effective Theory (CET) calculations, based on a perturbative treatment of nuclear interactions via meson exchange \cite{Epelbaum:2008ga,Machleidt:2011zz}, are able to quantitatively describe nuclear matter up to densities only slightly above $n_s$ \cite{Hebeler:2009iv,Hebeler:2010xb,Tews:2012fj,Holt:2016pjb,Drischler:2020yad}, while modern perturbative QCD (pQCD) becomes accurate only around $40n_s$ \cite{Kurkela:2009gj,Gorda:2018gpy,Gorda:2021kme,Gorda:2021znl}. In both cases, the quantity of the highest relevance for the physics of neutron-star matter is its equation of state (EoS), which can be given as the functional relationship between the pressure $p$ and energy (or baryon number) density $\epsilon$ (or $n$) in the medium.\footnote{In analytic calculations of thermodynamic quantities one typically obtains more than just the relation $p(\epsilon)$: both quantities as functions of the baryon chemical potential $\mu_\text{B}$. This fact can be taken advantage of in constraining the neutron-star matter EoS as has been recently demonstrated in \cite{Komoltsev:2021jzg}.} To describe this quantity --- and the complex dynamics related to the phase transition from hadronic to quark matter --- something more is clearly needed: a nonperturbative tool capable of describing the system in the limits of sizable coupling and baryon density.

%%%%%%%%%%%%%%%%%%%%%%%%%%%%%%%%%%%%%%%%%%%%%%%%%
%%%%%%%%%%%%%%%%%%%%%%%%%%%%%%%%%%%%%%%%%%%%%%%%%

The gauge/gravity duality --- or holography in short --- is a computational tool that has its origins on one hand in the study of so-called Dirichlet- or D-branes within string theory and on the other hand in conformal field theories in their large-$N$ limit (for a review, see e.g.~\cite{Aharony:1999ti}). In its original formulation, it provides a highly nontrivial relation between two very different physical systems: string theory in a 10-dimensional curved spacetime and a conformal and maximally supersymmetric quantum field theory living on its boundary \cite{Maldacena:1997re,Gubser:1998bc}. Over the years, the duality has, however, evolved into a practical tool for the description of strongly coupled field theories, owing to its mapping of the infinite-coupling and large-$N$ limit of the field theory to the tractable limit of classical supergravity on the stringy side. This has enabled numerous successful applications of the duality to e.g.~the physics of condensed matter theory \cite{Sachdev:2010ch} and heavy ion collisions \cite{Casalderrey-Solana:2011dxg}, and to the development of so-called bottom-up holography as a more phenomenological tool for particular physical systems (See, e.g., \cite{Erlich:2005qh,Gursoy:2007cb,Gursoy:2007er}).

Within heavy-ion physics, close in spirit to the topic of the review article at hand, holography has given rise to a number of qualitative and quantitative insights. To name a few of the most obvious ones, the universality of a small shear-viscosity-to-entropy ratio in a wide class of strongly coupled field theories nearly two decades ago \cite{Policastro:2001yc,Kovtun:2004de}, and the subsequent observation of the QGP nearly saturating this limit \cite{Romatschke:2007mq,Song:2010mg,Niemi:2015qia,Bernhard:2019bmu}, gave rise to the notion of the QGP acting as a `perfect liquid'. Somewhat later, a series of numerical studies of shock wave collisions in asymptotically AdS$_5$ spacetimes led to the realization that the onset of hydrodynamic behavior in heavy-ion collisions does not necessitate the thermalization of the medium, but that an isotropization or an even weaker `hydrodynamization' of the system suffices \cite{Chesler:2008hg,Heller:2012km,vanderSchee:2013pia}. This fact was, too, later confirmed using modern kinetic theory \cite{Heller:2016rtz}, but the seed to the insight was provided by holography.

While the above example applications by no means exhaust the list of insights holography has provided within heavy-ion physics, they serve the purpose of highlighting the two most common situations, where the gauge/gravity duality is found to be particularly valuable. First, it may reveal universal behavior in physical systems that would otherwise go unnoticed, which was clearly the case in the viscosity-to-entropy conjecture that has since then affected a number of subfields of physics (see, e.g., \cite{Schafer:2009dj}). Second, the way holography maps complicated field theory questions to ones in classical gravity has enabled progress in problems that were long deemed intractable; a prime example of these is thermalization in strongly coupled field theory, accessible via the study of shock wave collisions in asymptotically AdS$_5$ spacetimes \cite{Chesler:2010bi}.

%%%%%%%%%%%%%%%%%%%%%%%%%%%%%%%%%%%%%%%%%%%%%%%%%
%%%%%%%%%%%%%%%%%%%%%%%%%%%%%%%%%%%%%%%%%%%%%%%%%

Returning to the physics of dense QCD matter and neutron stars, we are clearly faced with a situation, where nonperturbative machinery is urgently needed, but due to the Sign Problem of lattice QCD no traditional field theory tool is available. At the moment, the state of the art in the field typically altogether bypasses the problematic region of neutron-star cores by either extrapolating the neutron-star matter EoS from the CET realm of $n\lesssim n_s$ upwards \cite{Hebeler:2013nza,Capano:2019eae}, or by interpolating between this region and the pQCD one at $n\gtrsim 40n_s$ \cite{Annala:2019puf,Annala:2017llu,Most:2018hfd,Annala:2021gom} (see also fig.~\ref{fig:epsmr}). Complementary to this approach, there have been numerous studies concentrating on various simplified models to dense QCD both in the nuclear and quark matter phases (see, e.g., \cite{Dexheimer:2009hi,Kaltenborn:2017hus}). Here, the results, however, often suffer from systematic uncertainties related to the incomplete description of the physical system that are difficult to quantitatively estimate. Given that our review mostly deals with the technical aspects of holographic models, we have decided to refrain from a more extensive discussion of other model approaches.

In light of all the above, it is natural to ask, whether holographic methods might offer the much-needed nonperturbative tool to attack the physics of dense QCD matter in the problematic region near the deconfinement transition. One may naturally worry about the effects of the typical $\Nc\to \infty$ approximation in particular on the hadronic phase of the theory, but the method's unique ability to describe strongly coupled fundamental-representation matter at high densities is at the same time extremely alluring. As is typical in applications of holography, one may choose between so-called top-down and bottom-up approaches, of which the former constitute more rigorous descriptions of field theories typically rather different from QCD while the latter offer somewhat more phenomenological descriptions of systems close to real nuclear and quark matter. Both approaches have been actively investigated during the past more than five years (see, e.g, \cite{Jarvinen:2021jbd} and references therein), and as we will detail below, they have produced valuable qualitative and quantitative insights into the physics of dense strongly interacting matter and neutron stars. 

The purpose of the review article at hand is twofold: first, we want to present the motivation for and the formalism of the holographic calculations that describe dense QCD matter, and second, we wish to review some of the most important results that this approach has provided thus far. To do so, we first introduce the reader to the field of dense QCD matter and neutron-star physics in section \ref{sec:setup}, and then move on to the formalism of dense holography in sec.~\ref{sec:denseholo}. The following two sections \ref{sec:thermodynamics} and \ref{sec:transport} then walk the reader through a number of recent results concerning the bulk thermodynamic and transport properties of (unpaired) quark matter, while sec.~\ref{sec:mergers} introduces new results relevant for studies of binary neutron-star mergers. Finally, in sec.~\ref{sec:openquestions} we present our (undoubtedly biased) vision of how holographic methods will continue to contribute to this expanding field of research in the coming years and in which physics questions they will likely make their most significant impact.

Having defined the main goals of the review, it is finally also worth mentioning a few interesting and closely related topics that we have decided to altogether skip here but that might be of interest to our readers. These include, e.g.,~the study of compact stars in alternative theories of gravity \cite{Barack:2018yly,Barausse:2020rsu}, efforts to discover traces of BSM physics using neutron-star observations \cite{Barack:2018yly,Barausse:2020rsu}, considerations of stellar solutions in the bulk of asymptotically anti de Sitter spacetimes \cite{deBoer:2009wk,Arsiwalla:2010bt}, as well as holographic approaches to heavy-ion \cite{Casalderrey-Solana:2011dxg,Brambilla:2014jmp,Ramallo:2013bua} and condensed matter physics \cite{Zaanen:2015oix,Hartnoll:2016apf}. The tools and methods used in holographic studies of dense QCD matter have often been first introduced in these neighboring fields, and we hope that these fields will in time also benefit from the holographic work that has recently been carried out in the context of compact-star physics.

%%%%%%%%%%%%%%%%%%%%%%%%%%%%%%%%%%%%%%%
%%%%%%%%%%%%%%%%%%%%%%%%%%%%%%%%%%%%%%%
\section{Theoretical and observational bounds for neutron-star matter}\label{sec:setup}
%%%%%%%%%%%%%%%%%%%%%%%%%%%%%%%%%%%%%%%
%%%%%%%%%%%%%%%%%%%%%%%%%%%%%%%%%%%%%%%

The existence of extremely compact stellar objects mostly consisting of neutrons was first proposed almost 90 years ago in 1933, when Walter Baade and Fritz Zwicky suggested such objects would be created in gravitational collapses related to the (then still mysterious) supernova explosions \cite{Baade:1934zex}. Some thirty years later, around 1967, Jocelyn Bell Burnell and Antony Hewish discovered periodic radio pulses from a source that was later understood to be a rapidly rotating and highly magnetized neutron star, i.e.~a pulsar \cite{Hewish:1968bj}. This marked the opening of a new subfield of astrophysics, which connects the study of nuclear matter to astrophysical observations: due to the extreme densities of neutron-star cores, these objects constitute a unique laboratory for matter at ultrahigh densities, not achievable in any Earth-based laboratory \cite{Lattimer:2004pg}. 

An obvious challenge in neutron-star observations stems from the fact that we are dealing with very small objects located far away from us: the diameter of an average neutron stars believed to be around 25km, while the closest known neutron stars are located hundreds of light years away. Traditionally, neutron-star observations have relied on detecting electromagnetic (EM) radiation from pulsars or on measuring the properties of binary systems where one component is a neutron star and the other e.g.~a white dwarf, which have more lately been complemented by observations of X-ray bursts from neutron-star surfaces \cite{Lattimer:2006xb}. As we shall explain in more detail below in sec.~\ref{sec:observations}, from these observations one can infer a host of neutron-star properties, including their spinning rates, surface temperatures, masses, and even radii. 

From the neutron-star properties listed above, the two that carry by far the largest amount of information about their inner structure are without doubt their masses and radii. This is so due to the fact that their  functional dependence on each other, the so-called neutron-star mass-radius (MR) relation, is intimately connected to the most important function characterizing the thermodynamic properties of neutron-star matter, its EoS. This connection follows from the Tolman-Oppenheimer-Volkov (TOV) equations of General Relativity (GR) that characterize hydrostatic equilibrium inside a nonrotating gravitationally self-bound object \cite{Tolman:1939jz,Oppenheimer:1939ne},
\begin{eqnarray}
\label{TOVeq}
 \frac{dp(r)}{dr}  &=& -\frac{(\epsilon(r) + p(r))}{r^2}  \frac{ (M(r) + 4\pi r^3 p(r)  )}{  ( 1-2M(r) /r )} \nonumber \\
\frac{dM(r)}{dr} &=& 4\pi r^2 \epsilon(r)\, .
\end{eqnarray}
Here, $p(r)$ and $\epsilon(r)$ stand for the pressure and energy density at a radial distance $r$ from the center of the star, while $M(r)$ corresponds to the gravitational mass confined within a sphere of the same radius. Upon specifying the EoS, i.e.~the unique functional relation between the pressure and energy density, these equations can be solved for the possible masses and radii of stable neutron stars, parameterized by the pressure in the center of a given star. This way, we obtain a one-to-one mapping between the EoS and the neutron-star MR-relation, so that any insights gained on one quantity can be immediately transferred to the other one.

\begin{figure}[!t]
    \centering
        \includegraphics[height=5.0cm]{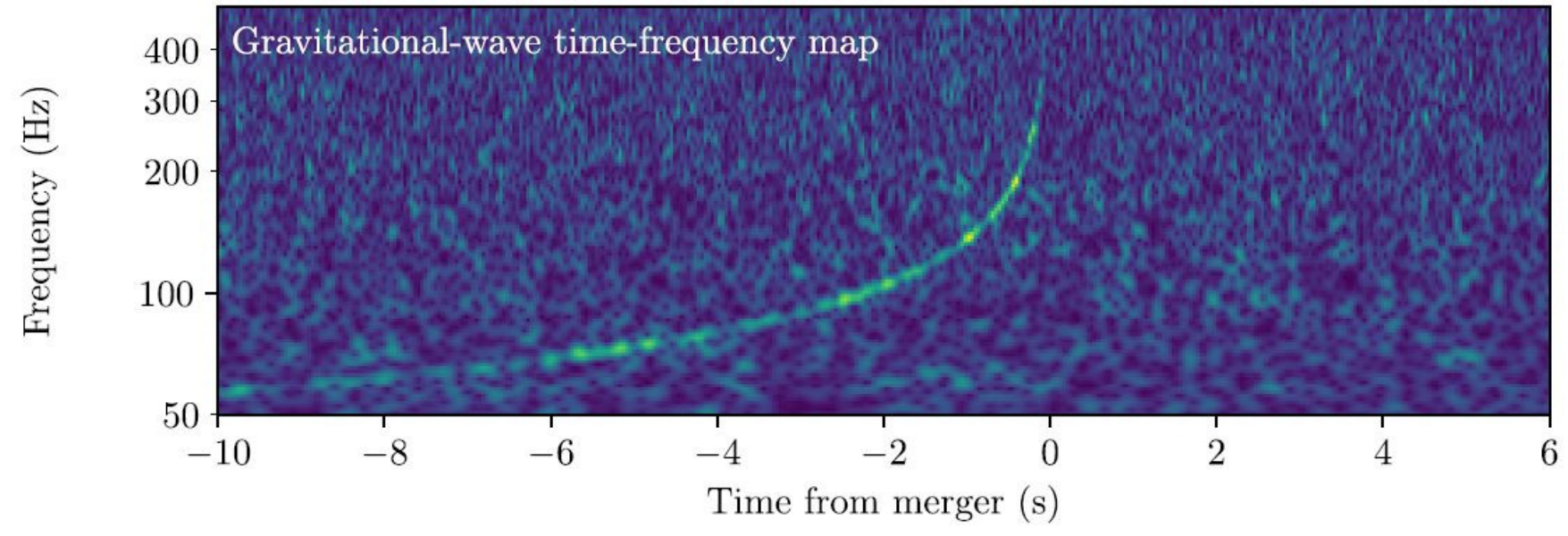} \\
    \caption{The GW time-frequency map from a combination of the LIGO-Hanford and LIGO-Livingston data for GW170817, taken from \cite{LIGOScientific:2017zic}.}
    \label{fig:gw1}
\end{figure}

Apart from the more traditional observations of quiescent neutron stars, the past few years have presented us an exciting opportunity to follow the final moments of a number of neutron stars through multimessenger, i.e.~combined gravitational wave (GW) and EM, observations of a number of binary neutron-star mergers. As we shall detail in sec.~\ref{sec:observations} below, already the first such recorded merger, GW170817, produced valuable information of two types. First,  LIGO and Virgo were able to measure the GW waveform corresponding to the inspiral phase of the merger of some $100$ seconds in duration \cite{TheLIGOScientific:2017qsa,Abbott:2018exr,Abbott:2018wiz} (see fig.~\ref{fig:gw1}), and soon thereafter a number of independent observatories detected signals across the EM spectrum that were confirmed to originate from the same part of sky as the GW one \cite{LIGOScientific:2017ync}. As we will explain below, both types of observations have been efficiently used to constrain the properties of the two stars taking part in the merger and to infer characteristics of the EoS of neutron-star matter.

%%%%%%%%%%%%%%%%%%%%%%%%%%%%%%%%%%%%%%%%%%%%%%%%%
%%%%%%%%%%%%%%%%%%%%%%%%%%%%%%%%%%%%%%%%%%%%%%%%%

In the remainder of this section, we will present a brief but relatively comprehensive introduction to what is presently known about the thermodynamic properties of neutron-star matter based on three key elements: first-principles calculations in QCD, relevant neutron-star observations, and a model-independent combination of these insights. Throughout these parts, we implicitly assume that all observations of compact stars made so far have been of neutron stars as opposed to e.g.~quark stars or twin stars, which is reflected also in the section title. We begin by reviewing state-of-the-art results in nuclear theory and perturbative QCD (pQCD) in sec.~\ref{sec:lim}, then move on to current observational methods and data in sec.~\ref{sec:observations}, and finally present the results of the so-called interpolation approach to the neutron-star matter EoS in sec.~\ref{sec:interpolation}. From these considerations, it will hopefully become clear, where input from novel methods such as holography is at the moment most urgently needed.

%%%%%%%%%%%%%%%%%%%%%%%%%%%%%%%%%%%%%%%
\subsection{Neutron-star matter: theory results}\label{sec:lim}
%%%%%%%%%%%%%%%%%%%%%%%%%%%%%%%%%%%%%%%

The quantities that characterize the collective properties of a thermal medium can be largely divided into two categories: first, bulk thermodynamic quantities relating observables such as pressure, energy density and baryon density to each other, and second, various transport coefficients, characterizing the response of the system to small departures from equilibrium. In this section, our goal is to present the current state-of-the-art results from first-principles machinery such as CET and pQCD, respectively applicable at moderately low and very high densities. For presentational simplicity, we will mostly concentrate on bulk thermodynamic quantities, where the task of comparing holographic predictions to existing results is considerably more straightforward. For transport properties, discussed in sec.~\ref{sec:transport}, one encounters pronounced sensitivity to the phenomenon of quark pairing in the in deconfined phase of QCD. Given that the true physical phase of cold and dense quark matter is only rigorously known at asymptotically high densities, this discussion quickly becomes highly convoluted and not fitting for this brief section of our review. Instead, we refer the interested reader to the excellent review \cite{Schmitt:2017efp} as well as the original reference \cite{Heiselberg:1993cr} where the current state-of-the-art pQCD results for the unpaired quark matter phase are derived.

Within bulk thermodynamic quantities, the most fundamental one is undoubtedly the partition function that can be used to express the pressure as a function of different chemical potentials (see, e.g., \cite{Kapusta:2006pm,Laine:2016hma}). For a quiescent ``old'' neutron star that has cooled down due to neutrino emission, it is a good approximation to assume that the medium is locally charge neutral and beta-equilibrated, although the former assumption can be relaxed in case of mixed phases, requiring a first order phase transition and a sufficiently low surface tension between the two (hadronic and quark matter) phases \cite{Glendenning:1992vb}. In this case, one can solve all but one of the otherwise independent chemical potentials as functions of a single one, often chosen as the baryon chemical potential. In the limit of negligibly low temperatures, the pressure then becomes a simple function $p(\mu_{\text{B}})$, from which it is straightforward to derive other forms, such as $p(\epsilon)$ or $p(n)$, perhaps more familiar from literature.

The outer layers of a neutron star include a gaseous atmosphere, a liquid ocean, and a solid crust, all corresponding to moderately low densities $n\ll n_s$. The atmosphere is a thin (from millimeters to tens of centimeters) layer of plasma on the surface of a neutron star, typically consisting of either hydrogen or helium, while the ocean is a somewhat thicker (some tens of meters) layer composed of a plasma of electrons and nuclei \cite{Potekhin:2014hja}. Accurately determining the composition of the atmosphere is a crucial ingredient of many observational analyses of neutron stars, including radius determinations (see, e.g.,~\cite{Nattila:2017wtj}), but the contributions of the atmosphere and ocean layers to the total mass of the star are typically negligible. The solid and considerably thicker (hundreds of meters) crust is on the other hand divided into two parts, with the outer layer corresponding to a lattice of increasingly neutron-rich nuclei with the dominant contribution to the pressure coming from an electron gas, similarly to the case of white dwarfs \cite{Chamel:2008ca}. In the inner crust, the system then reaches the so-called neutron-drip point, where neutrons begin to leak from the nuclei, continuing all the way to the crust-core interface where even lighter nuclei cease to exist. In these parts of the neutron star, corresponding to densities up to some 0.1-0.5$n_s$, the EoS can be determined with traditional tools of nuclear theory, significantly aided by the fact that the system is tightly constrained by experimental data (see, e.g., \cite{Fortin:2016hny} and references therein).

%%%%%%%%%%%%%%%%%%%%%%%%%%%%%%%%%%%%%%%%%%%%%%%%%
\begin{figure}[!t]
    \centering
        \includegraphics[height=6.8cm]{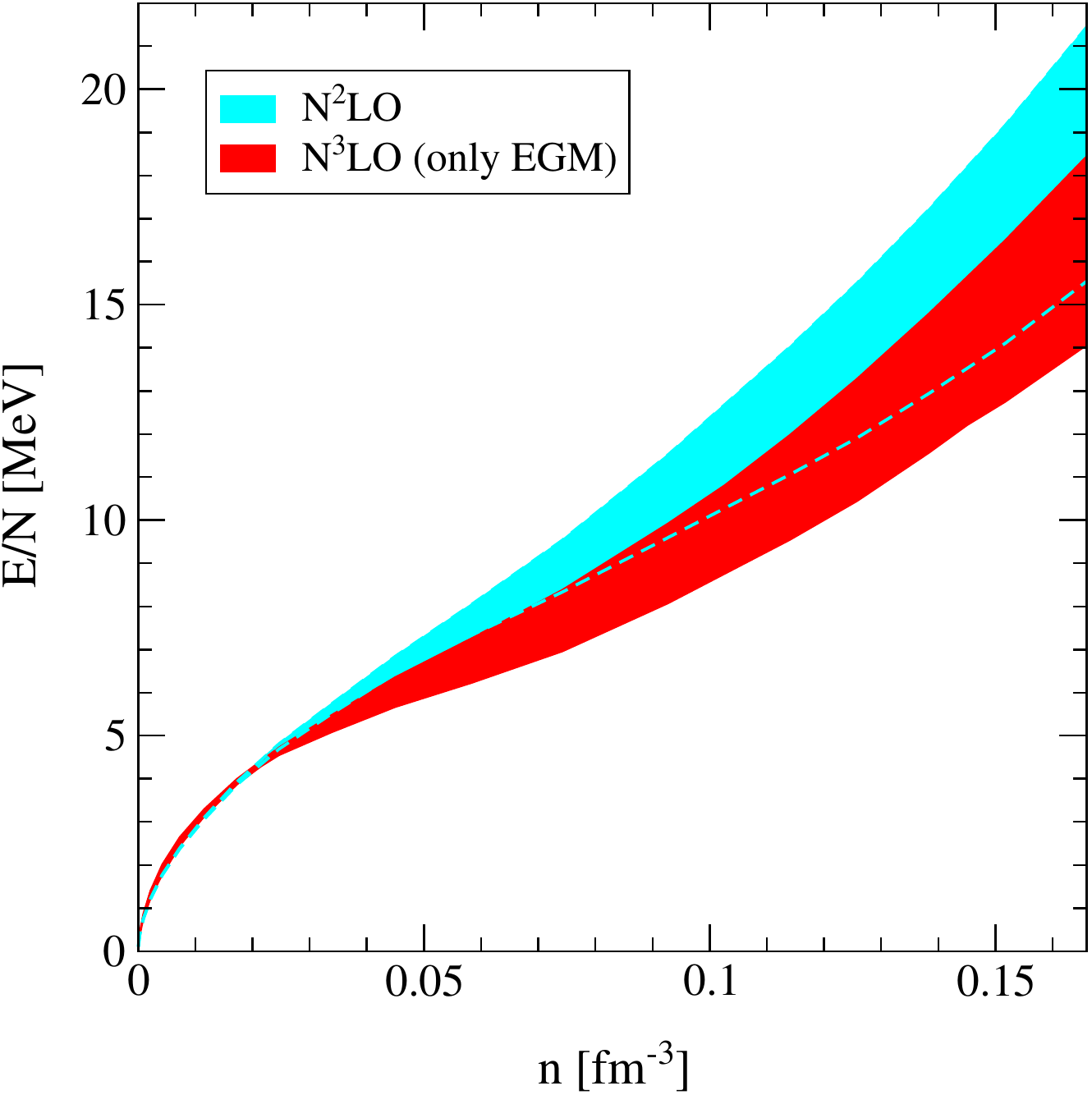}$\;\;\;\;\;\;\;$\includegraphics[height=7cm]{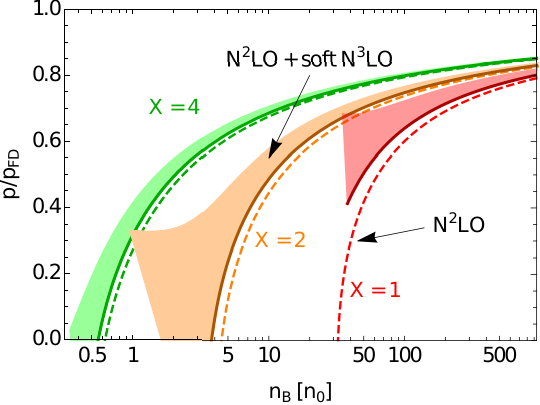} \\
    \caption{Left: Example results for the energy per baryon of pure neutron matter at sub-saturation densities, intimately related to the EoS \cite{Tews:2012fj}. The light blue and red bands parameterize the uncertainty of the result at NNLO and partial NNNLO, respectively. Right: The state-of-the-art partial NNNLO EoS of unpaired quark matter \cite{Gorda:2021kme}, with the dashed lines denoting the earlier NNLO result. Here, the parameter $X$ refers to the value of the (a priori unknown) renormalization scale parameter, while the colored bands parametrize the dependence of the result on an additional cutoff scale (see \cite{Gorda:2021kme} for details).}
    \label{fig:pressures}
\end{figure}
%%%%%%%%%%%%%%%%%%%%%%%%%%%%%%%%%%%%%%%%%%%%%%%%%

Proceeding to the neutron-star core, we finally encounter nuclear matter formed by individual nucleons (mostly neutrons), with the corresponding interactions typically described via meson exchange. Here, it becomes imperative to systematically account for the interaction corrections, for which the CET, dating back to the groundbreaking work of Weinberg \cite{Weinberg:1990rz}, offers a very useful tool. Decades of technically very involved calculations, reaching a partial Next-to-Next-to-Next-to-leading (NNNLO) in the CET powercounting and incorporating three-nucleon interactions \cite{Hebeler:2009iv,Hebeler:2010xb}, are currently able to quantitatively describe the EoS of both symmetric nuclear matter and pure neutron matter --- and interpolate between them --- up to densities exceeding the nuclear saturation density $n_s$ \cite{Tews:2012fj,Holt:2016pjb,Drischler:2020yad}. At this point, the systematic uncertainties in these computations begin to rapidly grow, and although some attempts have been made to reach densities as high as $2n_s$, most state-of-the-art results are only reported up to 1-1.5$n_s$ [for an example result, see fig.~\ref{fig:pressures} (left)]. For more details about this interesting topic, we refer the interested reader to a comprehensive recent review covering both the CET formalism and its application to the determination of the neutron-star matter EoS \cite{Drischler:2021kxf}.

Finally, in the the inner cores of neutron stars, possibly containing deconfined quark matter, baryon density reaches values up to 5-9$n_s$ depending on the physical EoS realized (see e.g.~\cite{Annala:2019puf,Annala:2021gom}). This region is outside the realm of any present-day first-principles theoretical tool, and it is indeed only at considerably higher densities, of the order of several tens of saturation densities, that one regains any theoretical control over the EoS of dense QCD matter. Here, it is a very different set of tools than the ones used at lower densities, namely those of resummed perturbative QCD, that allow one to determine the properties of the system in a weak coupling expansion, relying on the asymptotic freedom of the underlying theory \cite{Ghiglieri:2020dpq}. At present, the weak-coupling expansion of the pressure of dense quark matter is known up to a partial NNNLO, or $\alpha_s^3$ in the strong coupling constant, where a Hard Thermal Loop (HTL) resummation has been applied to the infrared (IR) sector of the theory \cite{Gorda:2018gpy,Gorda:2021kme,Gorda:2021znl}, building on the earlier results of \cite{Freedman:1976ub,Andersen:2002ey,Vuorinen:2003fs}. The uncertainty of this pQCD result becomes comparable to that of the CET EoS at $n\approx n_s$ around forty saturation densities, i.e.~considerably above the densities reached inside physical neutron stars [see fig.~\ref{fig:pressures} (right)]. As has been recently shown in \cite{Komoltsev:2021jzg} we will demonstrate in sec.~\ref{sec:interpolation} below, this however does not imply that the high-density EoS would not place important constraints on the neutron-star matter one at much lower densities.

%%%%%%%%%%%%%%%%%%%%%%%%%%%%%%%%%%%%%%%
\subsection{Neutron-star observations}\label{sec:observations}
%%%%%%%%%%%%%%%%%%%%%%%%%%%%%%%%%%%%%%%

In addition to the theoretical bounds on the neutron-star matter EoS reviewed above, this function --- as well as many other microscopic quantities describing the system --- can be efficiently constrained through neutron- star observations. In this section, we briefly review three different types of observational information, all of which provide information directly relevant for the EoS. These include neutron star mass and radius measurements, as well as multimessenger observations of neutron star mergers.

Apart from rotation frequencies that can be directly determined in any pulsar observation, the most easily accessible macroscopic property of a given neutron star is its mass. Even here, accurate measurements often require taking advantage of rather intricate properties of GR, such as the Shapiro delay, which refers to the slowing down of light rays when they propagate near a massive object \cite{Shapiro:1964uw}. For a highly inclined binary system formed by a millisecond pulsar and a companion (such as a white dwarf), observing this effect in radio pulses emitted by the pulsar enables a precise measurement of both the neutron star and its companion (see, e.g.~\cite{Demorest:2010bx}). This has led to some of the most accurate mass measurements of neutron stars, including the most massive neutron stars known, the pulsar J0740+6620 with a mass of $2.08\pm 0.07 M_\odot$ (one-$\sigma$ uncertainty limits) \cite{NANOGrav:2019jur}, and the previous record holders \cite{Demorest:2010bx,Antoniadis:2013pzd}. From the perspective of EoS determinations, by far the most important implication of these measurements is that they imply the near-certain existence of two-solar-mass neutron stars, which allows one to discard all EoS models that do not support such stars. Beyond this, mass measurements can give insights into the mass distribution of neutron stars \cite{Kiziltan:2013oja}, which however does not provide direct information about the microscopic properties of dense matter.

\begin{figure}[!t]
    \centering
        \includegraphics[height=6.2cm]{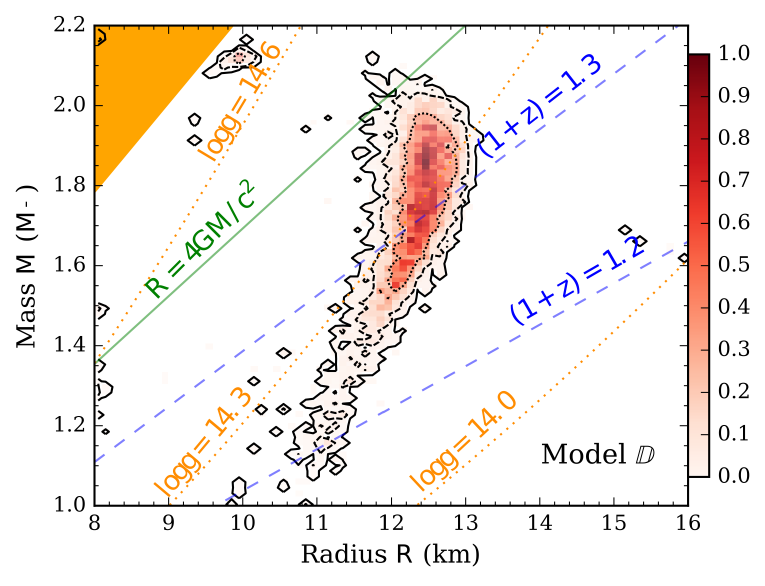}$\;\;\;$\includegraphics[height=6.0cm]{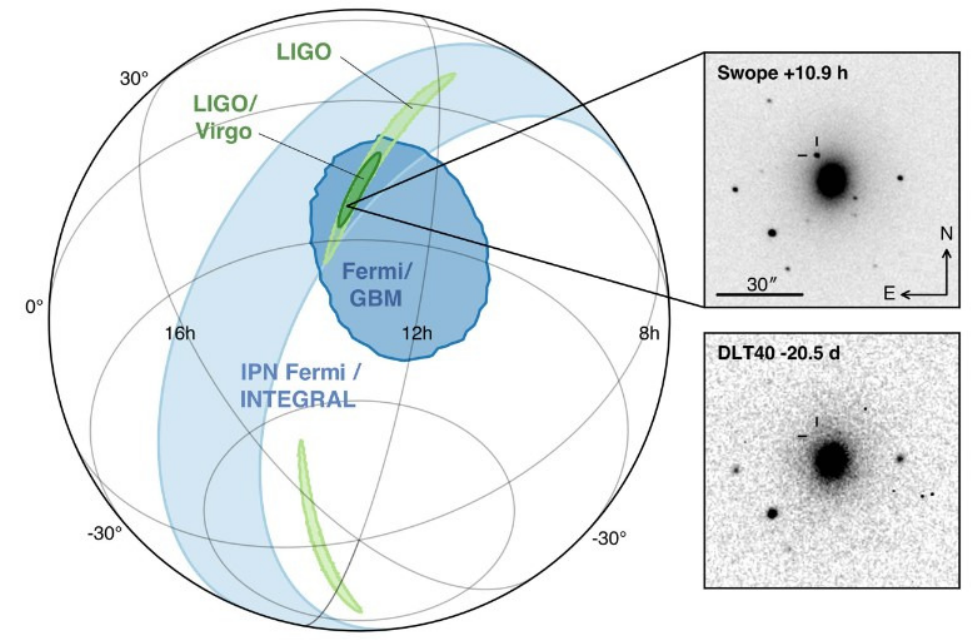}
    \caption{Left: Mass and radius posteriors for the neutron star 4U 1702-429, taken from \cite{Nattila:2017wtj}. Right: The localization of the GW, gamma-ray, and optical signals corresponding to the GW170817 event, taken from \cite{LIGOScientific:2017ync}. For more information on the results, we refer the interested reader to the original publications.}
    \label{fig:observations}
\end{figure}

While the accurate measurement of neutron-star masses can be very challenging, the determination of neutron-star radii is even more intricate and typically requires highly nontrivial astrophysical modeling \cite{Ozel:2016oaf}. Here, the past decade has witnessed very promising developments, originating in particular from observations of thermonuclear X-ray bursts in low-mass X-ray binary systems, where a neutron star is accreting matter from its companion \cite{Nattila:2017wtj} [for an example result, see fig.~\ref{fig:observations} (left)]. More recently, the NICER collaboration, which operates an X-ray telescope on the International Space Station, has used the pulse profile modeling technique to perform radius measurements, exploiting GR effects seen in X-rays emitted from the surface of a rotating neutron star \cite{Ozel:2015ykl}. With both techniques, one obtains a simultaneous measurement of a given star's mass and radius, which in some cases is complemented by independent mass measurements. Perhaps most strikingly, this has enabled setting a very nontrivial lower limit for the radius of the aforementioned massive pulsar J0740+6620 \cite{Miller:2021qha,Riley:2021pdl}, which can be used as a stringent constraint for the neutron-star matter EoS \cite{Raaijmakers:2021uju}.

On 17 August 2017, the detectors of the LIGO and Virgo GW observatories recorded a GW signal that lasted about 100 seconds and had features compatible with the inspiral phase of a binary neutron-star merger \cite{TheLIGOScientific:2017qsa}. After a delay of about 1.7 seconds, the GW signal was followed by a short gamma ray burst that was detected by the Fermi and INTEGRAL telescopes \cite{Goldstein:2017mmi,Savchenko:2017ffs}. These events were dubbed GW170817 and GRB170817A, respectively. Somewhat later, a number of observatories concentrated their efforts on the region of sky indicated by LIGO and Virgo, resulting in the detection of the astronomical transient AT 2017gfo some 11 hours later \cite{Valenti:2017ngx}. This observation was made across the EM spectrum, marking the birth of a new era of multimessenger astronomy. The confirmation that the object was a rapidly cooling cloud of neutron-rich material confirmed that the GW, gamma ray and EM signals had indeed all originated from the binary merger of two neutron stars \cite{LIGOScientific:2017ync}.

%%%%%%%%%%%%%%%%%%%%%%%%%%%%%%%%%%%%%%%%%%%%%%%%%
%%%%%%%%%%%%%%%%%%%%%%%%%%%%%%%%%%%%%%%%%%%%%%%%%

As briefly discussed in the beginning of this section, the GW170817 event produced two different types of observational input of great value to determinations of the neutron-star matter EoS. First, from the GW signal measured by LIGO and Virgo, describing the inspiral phase of the merger, the collaborations were able to place stringent limits on the tidal deformability of the stars. This quantity is defined as the coefficient $\Lambda$ in the relation
\begin{equation}
    Q_{ij}=-\Lambda \mathcal{E}_{ij}
\end{equation}
that describes the induced quadrupolar moment $Q_{ij}$ of a neutron star's gravitational field due to an external quadrupolar tidal field $\mathcal{E}_{ij}$  (see, e.g., \cite{Hinderer:2009ca} for a more detailed discussion of this physical quantity). In suitable dimensionless units, LIGO and Virgo were able to place a 90\%  of $70<\tilde{\Lambda}<720$ for a so-called binary-tidal-deformability parameter $\tilde{\Lambda}$ (see sec.~\ref{sec:mergers} for more details), which is a function of the masses and tidal deformabilities of the two stars, $M_1$, $M_2$, $\Lambda(M_1)$, and $\Lambda(M_2)$ \cite{LIGOScientific:2018hze}. For GW170817, the so-called chirp mass $M_\text{chirp}=1.186M_\odot$ has been determined very accurately, so the $\tilde{\Lambda}$ constraint can be put to efficient use by varying the less constrained mass ratio $q=M_2/M_1$ and discarding EoSs that do not satisfy $70<\tilde{\Lambda}<720$ for any viable value of $q$. The same constraint can equivalently be given for the tidal deformability of a $1.4M_\odot$ neutron star, in which case it takes the form $\Lambda(1.4M_\odot)\equiv \Lambda_{1.4}=190_{-120}^{+390}$ \cite{LIGOScientific:2018cki}. This has been demonstrated to be a very strong constraint for the EoS, and has led to discarding many otherwise viable model EoSs (see, e.g., \cite{Annala:2021gom} for a recent example).

Another piece of information gathered from the multimessenger observation of the GW170817 merger event is related to the presence of an EM counterpart, including the slightly delayed gamma ray burst \cite{LIGOScientific:2017ync}. It is commonly accepted that in particular the gamma ray burst signalled the presence of gravitational collapse to a black hole, which --- knowing the chirp mass of the event --- can be used to infer an upper limit for the maximal mass of stable neutron stars. In addition, the 1.7s delay between the GW and gamma ray signals has been interpreted as an indication for the presence of an intermediate state of either a hypermassive (more likely scenario) or supramassive (less likely) neutron star, respectively supported by differential and uniform rotation  \cite{Margalit:2017dij,Rezzolla:2017aly,Ruiz:2017due}. Again, this information can be used to efficiently discriminate between different proposed neutron-star matter EoSs by testing the maximal masses supported by non-rotating neutron stars and those rotating at the highest possible (``Kepler'') frequency. Assuming that a hypermassive neutron star was created in the GW170817 event would amount to setting the condition $M_\text{remnant}\geq M_\text{supra}$, while the case of a supramassive star would lead to the more conservative  $M_\text{remnant}\geq M_\text{TOV}$, with $M_\text{TOV}$ denoting the maximal mass of nonrotating neutron stars (see, e.g., \cite{Annala:2021gom} for a detailed discussion of this issue). These two conditions enable one to discard EoSs that produce $M_\text{TOV}$ values too high for gravitational collapse to have taken place in the relevant limit.

Finally, while a postmerger ``ringdown'' signal was not recorded for GW170817 due to the decreased sensitivity of the LIGO and Virgo instruments to the high frequencies involved in this part of the waveform, overcoming this barrier in a future merger event remains an intriguing prospect. The postmerger signal carries valuable information of the violent processes involved in the evolution of the system, and could, e.g., be used to probe the existence of a first order phase transition between nuclear and quark matter. This scenario has been studied via relativistic hydrodynamic simulations e.g.~in \cite{Most:2018eaw}.

In addition to the quantities discussed above, there are a number of other neutron-star properties that may become measurable in the future and that carry important information about the inner structure of the stars. An interesting subclass of these is formed by the moment of inertia $I$, the quadrupole moment of the mass distribution $Q$ and so-called Love numbers, typically denoted by $k_n$ \cite{Hinderer:2007mb}. Curiously, there exist a number of relations between these quantities that appear to be universal in the sense that they are obeyed by (nearly) all proposed neutron-star matter EoSs \cite{Yagi:2013bca}. This universality allows one to determine the values of other quantities from the measurement of one and in addition paves the way to nontrivial tests of GR through neutron-star observations. We refer the interested reader to the extensive review article \cite{Yagi:2016bkt} for further information on this interesting topic, which is somewhat outside the scope of our current review.

%%%%%%%%%%%%%%%%%%%%%%%%%%%%%%%%%%%%%%%
\subsection{Equation of State: combining theory with observations} \label{sec:interpolation}
%%%%%%%%%%%%%%%%%%%%%%%%%%%%%%%%%%%%%%%

Before proceeding to the main topic of our review, we want to briefly examine how far the above theoretical and observational insights get us in deciphering the properties of dense QCD matter in a model-independent fashion. This question is clearly of utmost importance in identifying where holographic methods can be expected to make their most significant contributions to the field of neutron-star physics.

%%%%%%%%%%%%%%%%%%%%%%%%%%%%%%%%%%%%%%%%%%%%%%%%%
\begin{figure}[!t]
    \centering
        \includegraphics[width=17.5cm]{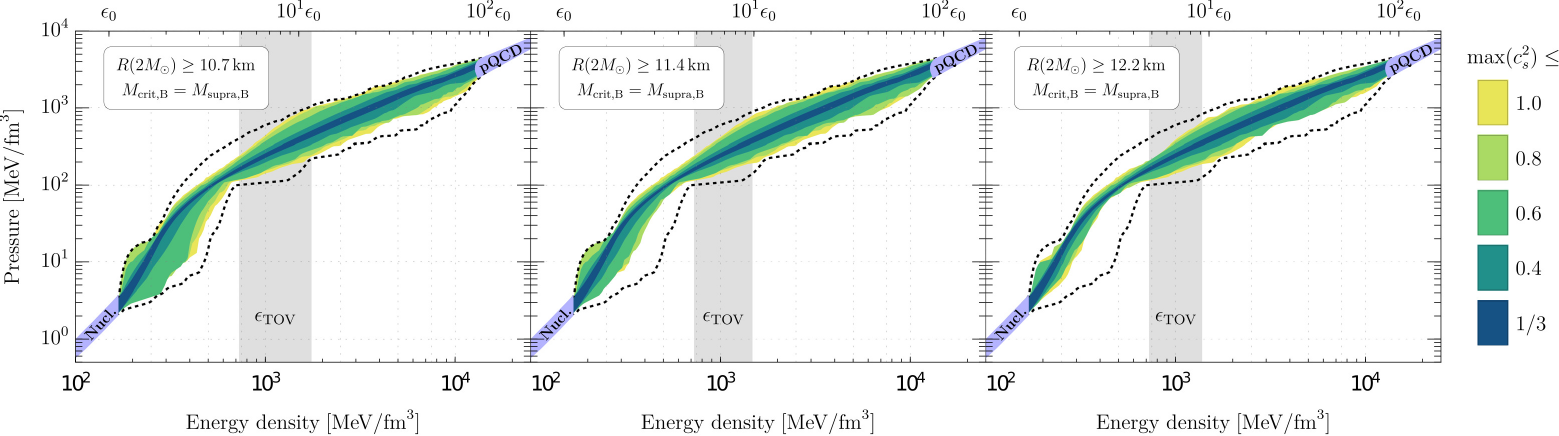} \\ \vspace{0.2cm} \includegraphics[width=17.5cm]{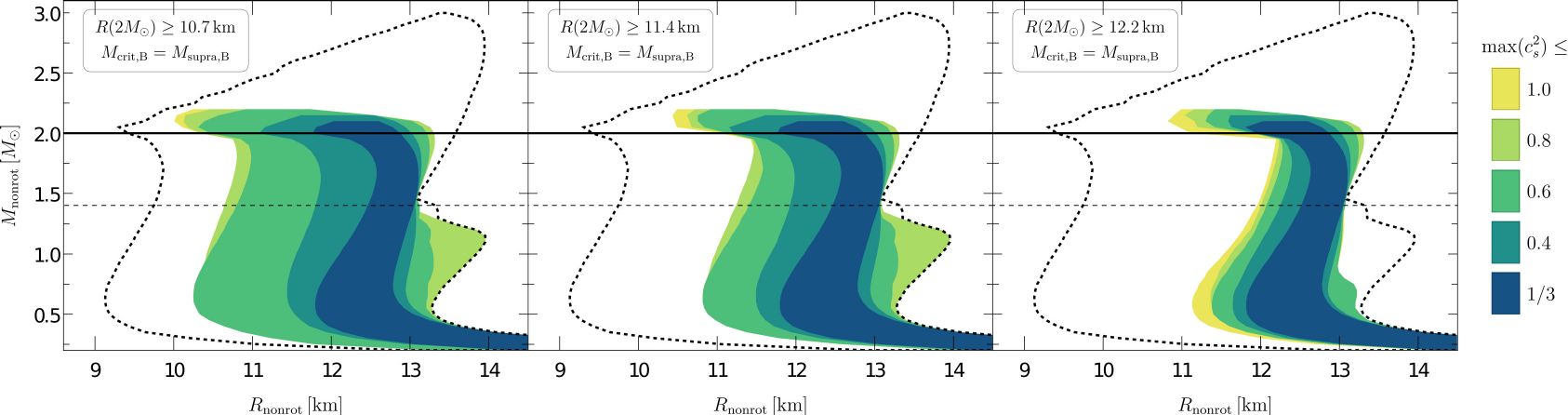} \\
    \caption{The allowed EoS and MR bands corresponding to the assumption of a hypermassive neutron star as an intermediate state in GW170817 and using the lower limits 10.7km, 11.4km, and 12.2km for the radius of a $2M_\odot$ neutron star. The figure is taken from \cite{Annala:2021gom}.}
    \label{fig:interp2}
\end{figure}
%%%%%%%%%%%%%%%%%%%%%%%%%%%%%%%%%%%%%%%%%%%%%%%%%

To put both the theoretical and observational results to optimal use, a natural way to proceed is to interpolate the EoS of dense beta-equilibrated QCD matter from low to high densities, ensuring that the CET and pQCD limits are met and that all the astrophysical constraints discussed above are satisfied. Such an approach requires the building of large ensembles of EoSs that include all possible physical behaviors of the quantity, which is most easily achieved by dividing the density interval from approx.~$n_s$ to $40n_s$ to multiple segments. On each of them, we then assume that the true EoS can be approximated by a simple ansatz function, such as a polytropic function in baryon density,  $p=a n^\gamma$, or that it can be integrated from a similarly simple ansatz for, e.g., the speed of sound squared.

Interpolation calculations of the type described above have been performed since the 2014 article \cite{Kurkela:2014vha}, which was motivated by earlier extrapolations of the CET EoS, carried out e.g.~in \cite{Hebeler:2013nza}. More recent works include \cite{Annala:2017llu,Most:2018hfd,Annala:2019puf,Annala:2021gom}, of which the most recent article \cite{Annala:2021gom} implemented the largest set of observational constraints so far. These include the existence of two-solar-mass neutron stars, the LIGO/Virgo constraints on the tidal deformability parameter $\tilde{\Lambda}$, the recent NICER results for the radius of the massive neutron star J0740+6620, and the presence of either a supramassive or a hypermassive neutron star as an intermediate state in the GW170817 merger.

In fig.~\ref{fig:interp2}, taken from \cite{Annala:2021gom}, we display EoS and MR bands corresponding to an ensemble, where one assumes the generation of a hypermassive neutron star in GW170817. In the three plots on both rows, one uses three different lower limits for the radius of PSR J0740+6620, ranging from the most conservative 2-$\sigma$ limit of the Amsterdam subgroup of NICER \cite{Raaijmakers:2021uju} to the most aggressive 1-$\sigma$ limit of the Maryland one \cite{Miller:2021qha}. Just as in fig.~\ref{fig:epsmr}, the color coding corresponds to the maximal value that the speed of sound squared reaches at any density, while the dashed lines correspond to results obtained with no information on the final state of the GW170817 merger or the radii of massive neutron stars. As these results suggest, while the EoS is starting to be rather strongly constrained, there still remains considerable freedom in particular in the crucial density interval between the CET regime (marked ``Nucl.''~in the figure) and the gray $\epsilon_\text{TOV}$ segment, where the centers of maximally massive neutron stars lie. This density range, where QCD is strongly coupled and where the deconfinement transition is expected to lie, is also clearly the one where holographic methods have the most to contribute.

%%%%%%%%%%%%%%%%%%%%%%%%%%%%%%%%%%%%%%%
%%%%%%%%%%%%%%%%%%%%%%%%%%%%%%%%%%%%%%%
\section{Holographic description of dense strongly interacting matter}\label{sec:denseholo}
%%%%%%%%%%%%%%%%%%%%%%%%%%%%%%%%%%%%%%%
%%%%%%%%%%%%%%%%%%%%%%%%%%%%%%%%%%%%%%%

Moving finally  to the main topic of our review, we will next introduce the holographic approach that will later be used to address questions concerning dense strongly interacting matter. With our ultimate focus in the physics of neutron stars, we will outline the most relevant building blocks needed to achieve a realistic framework able to describe matter in this extreme limit.

The early works applying holographic methods to describe compact stars typically constructed stellar solutions in the bulk of asymptotically anti de Sitter spacetimes \cite{deBoer:2009wk,Arsiwalla:2010bt}. The dual interpretation of these bulk solutions was in terms of phases of a degenerate Fermi gas. The activity then branched off towards condensed matter applications with such solutions better known as electron stars and clouds \cite{Hartnoll:2010gu,Hartnoll:2010ik,Puletti:2010de,Hartnoll:2016apf}. The body of work that aims to make contact with astrophysical compact stars is quite disparate from these. We will henceforth only focus on those results that use the AdS/CFT duality to model QCD-like matter at high density and do not attempt to interpret the corresponding gravity solutions in our daily life. In other words, the corresponding gravity solutions in, say, asymptotically AdS spacetimes do not have a direct relation to compact stars obtained as solutions of the TOV equations (\ref{TOVeq}) in asymptotically flat spacetimes.

%%%%%%%%%%%%%%%%%%%%%%%%%%%%%%%%%%%%%%%
\subsection{Ingredients of dense holography}\label{sec:whatisholo}
%%%%%%%%%%%%%%%%%%%%%%%%%%%%%%%%%%%%%%%

In this review we have an eye toward dense QCD. To this end, we first give a lightning review of the necessary building blocks that enable studying flavor dynamics using the gauge/gravity duality. We start by reminding the reader how the gauge/gravity duality incarnated and then gradually introduce further concepts in the field theory together with how they are dealt with on the dual gravity side. 

Let us start by recalling some basic aspects of the gauge/string correspondence. Its bedrock is the dual viewpoint of a stack of D-branes \cite{Polchinski:1995mt}. On one hand, the endpoints of open strings sweep out a hypersurface in a ten-dimensional spacetime. The dynamics of charged degrees of freedom attached to the endpoints are governed by a gauge field theory. On the other hand, D-branes themselves are massive objects and they bend the spacetime where they reside. A huge stack of D-branes can be replaced with a (classical) closed string theory living in the resulting curved geometry. The first concrete example realizing this notion is the reinterpretation of (3+1)-dimensional ${\cal N}=1$ SU($N$) super Yang-Mills (SYM) theory as a Type IIB string theory living in AdS$_5\times X^5$ equipped with self-dual five-form flux \cite{Maldacena:1997re}. If the Sasaki-Einstein manifold $X^5$ is the round five-sphere $S^5$, the dual field theory is the famous ${\cal N}=4$ superconformal SU($\Nc$) theory. Since then an infinite number of examples have been discovered; see, e.g., \cite{Martelli:2004wu}.

The gauge/string correspondence, or AdS/CFT duality, is a strong/weak duality in the sense that the gravitational theory is weakly curved when the effective coupling of the gauge theory is large and vice versa. The precise statement of the duality is that local (gauge-invariant) operators in the
field theory $\cO(x)$ map to fields $\phi(x,r)$ in the bulk gravitational theory with $r$ the radial coordinate, such that the quantum field theory (QFT) generating functional depending on a given source $\phi_0(x)$ dual to an operator $\cO$ is equal to the bulk partition function with the boundary condition $\phi(x,r\to\infty)=\phi_0(x)$ \cite{Witten:1998qj},
\be
 Z_\textrm{QFT}[\phi_0] = Z_\textrm{string\ theory}[\phi]\Big|_{\phi(r\to\infty)=\phi_0} \ .
\ee

What makes this duality useful is the remarkable feature that in the limit where the QFT possesses a large number of degrees of freedom and is strongly coupled, the dual description reduces to classical supergravity $Z_{\rm{string\ theory}}\approx Z_{\rm{SUGRA}}$. In other words, in this limit we only have to solve classical equations of motion in Einstein’s theory, perhaps including some additional fields. For a gauge field theory this limit is typically achieved when not only is $\Nc$ large, where $\Nc$ is the rank of the gauge group, but the `t Hooft coupling $\laYM = g_{\rm{YM}}^2 \Nc$ is also sizable.

\begin{figure}[t!]
\begin{center}
\includegraphics[width=0.65\textwidth]{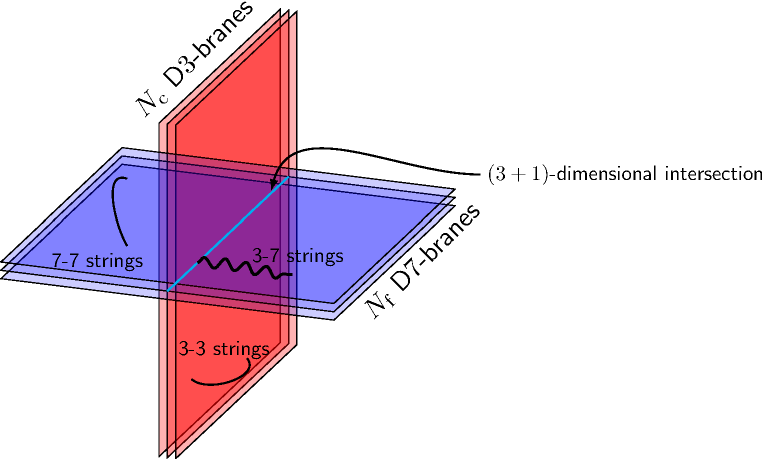}
\end{center}
\caption{Schematic illustration of the probe D3-D7 model. Open string degrees of freedom representing fundamental quenched quarks as attached to D7-branes at their endpoints. In the decoupling limit ($\Nc\to\infty$ with $\laYM$ and $\Nf$ fixed), the D3-branes are replaced by the geometry, where the adjoint degrees of freedom (3-3 strings) are represented by closed strings in the geometry and mesonic states (7-7 strings) by fluctuations of the flavor D7-branes. In this limit, the 7-7 strings do not interact with 3-7 or 3-3 strings anymore, which means that the group U$(\Nf)$ on the D7-branes becomes the global flavor group in the ambient (3+1)-dimensional SYM theory; see \cite{Erdmenger:2007cm} for more details.}\label{fig:d3d7}
\end{figure}

In order to extend the AdS/CFT correspondence to theories closer to dense QCD,  the minimal requirement is to be able to incorporate matter (flavor) degrees of freedom transforming in the fundamental representation of the gauge group. A standard procedure to incorporate flavor in the holographic correspondence is to include so-called flavor branes \cite{Karch:2002sh}. These extra branes fill the gauge theory directions and are extended also along the holographic coordinate. In the case of $\cN = 4$ SYM, whose gravity dual is obtained from the near-horizon geometry of a large stack of D3-branes, the much studied $\cN=2$ supersymmetric example entails flavor D7-branes intersecting color D3-branes in 3+1 dimensions of their worldvolume. We have depicted this particular brane intersection configuration in fig.~\ref{fig:d3d7}.

%%%%%%%%%%%%%%%%%%%%%%%%%%%%%%%%%%%%%%%%%%%%%%%%%%%%%%%%%%%%%%%%%%%%%%%%
%%%%%%%%%%%%%%%%%%%%%%%%%%%%%%%%%%%%%%%%%%%%%%%%%%%%%%%%%%%%%%%%%%%%%%%%

If the number $\Nf$ of D7-branes is small compared to the number $\Nc$ of D3-branes, one can adopt a so-called quenched approximation, in which the flavor branes are considered probes which do not alter the background geometry generated by the color branes. On the field theory side, quenching corresponds to disregarding the dynamical effects of quarks. This means that quantum effects produced by fundamentals are neglected and, accordingly, quarks are considered semi-classical dynamical sources that do not appear as closed loops in Feynman graphs. To go beyond the quenched approximation and to obtain solutions of gravity with brane sources that include the backreaction of the flavor branes when $\Nf \sim \Nc$ (see \cite{Nunez:2010sf} for a review), one generally needs to simplify the problem. One possibility is to introduce a smearing procedure which leads to a system of equations of motion with delocalized flavor sources which, in many cases, can be solved analytically. The solutions capture some important physical features which are related to the quantum effects of quarks, such as the effects of the fundamentals on the running of the couplings.

\begin{figure}[t!]
\begin{center}
\includegraphics[width=0.65\textwidth]{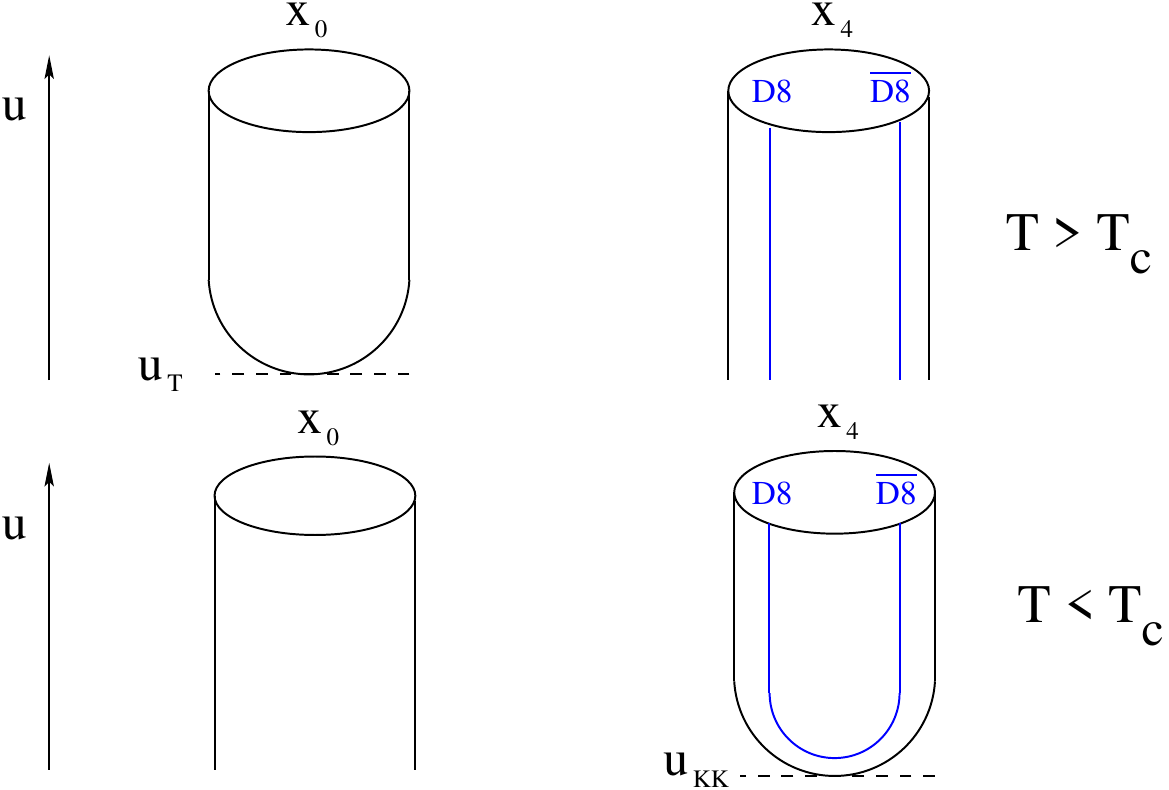}
\end{center}
\caption{The cylinder and cigar solutions in the Witten background for the deconfined ($T>T_c$) and confined ($T<T_c$) phases. Here $u$ is the holographic radial coordinate and $x_0$ and $x_4$ are the Euclidean time and the spatial directions compactified on circles of circumference $1/T$ and $2\pi/M_{KK}$, respectively. The relation between $u_{KK}$ and $M_{KK}$ can be determined by demanding the vanishing of the conical singularity of the background. The $\Nf$ probe D8-branes
($\overline{\text{D}8}$-branes) give rise to $\Nf$ right-handed (left-handed) massive quarks in the dual quantum field theory for non-antipodal brane embeddings (antipodal embedding gives massless quarks). The branes connecting in a ``U''-shaped configuration at $T < T_c$ can be interpreted as the geometrical realization of chiral symmetry breaking.}\label{fig:SSmodel}
\end{figure}

To map out the blueprint for flavor dynamics in holographic models, we recall how other physical features are encoded in the dual gravity description of the D3-D7 system. If the number of flavor branes is small $\Nf\ll \Nc$, the matter in the underlying gauge theory can be described by the dynamics of probe D7-branes in an ambient background spacetime. If we are further interested in the high temperature deconfining or plasma phase, then the spacetime geometry has a black hole \cite{Witten:1998zw}. The temperature and entropy density of the black hole are identified as the temperature and entropy density of the adjoint matter in the gauge theory, respectively. At low temperatures, and thus low energies, there can be a transition to a confining phase, with the exact details depending crucially on global symmetries and the relevant deformations of the gauge theory. In the vanilla example of probe D7-branes embedded in $AdS_5\times S^5$, corresponding to quenched fundamental matter in ${\cal N}=4$ SYM, this deconfinement-confinement transition happens at any non-zero temperature, no matter how small. For a Witten background \cite{Witten:1998zw} that is generated by D4-branes with one spatial dimension $x_4$ curled on a circle the deconfinement-confinement phase transition is happening at a temperature of the order of the inverse size of the radius of the circle. In the dual gravity side this is marked by the Hawking-Page transition from a geometry with a black hole to one without. The two geometries are doubly Wick-rotated cousins, where in the Euclidean geometry time $x_0$ is periodic with a radius equal to the inverse temperature. The circle of smaller radius collapses in the interior, so the transition is happening right when the two circles are of the same size; see figure~\ref{fig:SSmodel} for illustration.  Adding flavor in the Witten background by D8-$\overline{\textrm{D}8}$-brane pairs corresponds to the well-studied Sakai-Sugimoto model \cite{Sakai:2004cn,Sakai:2005yt,Rebhan:2014rxa}, but so long as the D8-brane pairs are treated as probes they will not affect the temperature, at which the deconfinement phase transition takes place. On the contrary, if the flavor branes do backreact on the geometry, they also alter the transition in a non-trivial way. From now on, when we have the deconfining phase of the gauge theory in mind, we associate this to the dual geometry with a black hole in it, irrespective of whether flavors are treated in a quenched or unquenched manner.

%%%%%%%%%%%%%%%%%%%%%%%%%%%%%%%%%%%%%%%%%%%%%%%%%%%%%%%%%%%%%%%%%%%%%%%%
%%%%%%%%%%%%%%%%%%%%%%%%%%%%%%%%%%%%%%%%%%%%%%%%%%%%%%%%%%%%%%%%%%%%%%%%

To describe matter at nonzero density in holography, one needs a non-trivial U(1) temporal gauge potential $A = A_t dt$ in the bulk theory. Per standard AdS/CFT dictionary, the chemical potential in the boundary theory is identified as the boundary value of this gauge field:\footnote{To be precise, this is for a vanishing value $A_t$ at the horizon; otherwise $\mu$ is the integral of the electric flux along the radial direction.}
\be
 A_t(r\to\infty) = \mu \ .
\ee
The sources of the gauge field are open string endpoints in the bulk and they can be either cloaked behind the black hole horizon or diluted on the worldvolume of the flavor branes.

We have now introduced two dimensionful quantities $T$ and $\mu$, with which we could already discuss the phases of the theories of interest. However, one can in addition break conformal invariance at the outset either explicitly, e.g.~by introducing masses for the fundamentals, or spontaneously, by not introducing any additional sources but making sure that  it is energetically more favorable to have non-trivial field profiles in the bulk spacetime that near the boundary give rise to non-zero vacuum expectation values for corresponding operators. For example, one can have phases with spontaneously broken chiral symmetry $\langle\bar\psi\psi\rangle\ne 0$ without the introduction of a bare mass for the fundamentals. Such a scenario can take place if the dual background geometry has an intrinsic energy scale such as the size of the circle in the above-mentioned Sakai-Sugimoto model \cite{Sakai:2004cn,Sakai:2005yt,Rebhan:2014rxa} or via an engineered logarithmic running of the coupling constant in the so-called improved holographic QCD \cite{Gursoy:2007cb,Gursoy:2007er} (see also \cite{Gubser:2008ny}) and its flavored incarnation V-QCD \cite{Jarvinen:2011qe}. In the latter, the letter V stands for the Veneziano limit of $\Nf/\Nc=\rm{fixed}$ while $\Nf,\Nc\to\infty$.

%%%%%%%%%%%%%%%%%%%%%%%%%%%%%%%%%%%%%%%
\subsection{Overview of different models for dense quark matter}\label{sec:holomodels}
%%%%%%%%%%%%%%%%%%%%%%%%%%%%%%%%%%%%%%%

Given the more general introduction to salient details of holography and the necessary ingredients to make contact with dense QCD, let us be more detailed in this subsection. We will also list, as well as explain the differences, of all models that have been used in the context of the {\emph{cold and dense}} environment typical of neutron-star interiors.

The dynamics of $\Nf$ flavor degrees of freedom in the field theory stem from studying a collection of $\Nf$ D$q$-brane actions of the type
\be\label{eq:Dpflavor}
 S_\mt{f} = T_q \int_{\Sigma_{q+1}} d^{q+1}\xi e^{-\phi}\sqrt{-\det(\hat g_{\mu\nu}+\hat B_{\mu\nu}+2\pi\alpha' F_{\mu\nu})} + T_q\int_{\Sigma_{p+1}} \sum_{n}\hat  C_{n}\wedge e^{\hat B + 2\pi\alpha' F} \ , 
\ee
where the Regge slope is related to the string length as $\alpha'=\ls^2$, the tension of the brane reads $T_q=(g_s(2\pi)^q \alpha'^{(q+1)/2})^{-1}$, and the multiplicative closed string coupling $g_s$ is related with the additive normalization of the dilaton field \cite{Polchinski:1998rr}. This action should be considered in the presence of gluonic degrees of freedom represented by Type IIA/IIB supergravity actions. In other words, the actions (\ref{eq:Dpflavor}) are treated as sources, feeding to the right-hand side of the Einstein field equations.

The first term in (\ref{eq:Dpflavor}) is the Dirac-Born-Infeld (DBI) action \cite{Dai:1989ua,Leigh:1989jq} and the second term is the Wess-Zumino (WZ) action \cite{Polchinski:1995mt}; see also lecture notes \cite{Polchinski:1996na,Bachas:1998rg}. Here the hats denote pullbacks of the (bulk) fields onto the $(q+1)$-dimensional worldvolume $\Sigma_{q+1}$ of the D$q$-brane with coordinates $\xi_i$, $i=0,\ldots,q$, while in the WZ term one restricts the sum involving the Ramond-Ramond potentials $C_n$, field strengths $F$, and NS-NS two-form fields $B$ in such a way that the resulting integrand is a $(q+1)$-form. There could also be curvature corrections entering in the WZ action  \cite{Bachas:1998rg}. In the rest of this subsection, the WZ term does not play any role as we are not considering nonzero magnetic fields. However, in Sec.~\ref{sec:holomodelsmore} a WZ term of this form is important, since one is interested also in solitonic solutions in the worldvolume of D-branes.

Of course, even in the absence of an NS-NS $B$-field and vanishing gauge fields $F=0$, no general solution to the Euler-Lagrange equations stemming from a collection of actions (\ref{eq:Dpflavor}) in the ten-dimensional supergravity background exist. This is due to various reasons: any non-coincident brane distribution leads to partial differential equations with Dirac delta functions appearing in, {e.g.}, Einstein equations, and for the non-Abelian $F\ne 0$ case of a coincident D$q$-brane stack the very definition of the DBI action is ambiguous (related to different possible trace prescriptions over flavor indices), to name a few. However, progress can be made by invoking different simplifications, or combinations thereof: 1) by considering a parametrically small number of flavor branes, the so-called probe limit $\Nf\ll \Nc$, in which case one can ignore the backreaction on the background geometry; 2) by considering only Abelian gauge fields and coincident brane distributions, in which case one can essentially take a sum of (\ref{eq:Dpflavor}); 3) by expanding the non-Abelian actions (\ref{eq:Dpflavor}) to low orders in $F^2$, the trace prescription of which becomes unambiguous; 4) by smearing the D$q$-branes over the transverse directions which renders the PDEs to ODEs; or 5) by ignoring the internal geometry, which we think of as a string-inspired bottom-up approach.

Remarkably, many of the  approaches mentioned above can be described in a unified manner \cite{Hoyos:2020hmq,Hoyos:2021njg}. An action that combines the different approaches consists of two separate pieces
\be\label{eq:ActionWithTachyon}
 S_\mt{total} = S_\mt{g} + S_\mt{f} \ ,
\ee
where $S_\mt{g}$ denotes the glue action and $S_\mt{f}$ is the flavor action. These two take the explicit forms
\bea
S_\mt{g} & = & \frac{1}{2\kappa_5^2}\int \d^5 x \sqrt{-g} \left( R - \frac{1}{2} \partial_\rho \phi \, \partial^\rho\phi - \pot(\phi,\chi) \right)  \label{eq:ActionDefsGluon} \\
S_\mt{f} & = & - \frac{\tens}{2\kappa_5^2} \int \d^5 x\, \Z(\phi,\chi) \sqrt{-\det \left( g_{\mu\nu} + \k(\phi,\chi) \, \partial_\mu \chi \, \partial_\nu\chi + \W(\phi,\chi) \, F_{\mu\nu} \right) } \ ,\label{eq:ActionDefsFlavor}
\eea
where $R$ is the 5D Ricci scalar, $\tens$ is the tension of the branes $\propto \Nf/\Nc$, and $\phi,\,\chi$ are scalar fields. The potential functions $V,\,\Z,\,\kappa,\,\W$ depend on these scalars and their exact interpretation varies with the model in question. Roughly, the field $\phi$ corresponds to the dilaton field which maps to the coupling constant in the field theory, while $\chi$ maps to the operator $\bar\psi\psi$, a non-trivial behavior of which leads to non-trivial chiral dynamics. We note that this action is written directly in five-dimensional language. To cast a model in this form requires a careful dimensional reduction down to five dimensions. In \cite{Hoyos:2020hmq,Hoyos:2021njg} this was obtained for the D3-D7 probe model. Explicitly,
\be
\textrm{D3-D7} : \qquad  \frac{\tens}{2\kappa_5^2} = 2 \pi^2\,  T_7 \,  \Nf \ , \ \phi=0 \ , V = -\frac{12}{L^2} \ , \ \W = 2 \pi \alphap \ , \ \k=1 \ , \ \Z = L^3\,\cos^3 \chi \label{eq:D3D7functions} \ ,
\ee
with the relation to the 't Hooft coupling $\laYM = 4\pi g_s \Nc$ taking the form $\frac{L^4}{\ls^4} = \laYM$. The scalar field $\chi=\chi(r)$ depends on the radial direction and can be identified with the azimuthal angle $\theta(r)$ of the internal $S^3$ space. The other radially dependent field is the temporal gauge potential $F_{rt} = A_t'(r)$,  which is crucial to model non-zero density of flavors in the field theory. The probe D3-D7 system at finite baryon density was first studied in \cite{Kobayashi:2006sb,Mateos:2007vc}.

There is also a refinement of the probe D3-D7 model, which adds a phenomenological component  \cite{Evans:2011eu,BitaghsirFadafan:2019ofb}. The idea there is to have a running quark mass to give rise to a richer phase diagram that would resemble QCD better. The approach includes an effective dilaton that controls the running of the anomalous dimension of the quark bilinear. Interestingly,  this results in a chirally broken finite density phase which appears at intermediate chemical potentials. Similar chirally broken deconfined phases are also possible in V-QCD but this phase turns out to be subdominant at intermediate densities when the V-QCD model is carefully tuned to match onto lattice QCD data at vanishing chemical potential.

The D3-D7 model as laid out above can only be trusted up to first order in a parameter $\tens\propto\Nf/\Nc$.  Since the dual field theory for AdS$_5\times S^5$ without flavor is conformal, the supposition is that the addition of extra degrees of freedom will render the beta function positive. This indeed happens and leads to a pathological UV limit signaled by the appearance of a Landau pole. On the gravity side, this amounts to complications, too, as one needs to set up the holographic dictionary at a finite cut-off surface \cite{Bigazzi:2009bk,Bigazzi:2011it,Bigazzi:2011db,Bigazzi:2013jqa,Faedo:2016cih}. Nevertheless, similarly to QED, one can have meaningful discussions of the IR physics even with pathological UV behavior assuming that a scale separation exists. This scale separation is controlled by the combination $\laYM \Nf/\Nc$, i.e.~the weight factors for the internal flavor loops. This implies that there would be an obstruction to considering $\Nc$ and $\Nf$ to be of the same order. One can, however, determine the effects of unquenched quarks via an expansion in $\laYM \Nf/\Nc$. Indeed, several works have considered backreacted D3-D7 systems in the smeared approximation \cite{Bigazzi:2009bk,Bigazzi:2011db,Magana:2012kh,Faedo:2016cih} also at finite density \cite{Bigazzi:2011it,Cotrone:2012um,Bigazzi:2013jqa,Faedo:2017aoe}. This is an important road to follow as the quenched approximation is expected to break down at low temperatures and finite density \cite{Mateos:2011bs,Faedo:2017aoe}.

Besides flavoring the AdS$_5\times X^5$ with $X^5=S^5$ in the D3-D7 model discussed above, the top-down flavored configurations resembling QCD have been considered in the confining cascading Klebanov-Witten theory \cite{Klebanov:1998hh} with D3-branes placed not in flat space but at the tip of the Calabi-Yau cone over the 5d Sasaki-Einstein manifold $X^5=T^{1,1}$. The conifold can be resolved and deformed, the corresponding theories are called Klebanov-Tseytlin \cite{Klebanov:2000nc} and Klebanov-Strassler \cite{Klebanov:2000hb} theories. The works studying flavors in these geometries include quenched approximations with probe D7-branes \cite{Karch:2002sh,Ouyang:2003df,Kuperstein:2004hy,Sakai:2003wu,Dymarsky:2009cm} as well as extensions to unquenched massless flavors in  \cite{Benini:2006hh,Benini:2007gx,Bigazzi:2009bk,Bigazzi:2011db} and massive ones in \cite{Bigazzi:2008qq,Bigazzi:2008zt,Bigazzi:2009bk}, where one uses the smearing approximation to backreact the D7-branes \cite{Nunez:2010sf}. 

So far, none of the backreacted top-down constructions have been considered in the high density and small temperature context of neutron stars. In principle, one could make general remarks, if one were able to write down the dimensionally reduced action in the form of eq.~(\ref{eq:ActionWithTachyon}). Alas, dimensional reduction down to five dimensions of backreacted systems is not easy, especially when the unquenched flavors are massive. Moreover, to date no examples exist where massive flavors are treated at finite temperature, let alone at also finite density.

Another popular approach in holographic QCD is to consider pairs of D8-$\overline{\text{D8}}$-branes in the Witten background, {i.e.}, the Sakai-Sugimoto model introduced earlier. The dual field theory in this case is five-dimensional, and there is no UV fixed point. Besides this, the Sakai-Sugimoto model is probably the closest top-down holographic model to planar QCD, as one can describe in very simple geometric terms many physical properties such as confinement, chiral symmetry breaking, and the confinement-deconfinement phase transition \cite{Rebhan:2014rxa}. However, since the model is intrinsically non-supersymmetric, finding a backreacted solution to the geometry is complicated. Works in this direction include \cite{Burrington:2007qd,Bigazzi:2014qsa}, though so far only at small baryon densities. To this end, all works at high density feature with quenched quarks. In addition to this, to describe deconfined quark matter, one needs to consider an uncontrolled extrapolation from the high temperature phase down to small temperatures, if one is interested in quark matter in the regime relevant for neutron stars. On the contrary, however, one can directly consider baryons within the Sakai-Sugimoto model at vanishing temperature. This leads to very realistic equations of state for the nuclear matter phase and thereby also to realistic compact stars \cite{Kovensky:2021kzl} as will be later discussed in this review.

In a bit more detail, the Sakai-Sugimoto model consists of $\Nc$ D4-branes in Type IIA string theory wrapping a circle with antiperiodic boundary conditions for fermions, and then $\Nf$ probe D8-branes at a point on the circle as well as $\Nf$ probe $\overline{\text{D8}}$-branes at another point on the circle. At energies well below the Kaluza-Klein scale, the spectrum on the D4-branes is precisely that of massless four-dimensional “QCD”  SU$(\Nc)$ Yang-Mills theory with $\Nc$ colors of gluons and $\Nf$ flavors of quarks. The holographic dual geometry description emerges as usual in the large-$\Nc$ limit, in which case the D4-branes are replaced by their near-horizon supergravity background. The action for the probe branes in this case is (the collection of) DBI actions in (\ref{eq:Dpflavor}) with $q=8$.  One crucial difference to the probe D3-D7 model is that the dilaton is not constant but depends on the holographic radial coordinate. It actually grows and diverges towards the boundary of spacetime which is the manifestation of the fact that there is no UV fixed point in the boundary field theory. The Sakai-Sugimoto model at finite baryon density has been considered, {e.g.}, in \cite{Bergman:2007wp,Horigome:2006xu,Davis:2007ka,Rozali:2007rx,Kim:2007zm,deBoer:2012ij,Li:2015uea,BitaghsirFadafan:2018uzs,Kovensky:2021kzl} and at finite isospin density, {e.g.}, in \cite{Parnachev:2007bc,Aharony:2007uu,Kovensky:2021ddl}.

In addition to the above, also other probe D$p$-D$q$ systems \cite{Arean:2006pk,Ramallo:2006et,Myers:2006qr,Mateos:2007vn,Jokela:2015aha,Itsios:2016ffv} have been investigated in the context of high density and low temperature, with results applied to the physics of compact stars \cite{Burikham:2010sw,Kim:2014pva}. As long as $p\ne 3$, similarly to the Sakai-Sugimoto model ($p=4$), the dual bulk spacetime is not asymptotically anti de Sitter, {i.e.}, there is no UV fixed point in the field theory. Nevertheless, such D$p$-D$q$ configurations provide a useful testing ground for ideas born out of other holographic frameworks, and, of course, vice versa.

Let us then switch to models which are more phenomenological and hence further relax the  string theory constraints. The most developed of these is arguably the V-QCD model. Being five-dimensional, this model assumes the form of the action (\ref{eq:ActionWithTachyon}) directly. In the chirally symmetric situation, this action reads
\be\label{eq:VQCDaction}
 S_\mt{V-QCD}   =  \frac{1}{2\kappa_5^2}\int \d^5 x \sqrt{-g} \left( R - \frac{4}{3} \frac{(\partial \la)^2}{\la^2}  - \pot(\la) \right)   - \frac{\Nf}{2\Nc\kappa_5^2} \int \d^5 x\, V_{{\mt{f}}0}(\la) \sqrt{-\det \left( g_{\mu\nu}  + \W(\la) \, F_{\mu\nu} \right) } \ . 
\ee
Note that here the scalar $\phi$ has been redefined, $\lambda = e^{\sqrt{3}\phi/\sqrt{8}}$, since in this normalization the source term of the dilaton $\lambda$ at the boundary matches the 't Hooft coupling in QCD at two loops. The dilaton here depends on the holographic coordinate, which corresponds to the running coupling in V-QCD, while in the D3-D7 model above it is constant. Moreover, unlike in the D3-D7 model, here we assume the limit $\Nc,\Nf\to\infty$ while keeping $\tens=\Nf/\Nc$ fixed, which means that the flavor part (second term) is fully backreacted with the gluon sector (first term). The V-QCD model without flavors is called improved holographic QCD (IHQCD) which is a gravity dual to pure Yang-Mills theory in the large-$\Nc$ limit \cite{Gursoy:2007cb,Gursoy:2007er,Gursoy:2010fj}.

As mentioned above, in (\ref{eq:VQCDaction}) we have set the (tachyon) scalar field $\chi$ appearing in (\ref{eq:ActionDefsFlavor}) to zero. This field is dual to the chiral condensate $\bar \psi\psi$, whose boundary value is related to the masses of the quarks; see \cite{Bigazzi:2005md,Casero:2007ae,Iatrakis:2010zf,Iatrakis:2010jb,Jarvinen:2015ofa} for discussions specifically in the V-QCD and \cite{Bergman:2007pm,Dhar:2007bz,Dhar:2008um,Jokela:2009tk} in the Sakai-Sugimoto models, respectively. In the Sakai-Sugimoto model, this amounts to considering non-antipodal D8-brane embeddings.

The equations of motion following from (\ref{eq:VQCDaction}) are downright gross, and there are no known analytic solutions given the rather complicated potentials which will be discussed below. One can analyze the EoM asymptotically in the IR part of the geometry \cite{Hoyos:2021njg} and fields close to the boundary to yield observables \cite{Alho:2012mh,Alho:2013hsa}. Only the temporal gauge field $A_t$ can be eliminated from the system as in D3-D7 model, as it is a cyclic coordinate in the action. In the limit of small temperatures, keeping $\mu$ finite, the IR part of the geometry resembles a small black hole, and precisely at the $T=0$ point there is an AdS$_2$ fixed point; see \cite{Hoyos:2021njg} for a discussion of the solution space thereof and \cite{Alho:2013hsa} for a discussion of the fluctuations.

%%%%%%%%%%%%%%%%%%%%%%%%%%%%%%%%%%%%%%%%%%%%%%%%%%%%%%%%%%%%%%%%%%%%%%%%
\begin{figure}[t!]
\begin{center}
\begin{tabular}{cc}
\includegraphics[width=0.42\textwidth]{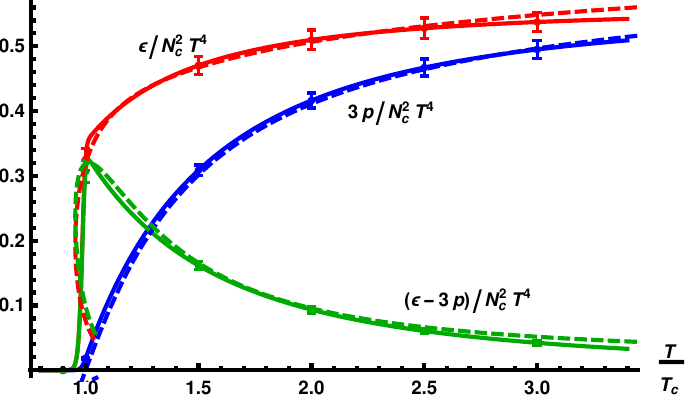}  & \includegraphics[width=0.42\textwidth]{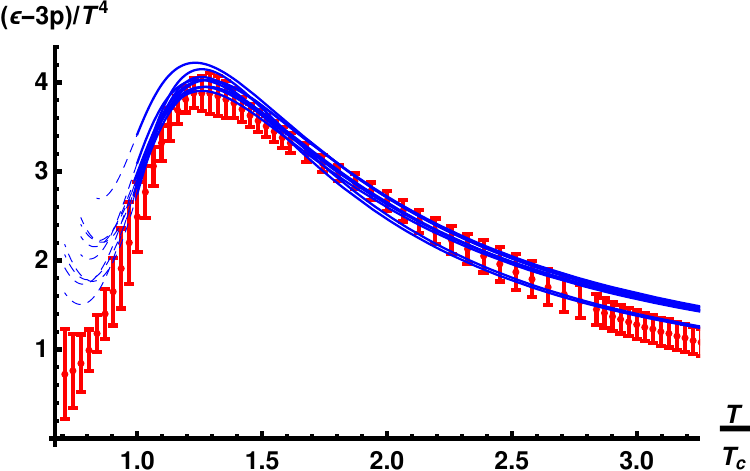}
\end{tabular}
\end{center}
\caption{Fitting $V(\la)$ to  large-$\Nc$ pure Yang-Mills lattice data for thermodynamic quantities (left) and $V_{f0}(\la)$ to QCD lattice data for the interaction measure (right). The red, blue, and green curves on the left correspond to the energy density, pressure, and interaction measure, respectively. On the left, the solid curves and error bars stand for the lattice data \cite{Panero:2009tv}, and the dashed curves for fits. On the right, the red dots and error bars show the lattice data \cite{Borsanyi:2013bia}, while blue curves are the fits. Both figures are adapted from \cite{Jokela:2018ers}.}\label{fig:VQCDthermofits}
\end{figure}
%%%%%%%%%%%%%%%%%%%%%%%%%%%%%%%%%%%%%%%%%%%%%%%%%%%%%%%%%%%%%%%%%%%%%%%%

\begin{figure}[t!]
\begin{center}
\begin{tabular}{cc}
\includegraphics[width=0.5\textwidth]{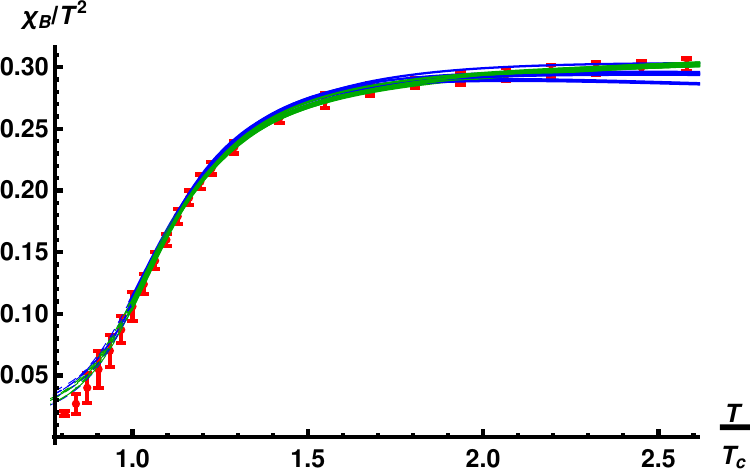} 
\end{tabular}
\end{center}
\caption{Fitting $\W(\la)$ to QCD lattice data \cite{Borsanyi:2011sw} for the baryon number susceptibility $\chi_B=(\partial^2 p/\partial\mu^2)|_{\mu=0}$ in V-QCD. The red dots and error bars show the lattice data, and green and blue curves are the fits. The figure is adapted from \cite{Jokela:2018ers}.}\label{fig:VQCDthermofits2}
\end{figure}

The potentials in (\ref{eq:VQCDaction}) are chosen to match several qualitative and quantitative features of QCD, namely, confinement and asymptotic freedom both for weak \cite{Jarvinen:2011qe,Gursoy:2007cb} and strong \cite{Ishii:2019gta,Jarvinen:2011qe,Gursoy:2007er,Arean:2013tja,Jarvinen:2015ofa,Arean:2016hcs} Yang-Mills coupling. The selection of the analytic forms of the potentials is explained at length in a concurrent review \cite{Jarvinen:2021jbd}, and their precise definitions are collected in Appendix C of \cite{Hoyos:2021njg}. It is very crucial to keep in mind here  that while the analytic forms of the potentials are very stiff, there are many free parameters due to the model being phenomenological. The remaining degrees of freedom are heavily constrained by a precise fit to lattice data of the thermodynamics in the glue sector \cite{Gursoy:2009jd} and in the flavor sector \cite{Jokela:2018ers} in pure Yang-Mills and (2+1) flavor QCD theories, respectively. In fig.~\ref{fig:VQCDthermofits}, we show the results of the fitting to the thermodynamics of gluon and flavor data, while in fig.~\ref{fig:VQCDthermofits2} we display the fitting to the baryon number susceptibility. Especially the latter case is important in order to have a meaningful base to expect reasonable results when extrapolating to higher densities. The resulting EoS for quark matter is not completely unambiguous as the fitting at small chemical potentials is not restrictive enough; in particular, the normalization of the gauge field is not determined by the first cumulant $\chi_B=(\partial^2 p/\partial\mu^2)|_{\mu=0}$ \cite{Chesler:2019osn}. The spread of all realistic possibilities is, however, well represented by the choice of three distinct potentials, which will in the following be called soft (Pot.~{\bf{5b}}), intermediate (Pot.~{\bf{7a}}), and stiff (Pot.~{\bf{8b}}) \cite{Jokela:2018ers}. What we mean by realistic possibilities has to do with comparison to known theoretical and astrophysical bounds on the EoS at finite densities but very low temperatures, to which we return in the next section.

%%%%%%%%%%%%%%%%%%%%%%%%%%%%%%%%%%%%%%%%%%%%%%%%%%%%%%%%%%%%%%%%%%%%%%%%
%%%%%%%%%%%%%%%%%%%%%%%%%%%%%%%%%%%%%%%%%%%%%%%%%%%%%%%%%%%%%%%%%%%%%%%%

Other well-studied bottom-up models for QCD are based on Einstein-Maxwell-scalar theories \cite{Gubser:2008ny,DeWolfe:2010he,Critelli:2017oub}. They can be viewed as Taylor expanded versions of eq.~(\ref{eq:ActionWithTachyon}) with vanishing $\chi=0$, as one expands the flavor brane action up to second order in the gauge field strength $F^2$. In a string theory setting, this means that one expands up to second order in $\alpha'=\ls^2$; to exemplify, see the action for the D3-D7 setup (\ref{eq:D3D7functions}). This exercise would lead to the action 
\be\label{eq:EMDaction}
 S = \frac{1}{2\kappa_5^2}\int \sqrt{-g}\left( R -\frac{1}{2}\partial_\mu\phi\partial^\mu\phi-V(\phi)-f(\phi)F^2/4\right) \ .
\ee
It is worth noting that this action has also been fitted to lattice data \cite{Gubser:2008ny}. The refined version of the original model introduced in \cite{Gubser:2008ny} is given in \cite{Grefa:2021qvt}, with the potentials appearing in (\ref{eq:EMDaction}) reading
\bea
 V(\phi) & = & -12\cosh(0.63\phi)+0.65\phi^2-0.05\phi^4+0.003\phi^6 \\
 f(\phi) & = & \frac{1}{(1+c_3)\cosh(c_1\phi+c_2\phi^2)}+\frac{c_3}{(1+c_3)\cosh(c_4\phi)} \ ,
\eea
where $c_1=0.27,c_2=0.40,c_3=1.70,c_4=100$, and $\kappa_5^2 = 11.56$. Other parametrizations for the potentials also lead to good fits to lattice data \cite{Knaute:2017opk,Cai:2022omk}. However, the low-temperature and high-density regime has received little attention, with a notable exception being \cite{Mamani:2020pks}, where $f(\phi)=1$ is used.

All of the above models deal with spatially homogeneous and isotropic phases, which are viewed as duals of a deconfined quark matter phase. One other underlying assumption is that quark matter is considered to be unpaired, {i.e.}, we do not expect any pairing mechanism to be present. It is, however, quite likely that at extremely high densities and low temperatures the dominant phase of holographic quark matter would be something resembling the color superconducting phases expected to occur in real QCD \cite{Alford:1998mk,Brambilla:2014jmp,Alford:2007xm}. In holography, several works have addressed this extension \cite{Chen:2009kx,Basu:2011yg,Rozali:2012ry,BitaghsirFadafan:2018iqr,Faedo:2018fjw,Henriksson:2019zph,Henriksson:2019ifu,Faedo:2019jlp,Ghoroku:2019trx,BitaghsirFadafan:2020otb}, and they, too, can  be roughly divided to two categories: top-down and bottom-up. The mechanism behind obtaining paired phases, which we call finite-density Higgs phases, is to envision pulling out a substack of color branes from the original stack of $\Nc$ D3-branes. This pulling out gives rise to gluon masses and should happen spontaneously, {i.e.}, when the baryon chemical potential is large enough so that the D3-branes are nucleated at a finite value of the bulk radial coordinate \cite{Henriksson:2019zph,Henriksson:2019ifu}. This position is related to the pairing gap. The mechanism higgses the gauge group SU$(\Nc)\to $SU$(\Nc-n)\times $U$(n)$ for $n$ coincident branes localized outside the original stack, so there are new phases with spontaneously broken color symmetry. One should note, however, that this type of a phase is not to be understood to be on equal footing with the standard description of a color superconducting phase in QCD, because brane nucleation can occur at large chemical potential even without any clearly distinguished baryonic or quark degrees of freedom. It is also not clear if one should invoke descriptions involving Fermi surfaces at strong coupling.
In both cases, though, there is a spontaneous breaking of the gauge group. In bottom-up scenarios, one adds complex non-color singlet scalar fields to the Einstein-Maxwell actions ({\ref{eq:EMDaction}) \cite{BitaghsirFadafan:2018iqr,Ghoroku:2019trx}, and at low enough temperatures these scalars develop vacuum expectation values.

%%%%%%%%%%%%%%%%%%%%%%%%%%%%%%%%%%%%%%%
\subsection{Holographic models for nuclear matter}\label{sec:holomodelsmore}
%%%%%%%%%%%%%%%%%%%%%%%%%%%%%%%%%%%%%%%

Having discussed the deconfined quark matter phase above, let us next briefly explore the nuclear matter phase and simultaneously give an overview of how nucleons are treated in holography. Nucleons, as all baryons, are heavy objects in the large-$\Nc$ limit, and their treatment is not completely straightforward.  To understand the asymmetry and large hierarchy between mesonic and baryonic states in string theory, let us first recall how finite charge density emerges. 

Above, we explained that to have nonzero electric flux, this quantity needs to be sourced by some objects in the bulk. We can think of them as open string endpoints on D-branes. In a typical situation, and in the applications we have in mind in this review, as the two endpoints carry opposite charges, the strings should be stretched between two different D-branes to lead to a state of nonzero density. The situation in the above examples corresponds to cases where one of the endpoints of the open string is on a flavor D$q$-brane while the other endpoint is hidden behind the horizon. The field theory interpretation of this is a heavy quark traversing a plasma. At finite density, the open strings will pull the flavor brane down to the horizon, and all sources of electric flux are cloaked \cite{Kobayashi:2006sb}.\footnote{In the presence of Wess-Zumino terms, the brane may stay above the horizon as a magnetic field can effectively eat up the electric flux leading to a vanishing electric displacement field at the lower end of the flavor brane. This leads to the existence of finite-density incompressible states with a gap for charged excitations even in the deconfining phase \cite{Bergman:2010gm}.}

Another possibility for the open string endpoints is to find another D-brane that wraps all \cite{Witten:1998xy,Imamura:1998gk} (or some \cite{Camino:2001at}) internal directions of the geometry. In the field theory directions, this brane looks pointlike and is called a ``baryon vertex'' \cite{Witten:1998xy,Callan:1998iq,Callan:1999zf,Brandhuber:1998xy,Craps:1999nc,Camino:1999xx}. It is a dynamical problem to determine where the brane dual to the baryon vertex wishes to settle down: to sit on a flavor brane, drop through the horizon, or lie somewhere in between. Each baryonic brane has $\Nc$ strings attached to it, but a random configuration is not a solution of the equations of motion. There is a net force \cite{Callan:1999zf} that pulls the baryons onto the flavor D$q$-brane, and the wrapped brane renders into a soliton configuration in the worldvolume \cite{Sakai:2004cn,Hata:2007mb,Hashimoto:2008zw,Seki:2008mu}. This solution can be obtained in holographic models when the WZ term (\ref{eq:Dpflavor}) is nonvanishing.\footnote{For example, in the Sakai-Sugimoto model the RR potential $\hat C_3\ne 0$ leads, after partial integration and subsequent integration over the internal directions, to a topological term $\int_{\Sigma_9}\hat C_3\wedge F^3\propto -\int_5 A\wedge F\wedge F$ which thereby sources the baryon charge.} This setup will then correspond to  baryonic matter on the field theory side. 
For a recent in depth review especially on the holographic treatment of baryonic matter, see \cite{Jarvinen:2021jbd}.

In addition to being non-perturbative objects (recall that the tension of the branes is $\propto 1/g_s$), holographic baryons are also parametrically heavy because they are bound states of $\Nc$ strings \cite{Witten:1998xy,Imamura:1998gk}. In the string theory description at large-$\Nc$, however, mesonic operators are constructed from light fields: they correspond to small fluctuations of flavor branes and can be described in terms of single open strings. There is therefore a large mass hierarchy between mesons and baryons in the standard holographic description, which is an artifact of the large-$\Nc$ limit and needs to be borne in mind, recalling that such a hierarchy is not present in real QCD \cite{Sonnenschein:2016pim}. Interestingly, there is an alternative large-$\Nc$ limit that one could consider to construct baryonic operators \cite{Hoyos-Badajoz:2009zmh}, which alleviates this tension \cite{Hoyos:2016ahj}, but it has not yet been considered in the neutron-star context.

Let us return back to the usual large-$\Nc$ limit and consider baryons as solitonic “instanton” configurations of gauge fields in the bulk, similar to BPST instantons in Yang-Mills theory \cite{Belavin:1975fg,Witten:1998xy}. The disparity is that the time coordinate is replaced by the holographic coordinate $r$ so that the baryon is a soliton (localized in space) rather than an instanton (localized in time). Baryons treated this way have been studied in the bottom up QCD model \cite{Pomarol:2008aa,Panico:2008it} and in the Sakai-Sugimoto model \cite{Hong:2007kx,Hata:2007mb,Hong:2007dq,Kim:2007zm,Hashimoto:2008zw,Pinkanjanarod:2020mgi,Pinkanjanarod:2021qto,Burikham:2021xpn} as well with corrections beyond the instanton solution therein \cite{Cherman:2009gb,Cherman:2011ve,Bolognesi:2013nja,Rozali:2013fna}. Multi-soliton solutions have also been considered in the Sakai-Sugimoto model \cite{Kaplunovsky:2012gb,deBoer:2012ij,Kaplunovsky:2015zsa,Preis:2016fsp,Baldino:2017mqq,BitaghsirFadafan:2018uzs}.
The holographic solitons are topologically protected and carry the same winding number $\Pi_3(\rm{SU}(\Nf))=\mathbb{Z}$ as the skyrmions, in fact if one writes down the action for the wrapped branes one finds that it bares close resemblance to the Skyrme model \cite{Sakai:2004cn,Sakai:2005yt}.

Within Sakai-Sugimoto, a rough model for physics of QCD at finite density was constructed early on in \cite{Bergman:2007wp,Rozali:2007rx,Kim:2007zm}, where the phase diagram was explored and shown to have features expected for QCD. Also, phases resembling quarkyonic phases suggested to occur in large-$\Nc$ QCD \cite{McLerran:2007qj,Andronic:2009gj,Fukushima:2013rx,Philipsen:2019qqm}, with recent applications in neutron-star context \cite{McLerran:2018hbz,Jeong:2019lhv,Sen:2020peq,Duarte:2020xsp,Zhao:2020dvu}, may be possible in the Sakai-Sugimoto model \cite{deBoer:2012ij,Kovensky:2020xif}. At low densities, a simple approximation of treating baryons as instantons with possible two-body interactions is reasonable. However, at higher densities many-body interactions become increasingly important and at neutron-star densities even the holographic description is involved. Even worse, at the large-$\Nc$ limit that holography assumes, the form of nuclear matter is not a fluid as expected for $\Nc=3$ but a crystal, a lattice of BPST instantons, with non-trivial transitions between phases as a function of density \cite{Rho:2009ym,Kaplunovsky:2010eh,Kaplunovsky:2012gb,Kaplunovsky:2013iza,Jarvinen:2020xjh}. The emergence of crystalline structure owes to two effects. First of all, the baryonic instantons squat: they are nonrelativistic not only because they are heavy $\propto \Nc$ but that their kinetic energy scales as $\propto\Nc^{-1}$. Second, the size of the instantons scale as $\propto (\sqrt\laYM M_{KK})^{-1}$ which means that at infinite coupling limit one cannot neglect stringy effects and one should consider finite-$\laYM$ corrections. Infinite coupling is insufficient to see the repulsive core of the instantons.

In nuclear matter, as described by QCD, at small distances we expect two instantons to overlap at high density as based on phenomenological studies (see e.g.~\cite{Zhitnitsky:2006sr} for discussion). Similarly for holographic nuclear matter, for distances ($\ll (\sqrt\laYM M_{KK})^{-1}$) we assume that due to repulsive interactions \cite{Hashimoto:2009ys}, the instantons begin to populate also the holographic radial direction $r$ and the lattice emerges \cite{Kaplunovsky:2012gb} (in addition to field theory directions). The crystalline phases disappear in the strict $\laYM\to\infty$ limit though and one can follow the pointlike instanton description as first laid out in \cite{Bergman:2007wp}.

In \cite{Rozali:2007rx} another approximation was proposed. There the instanton structure was replaced by a homogeneous non-Abelian field, motivated by dense BPST instanton configurations which are smeared over spatial directions. In addition to considering the original antipodal embeddings in the Sakai-Sugimoto model \cite{Rozali:2007rx}, this approach has been extended to non-antipodal ones in \cite{Li:2015uea} and to the case with  isospin chemical potential in \cite{Kovensky:2021ddl,Kovensky:2021kzl}. See also \cite{Elliot-Ripley:2016uwb} for a slightly different smearing approach. The homogeneous approximation has also been applied in the V-QCD model  \cite{Ishii:2019gta,Jokela:2020piw,Jokela:2021vwy}. Interestingly, the approach leads to realistic neutron-star physics as will be discussed in subsection \ref{sec:realistic}.

%%%%%%%%%%%%%%%%%%%%%%%%%%%%%%%%%%%%%%%
%%%%%%%%%%%%%%%%%%%%%%%%%%%%%%%%%%%%%%%
\section{Building compact stars with a holographic Equation of State}\label{sec:thermodynamics}
%%%%%%%%%%%%%%%%%%%%%%%%%%%%%%%%%%%%%%%
%%%%%%%%%%%%%%%%%%%%%%%%%%%%%%%%%%%%%%%

Next, we proceed to discuss what we have learned from using the holographic approach to model dense QCD matter, and apply these lessons to the physics of compact stars. We start by discussing zero-temperature physics at high densities, the deconfinement phase of quark matter and illustrate that holographic approach is also useful in this regime. We then continue to outline some lessons that we have learned from holographic modeling of dense strongly interacting QCD matter using holography. Final subsection discusses state-of-the-art holographic constructions that give realistic neutron stars in the light of all available astrophysical and theoretical constraints.

%%%%%%%%%%%%%%%%%%%%%%%%%%%%%%%%%%%%%%%
\subsection{Proof of concept}\label{sec:thermoTzero}
%%%%%%%%%%%%%%%%%%%%%%%%%%%%%%%%%%%%%%%

The first work applying AdS/CFT methods to the study of neutron-star matter is invoked the D3-probe-D7 system \cite{Hoyos:2016zke}. Paradoxically, this is likely one of the simplest models and yet, as we will see below, gives surprisingly realistic results. The authors of \cite{Hoyos:2016zke} were able to demonstrate that the resulting EoS is in the same ballpark as those coming from naive extrapolations of CET results to higher densities. 

%%%%%%%%%%%%%%%%%%%%%%%%%%%%%%%%%%%%%%%%%%%%%%%%%%%%%%%%%%%%%%%%%%%%%%%%%%%%%%%%%
\begin{figure}[t!]
\begin{center}
\begin{tabular}{cc}
\includegraphics[width=0.38\textwidth]{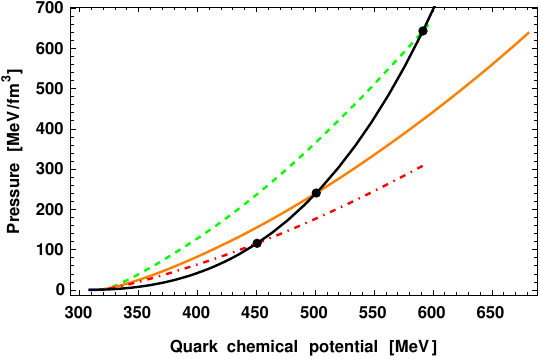}  & \includegraphics[width=0.54\textwidth]{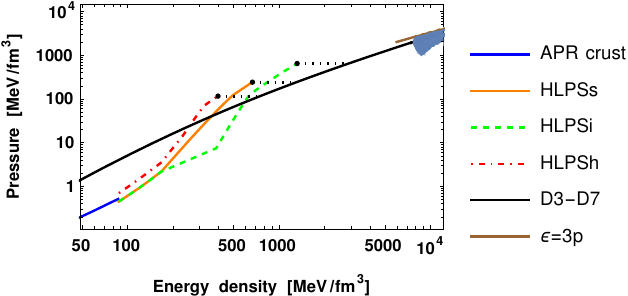}
\end{tabular}
\end{center}
\caption{Left: The colored curves correspond to nuclear matter equations of state from \cite{Hebeler:2013nza}, the orange being the typical one, while red and green variants corresponding to the most stiff and soft variants, respectively. The black curve is the EoS from D3-D7 model. The dots represent the merging point where one assumes holographic EoS on the high density phase. Right: The same curves but in the pressure-energy density plane. The first order phase transition corresponds to the constant pressure dashed line segments. We note that the D3-D7 EoS curve enters the blue band on the right as dictated by the perturbative QCD constraints \cite{Kurkela:2009gj}.}\label{fig:D3D7EoS}
\end{figure}
%%%%%%%%%%%%%%%%%%%%%%%%%%%%%%%%%%%%%%%%%%%%%%%%%%%%%%%%%%%%%%%%%%%%%%%%%%%%%%%%%

In the D3-D7 system, the $T=0$ EoS one can be solved analytically\footnote{It is also possible to obtain temperature corrections analytically \cite{Karch:2009eb,Ammon:2012je,Hoyos:2021njg}.} and reads  \cite{Karch:2007br,Hoyos:2016zke}
\be\label{eq:D3D7EOS}
 \epsilon = 3p + 4m^2\sqrt{f_0} \sqrt{p} \ ,
\ee
where $m$ can be interpreted as the constituent quark mass and $f_0=\frac{\Nc\Nf}{4\gamma^3\laYM}$ with $\gamma=\Gamma(7/6)\Gamma(1/3)/\sqrt\pi$. To connect with real QCD, one should extrapolate $\Nc\to 3$ and set $\Nf=3$. In \cite{Hoyos:2016zke}, the authors also chose the value of the 't Hooft coupling $\laYM\approx 10.74$ so that in the limit of asymptotically high densities the pressure matches with that of pQCD, i.e.~the free Fermi gas limit. One important underlying assumption is that all quark species have the same mass, in which case beta equilibrium and charge neutrality are obtained when the three chemical potentials agree. Both the pressure $p$ and energy density $\epsilon$ are then functions of a single quark chemical potential $\mu$. fig.~\ref{fig:D3D7EoS} illustrates that by judiciously choosing the value $m\approx 308.55\,\text{MeV}$, one finds that holographic quark matter, as modeled via the D3-D7 system, passes non-trivial consistency tests: the pressure and the EoS are compatible with CET results at low densities and they match onto constraints set by pQCD at high densities.

Given the success of describing quark matter in neutron-star conditions, one can ask whether the holographic results might have predictive power in the context of neutron stars. To this end, the authors of \cite{Hoyos:2016zke} chose a hybrid approach, using existing EoSs for nuclear matter, such as those extrapolated from CET \cite{Hebeler:2013nza} (denoted by HLPSs, HLPSi, and HLPSh in  fig.~\ref{fig:D3D7EoS}) at low densities and then matching them to the holographic EoS at some intermediate density where the two pressures coincide. This leads to strongly first order phase transitions (marked as black dots in fig.~\ref{fig:D3D7EoS}) between the two. In fact, all existing holographic models to date, matched this way onto ``traditional'' nuclear matter models, lead to strongly first order phase transitions at least if all astrophysical constraints discussed in sec.~\ref{sec:observations} are taken into account. This seems to be one overarching prediction of many holographic models: if quark matter is described using a holographic model, there are no quark matter cores inside quiescent neutron stars. This statement is best understood by plugging in the EoS from eq.~(\ref{eq:D3D7EOS}), matched onto the HLPS curves shown in fig.~\ref{fig:D3D7EoS}, into the TOV equations (\ref{TOVeq}). What one finds are mass-radius relations depicted in fig.~\ref{fig:MRcurvescut}: the MR-curves follow the corresponding ones of \cite{Hebeler:2013nza} up to the point when they sharply bend down. This point is in one-to-one correspondence with the critical densities of the first-order phase transitions. Stars residing on this black branch are unstable, as there exists an instability with respect to radial oscillations \cite{Glendenning:1997wn}. The fact that the transitions are strong can to some extent be attributed to the matching procedure, and is ultimately a consequence of the holographic EoSs being quite soft relative to the nuclear matter ones at the matching densities. Holographic models therefore suggest that the maximum mass of neutron stars are set by the location of the deconfinement phase transition.

\begin{figure}[t!]
\begin{center}
\begin{tabular}{cc}
\includegraphics[width=0.9\textwidth]{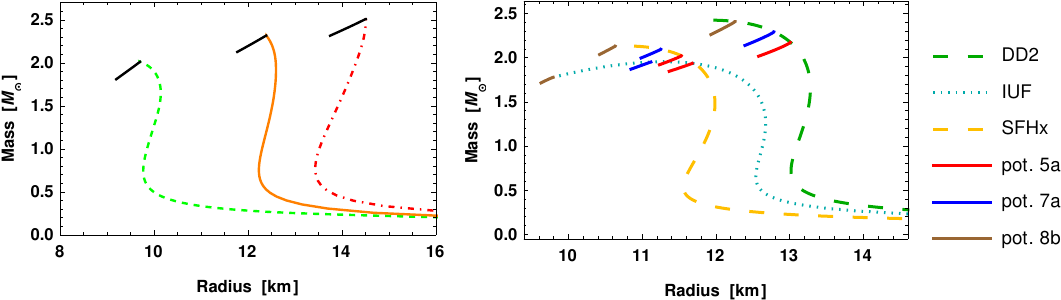}  
\end{tabular}
\end{center}
\caption{Left: Self-gravitating compact stars depicted as solutions of the TOV equations with the hybrid EoSs of \cite{Hoyos:2016zke} described in the main text, leading to no stable quark-matter cores. The color coding follows that of fig.~\ref{fig:D3D7EoS}.  Figure adapted from \cite{Hoyos:2016zke}. Right: The same result, but this time with hybrid EoSs based on merging nuclear matter model EoSs with three representatives of the V-QCD model. Again, no stable quark-matter cores are possible due to sizable latent heat associated with the phase transition. Figure adapted from \cite{Chesler:2019osn}.}\label{fig:MRcurvescut}
\end{figure}

In fig.~\ref{fig:MRcurvescut}, we also show a figure obtained using the V-QCD EoS matched onto nuclear matter model EoSs\footnote{One can extract the equations of state of DD2 \cite{Hebeler:2013nza}, IUF \cite{Hempel:2009mc,Fattoyev:2010mx}, and SFHx \cite{Steiner:2012rk} used in fig.~\ref{fig:MRcurvescut} from the online data base CompOSE \cite{Typel:2013rza}. One only needs to properly implement $\beta$-equilibrium and charge neutrality constraints to compare with V-QCD. The authors of \cite{Chesler:2019osn} have also supplemented the numerical data in the arXiv submission for non-zero electron fraction $Y_q\ne 0$ and $T\ne 0$.} to illustrate the similarities with the above solutions based on the hybrid D3-D7/HLPS EoSs. More massive stars can be obtained by considering rapidly spinning configurations using the V-QCD EoS \cite{Demircik:2020jkc}, leading to the interesting suggestion that the $M = 2.59^{+0.08}_{
-0.09}M_\odot$ object of GW190814 \cite{LIGOScientific:2020zkf} could have been a neutron star, void of a quark matter core. If one implements the IR running of the mass parameter $m$ in the probe D3-D7 model, the phase transition generically gets slightly milder \cite{BitaghsirFadafan:2019ofb} and for a select window of parameters can lead to stable quark-matter cores \cite{BitaghsirFadafan:2020otb}. Considerations in Einstein-Maxwell-scalar theories as proxies for holographic deconfined quark matter \cite{Mamani:2020pks} do not lead to stable quark-matter cores even for the most massive stars.

%%%%%%%%%%%%%%%%%%%%%%%%%%%%%%%%%%%%%%%
\subsection{Lessons from the deconfined phase}\label{sec:forward}
%%%%%%%%%%%%%%%%%%%%%%%%%%%%%%%%%%%%%%%

In the previous subsection, we found out that holographic QCD models can be compatible known theoretical bounds in the neutron-star environment. The obvious follow-up question is the following: since holographic models can be used to determine the EoS also in the no-man's land outside the respective regions of applicability of both CET and pQCD,
\begin{enumerate}
    \item can they be tuned to be compatible with different astrophysical bounds? 
    \item do they have real predictive power?
\end{enumerate}
The answer to both of these questions is indeed affirmative, and will be discussed in due course. Let us, however, first take a digression and dwell on a couple of novel aspects in which holography has already offered lessons beyond the suggestion that the maximum mass of the neutron stars is likely set by the deconfinement transition.

First, we consider the simple EoS of (\ref{eq:D3D7EOS}) alone, {i.e.}~without merging it with any non-holographic EoS. It turns out that in this case, one can solve for the TOV equations analytically in a perturbative manner. This is curious since there are not many examples of EoSs which, after feeding into the TOV equations (\ref{TOVeq}), would lead to analytic results. The best-known analytic example was provided by Buchdahl \cite{1967ApJ...147..310B}, the EoS of which reads
\be\label{eq:BuchdahlEoS}
 \epsilon = -\sqrt 5 p + 12\sqrt{p_\ast p} \ ,
\ee 
where $p_\ast$ is a parameter. There is clearly an interesting resemblance between eqs.~(\ref{eq:BuchdahlEoS}) and (\ref{eq:D3D7EOS}), inviting an analytic continuation between the respective solutions of the TOV  equations, but thus far attempts in this direction have failed. Other known EoSs that provide analytic solutions are single polytropic ones,\footnote{In addition, analytic solutions to TOV equations can be obtained by using constant energy density $\epsilon=\text{const.}$ (incompressible fluid) as well as energy densities given via radial profiles in Tolman IV and VII solutions \cite{PhysRev.55.364}. Of these, Tolman VII is the most realistic one, while in the Tolman IV solution one has a nonvanishing surface density, a situation typical of stars made of strange quark matter \cite{doi:10.1063/1.4909560}.} with polytropic index $\gamma$,
\be\label{eq:polyEoS}
 \epsilon \propto p^{1/\gamma} \ .
\ee
These will lead to solutions, where the radius $R$ of the star scales with the mass $M$ as
\be\label{eq:polyRM}
 R \propto M^{\frac{\gamma-2}{3\gamma-4}} \ .
\ee

At low densities or pressures, the D3-D7 EoS falls into the class of eq.~(\ref{eq:polyEoS}) with $\gamma=2$. This means that a star made purely out of a miniscule amount of exotic quark matter, a nugget \cite{Bodmer:1971we,Farhi:1984qu,Witten:1984rs}, will possess a non-zero radius. This observation provides a strikingly different mass-radius curve than those usually associated with exotic quark stars (see review \cite{Weber:2004kj} for a discussion of a variety of model EoSs and \cite{Fraga:2001id} for perturbative QCD): the MR-curve does not originate from the origin of the MR-plane. The mass-radius relationship for exotic quark stars composed of matter as modeled solely using the D3-D7 EoS (\ref{eq:D3D7EOS}) reads perturbatively \cite{Annala:2017tqz}
\be
 M(R) = \frac{M_0}{c_0}\left[\frac{R_0-R}{R_0} -\frac{c_1}{c_0}\left(\frac{R_0-R}{R_0}\right)^2 +\ldots\right] \ ,
\ee
where $c_0\approx 1.853$, $c_1\approx 2.948$, $R_0 = \pi/k$, $M_0 = c^2 R_0/G$, and $k^2 = 32\pi f_0m^4 G/c^4$. This perturbative expansion matches very accurately with the numerical results at smallish $(R-R_0)/R_0$. The analytic framework is robust enough to also allow for the computation of the electric Love number $k^{el}_2$ (related to the tidal deformability), quadrupole moment of mass distribution $Q$, and moment of inertia $I$ in perturbative series in compactness $C=GM/(c^2R)$:
\bea
 \bar\Lambda & = & \frac{2}{3C^2}k^{el}_2 \approx \frac{2}{3C^2}\left[0.260-1.994C^{-2}\right]  \label{eq:D3D7love}\\
 \bar Q & = & -\frac{M}{I^2}\frac{Q}{\Omega^2/c^2} \approx 0.261 C^{-2} \label{eq:D3D7Q}\\
 \bar I & = &  \frac{c^4}{G^2 M^3} I \approx -30.35 C  \  .\label{eq:D3D7I}
\eea
Here, the quantities on the left are dimensionless and $\Omega$ is the angular velocity of the star. The perturbative expressions for the quantities $I$ and $Q$ require one to consider stars rotating with a small angular velocity $\Omega$, and all numerical factors are approximations of closed formulas. Higher order terms as well as other electric and magnetic Love numbers are also available in \cite{Annala:2017tqz}. 

The quantities (\ref{eq:D3D7love})-(\ref{eq:D3D7I}) are scaled judiciously with the properties of the stars to make contact with quasi-universal I-Love-Q relations \cite{Yagi:2016bkt}. The reason to be interested in these quantities is that their ratios are largely insensitive to the EoS of stellar matter as suggested by the I-Love-Q relations. Indeed, even EoS following from the holographic modeling at all densities lead to results which satisfy these relations up to few per cent. The relations have been scrutinized also in the V-QCD model \cite{Jokela:2020piw}; see also \cite{Burikham:2021xpn} for another holographic example. However, there is a catch as first pointed out in \cite{Annala:2017tqz}. The relations {\emph{can}} be violated by a lot (exemplified in \cite{Annala:2017tqz} to 20\% level), if there is a sharp phase transition happening at lowish density, corresponding to radii not far away from the surface of the star. This is a generic result and not in any way tied with holographic modeling, as one can mimic any theory by simply forgetting about the microscopic origin and picking an EoS by hand \cite{Alford:2004pf}. Indeed, this possibility for the violation was soon also realized in non-holographic frameworks \cite{Sieniawska:2018zzj,Han:2018mtj} as well.\footnote{The I-Love-Q relations were already previously known to break down for stars with large rotation rates \cite{Doneva:2013rha} or in the presence of magnetic fields \cite{Haskell:2013vha}.} See also the recent work \cite{Tan:2021nat}, which studies generalizations of universal relations in binary systems in the presence of first order phase transitions.

Interestingly, if one entertains the possibility of having an intermediate phase of exotic matter in the crust region of a neutron star with a subsequent transition back to nuclear matter towards denser regions, as, {e.g.}, described by the D3-D7 model merged with HLPS EoSs with judiciously chosen parameter $m$, one cannot easily rule out the existence of such stars \cite{Annala:2017tqz}. They are compatible with known astrophysical bounds such as the tidal deformability bound provided by the LIGO/Virgo collaboration \cite{TheLIGOScientific:2017qsa} as inferred from the event GW170817. We note that qualitatively similar compact stars can be obtained using the MIT bag model supplemented with quark pairing effects \cite{Alford:2004pf,Alford:2002rj,Zdunik:2012dj}.

Absolutely stable strange quark matter exists if one can find an s quark chemical potential, for which the s quark density is nonzero and the energy per baryon is lower than for nuclear matter at vanishing pressure \cite{Bodmer:1971we,Farhi:1984qu,Witten:1984rs}. The existence of quark stars, i.e.~compact star solutions built from such stable quark matter, is considered very unlikely, but has not been convincingly ruled out either. If one constructs stars made of quark matter provided by probe D$p$-D$q$-brane intersections, including the probe D3-D7 model, the results are typically incompatible with observed compact stars, often featuring e.g.~unrealistically large radii \cite{Kim:2014pva,Burikham:2010sw}. It is thus safe to conclude that if quark stars built from an EoS resembling the holographic candidates exist at all, they at least do not form the majority of all compact stars in existence.

Before moving away from the holographic description of (unpaired) deconfined quark matter, we finally briefly return to the issue that ended the previous subsection and was illustrated in fig.~\ref{fig:MRcurvescut}. If it is the case that the holographic EoS for the quark-matter phase is universally very soft, then what can holography could offer beyond the prediction of the absence of stable quark cores? Even worse, as some evidence has already accumulated for stable quark-matter cores in the most massive stars \cite{Annala:2019puf}, one may wonder if the holographic approach is already doomed. It turns out that the answer to the latter concern is quite surprising and singles out holographic scenarios from other nuclear matter models, as we will discuss in the following subsection. Here, we will however first address the first concern.

There are two important caveats to keep in mind. First, we have so far only focused on unpaired quark-matter phases, and second, even if quiescent neutron stars did not have quark-matter cores, deconfined matter may well be produced in other dynamical settings. In fact, it has already been argued that during the neutron-star coalescence, the temperatures and densities may well reach values large enough so that one ends up exploring the deconfining regimes of the QCD phase diagram. Should a strongly first order phase transition exist, imprints thereof would likely be seen in GW data corresponding to the ringdown phase \cite{Most:2018eaw,Bauswein:2018bma,Chesler:2019osn,Ecker:2019xrw}. In this context, it is important to realize that if quark matter indeed emerged during a merger, one would need to be able to describe matter in an out-of-equilibrium setting. At the moment, holography provides the only first-principles field theoretical framework able to make predictions in such a strongly coupled and time-dependent environment. To this end, we will in sec.~\ref{sec:transport} describe the first such predictions, corresponding to the transport properties of dense quark matter slightly perturbed from equilibrium.

Returning to the question of quark pairing, one may naturally ask, what are the most significant physical effects we have neglected so far by only considering unpaired quark matter? At the very least the transport properties of paired matter would likely be wildly different, but one can also ponder if the equilibrium properties would be affected. The standard perturbative lore says that thermodynamic quantities, such as energy, pressure, and the speed of sound are not very sensitive to quark pairing  at high chemical potentials, and that the pairing contributions will be subleading relative to those from unpaired colors and flavors \cite{Alford:2007xm}. However, we also know that at core densities, QCD is strongly coupled and there is no guarantee that there is any meaningful notion of Fermi surfaces. Could the ``paired'' holographic quark-matter phase be stiff $c_s^2>1/3$, which would then allow stable quark-matter cores? Some indications towards this have been recently seen in an adjusted D3-D7 model, where color superconducting phases were included \cite{BitaghsirFadafan:2018iqr,BitaghsirFadafan:2020otb}. The paired quark-matter phase was found to be stiff enough so that stable quark cores were reached.

Interestingly, in the early days of applied holography there was a conjecture on an upper bound for the speed of sound in any physical system given by the conformal value $c_s^2\leq 1/\sqrt 3$, as all known examples of asymptotically AdS$_5$ backgrounds obeyed this limit  back then \cite{Cherman:2009tw}, with dynamically unstable solutions \cite{Buchel:2009ge,Buchel:2010wk} being the only known counterexamples. There exist, however, several simple counterexamples to this conjecture. Among these there are those that do not have a four-dimensional conformal fixed point in the UV, and thus do not correspond to ordinary renormalizable field theories in four dimensions. Such  ``trivial'' examples include (3+1)d probe intersection models D$4$-D$6$, D$5$-D$5$, and the D$4$-D$8$ Sakai-Sugimoto model \cite{Kulaxizi:2008jx}; see \cite{Jokela:2015aha,Itsios:2016ffv} for details. Another class of models are those with nonrelativistic backgrounds having Lifshitz scaling \cite{Kachru:2008yh,Hoyos-Badajoz:2010ckd,Jokela:2016nsv,Taylor:2015glc}. The violation of the bound is again in some sense trivial, since the dual field theory is nonrelativistic.

The works \cite{Hoyos:2016cob,Ecker:2017fyh} provided a family of finite-density solutions in holographic models with UV fixed points. The conclusion was that there is no universal bound for the speed of sound in holographic models dual to ordinary four-dimensional relativistic field theories. This came as good news for everyone wishing to build realistic holographic models for high-density nuclear or quark matter, as common lore in nuclear physics states that significantly higher speeds of sound are likely realized inside neutron stars. Thereafter, the ``violation'' of the bound was also reported in V-QCD in the nuclear matter phase \cite{Ishii:2019gta,Jokela:2020piw}, in the phenomenologically adjusted D3-D7 model \cite{Evans:2011eu,BitaghsirFadafan:2020otb}, in the presence of a magnetic field \cite{Grozdanov:2017kyl,Gursoy:2017wzz,Gursoy:2021efc}, and in theories with multitrace deformations in the absence of chemical potential \cite{Anabalon:2017eri}.

%%%%%%%%%%%%%%%%%%%%%%%%%%%%%%%%%%%%%%%
\subsection{Building a realistic neutron star}\label{sec:realistic}
%%%%%%%%%%%%%%%%%%%%%%%%%%%%%%%%%%%%%%%

Having discussed the holographic description of the deconfined quark-matter phase at length above, let us next study how the  nuclear matter phase can be described in a holographic setting. Recall that (unpaired) quark matter as described by various holographic models is rather soft, meaning that the stiffness of the EoS as measured by the rate $c_s^2=(\partial p/\partial\epsilon)_s$ is small. This had the effect that isolated neutron stars cannot have stable quark-matter cores, as when even a tiny fraction of quark matter appears in the core of a star, it immediately collapses into a black hole.

Interestingly, this result is at tension with recent results that most massive stars should have sizable stable quark-matter cores, unless the above rate, conventionally called the speed of sound, takes very large values $c_s^2>0.7$ \cite{Annala:2019puf}. In this subsection, we argue that this is in fact great news for the holographic models and show that bypassing some of the underlying assumptions in the analysis of \cite{Annala:2019puf} is a distinctive characteristic of holographic models.

To incorporate the nuclear matter phase in a holographic setting, let us first recall that at low densities (around and below the nuclear saturation density $n_s$), one can systematically take interactions between nucleons into account in an effective field theory setup. However, when the densities considerably exceed the nuclear saturation density, nucleons start to overlap, with the convergence of the perturbative expansion suffering. One can thus ask, if it would be better to switch to a holographic description already at densities slightly above $n_s$, as an expansion around an infinitely strongly coupled limit might in fact provide a better description. 

One expects that the nuclear matter phase emerges through solitonic solutions in the holographic models, but as discussed in the previous section, simultaneously treating several solitons is complicated on the bulk gravity side. At higher densities one can, however, ask if treating the solitons as a homogeneous configuration would be feasible. In this regime, the QCD coupling constant is certainly still sizable, and the holographic treatment thus remains well motivated.

%%%%%%%%%%%%%%%%%%%%%%%%%%%%%%%%%%%%%%%%%%%%%%%%%%%%%%%%%%%%%%%%%%%%%%%%
\begin{figure}[t!]
\begin{center}
\begin{tabular}{cc}
\includegraphics[width=0.9\textwidth]{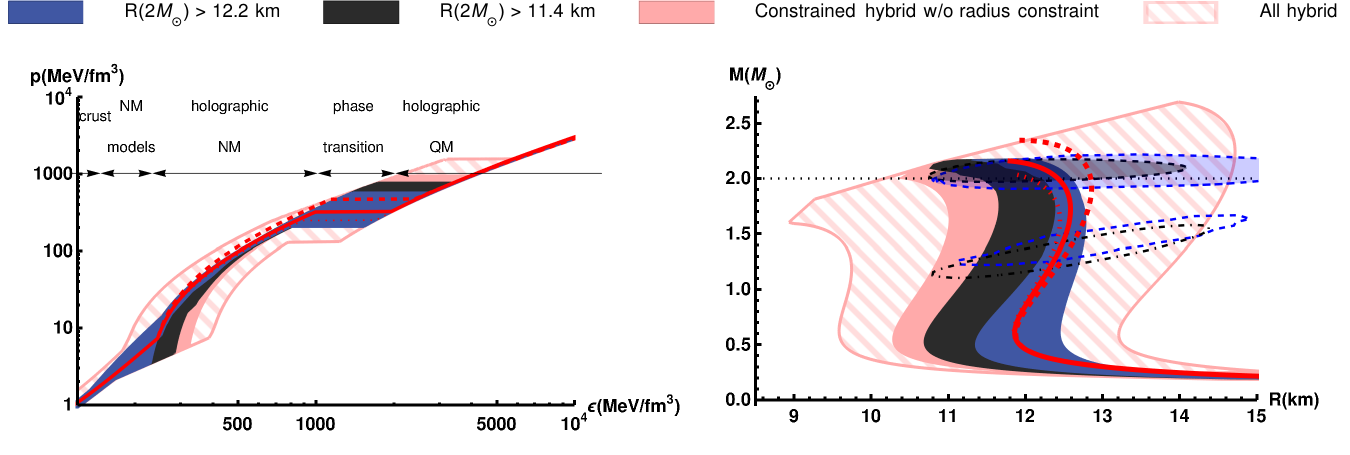} 
\end{tabular}
\end{center}
\caption{Left: The EoS bands spanned by the hybrid EoSs subject to different constraints. Right: The mass-radius curves constructed using the EoSs of the left plot. The striped bands correspond to all of the hybrid EoSs, the light red bands to the ones not subject to the NICER radius constraints, and the blue ones to the most stringent NICER radius constraint provided by Miller et al. \cite{Miller:2021qha}. Also shown are the V-QCD (APR) hybrid EoS curves on red. Figure adapted from \cite{Jokela:2021vwy}.}\label{fig:hybrid}
\end{figure}
%%%%%%%%%%%%%%%%%%%%%%%%%%%%%%%%%%%%%%%%%%%%%%%%%%%%%%%%%%%%%%%%%%%%%%%%

In the following, we  review some recent results stemming from ``hybrid'' constructions implemented explicitly in the V-QCD model \cite{Ecker:2019xrw,Jokela:2020piw,Jokela:2021vwy}. The idea is that at very low densities one uses nuclear matter models that are still allowed by astrophysical constraints, but when the densities exceed values in the ballpark of $n\sim 1.5n_s$ one continuously switches over to the holographic model. To this end, the dense NM phase on the holographic side is chosen using the homogeneous approximation \cite{Ishii:2019gta}. At even higher densities, one uses the {\emph{same}} V-QCD model to describe the deconfined (unpaired) quark-matter phase; that is, the deconfined-confined phase transition is described within the same model. The transition turns out to be a strongly first order one over a variety of different weakly coupled nuclear matter models and where the jump to the holographic modeling is implemented \cite{Jokela:2020piw,Demircik:2020jkc,Jokela:2021vwy}. The implication is that no stars have stable quark-matter cores. 

The construction of choosing the hybrid EoSs is depicted in fig.~\ref{fig:hybrid}, where the largest striped band consists of all physically sensible EoSs. We has not chosen any particular model but instead considered a family of EoSs \`a la \cite{Kurkela:2014vha,Annala:2017llu,Fraga:2015xha}. The thermodynamic functions are arbitrary as long as the following conditions are met: the pressure $p$ and the chemical potential $\mu$ need to be continuous, the thermodynamics has to be consistent (meaning $\partial_\mu p=n>0,\partial^2_\mu p>0$), and the speed of sound as measured by the rate $c_s^2=(\partial p/\partial\epsilon)_s$ in ideal hydrodynamics cannot exceed unity anywhere. Furthermore, we demand that at low density the EoS should be consistent with phenomenology and conform with CET predictions, while at very high density the EoS should agree with pQCD. {\emph{Any}} EoS following from holographic modeling or elsewhere has to lie inside this striped band.

In the light red bands of fig.~\ref{fig:hybrid}, two observational constraints are assumed: 1) the EoS should be stiff enough to sustain at least two-solar-mass stars, $M_{max}\geq 2M_\odot$ and 2) the tidal deformability of a slowly rotating neutron star should conform with the updated LIGO/Virgo constraint \cite{Abbott:2018exr}: $580>\Lambda_{1.4}>70$. The blue and black bands in addition assume that the EoSs respect the two independent {NICER} teams' (Riley et al. \cite{Riley:2021pdl} and Miller et al. \cite{Miller:2021qha}) 1-$\sigma$ estimates for the equatorial radius of PSR J0470+6620, respectively; see \cite{Jokela:2021vwy} and Sec.~\ref{sec:observations} for a discussion on these specific numbers. Notice that older {NICER} results \cite{Riley:2019yda,Miller:2019cac} for the lighter PSR J0030+0451 that constrain neutron stars with masses $\sim 1.4M_\odot$ are also included in fig.~\ref{fig:hybrid}. They do not have a significant effect, as the most stringent constraints stem from
restricting the radii of two-solar-mass stars as inferred from observing PSR J0740+6620, and the EoSs that satisfy the newer Miller et al.~constraint \cite{Miller:2021qha} also satisfy the older radius constraints.

It is important to note that, as seen in fig.~\ref{fig:hybrid}, there are several hybrid EoSs that are consistent with all known astrophysical constraints. One example is offered by hybrid EoSs that at low densities correspond to the APR EoS \cite{Akmal:1998cf}, describing zero-temperature nucleonic $npe\mu$ matter in $\beta$-equilibrium, and are matched at $n_{tr}/n_s = 1.6$ with the holographic model. These EoSs have been submitted to the online repository CompOSE (CompStar Online Supernovae Equations of State online service, \url{https://compose.obspm.fr/} ~\cite{Typel:2013rza}) and are called V-QCD(APR) soft (using Pot.~{\bf{5b}}): dotted curve; V-QCD(APR) intermediate (using Pot.~{\bf{7a}}): solid curve; V-QCD(APR) stiff (using Pot.~{\bf{8b}}): dashed curve. The options arise from the fact that holographic quark matter is not completely unambiguous as the fitting of the holographic model to lattice QCD at low densities is not restrictive enough. The three potentials {\bf{5b,7a,8b}} represent a good selection of all realistic possibilities; see Sec.~\ref{sec:holomodels} for further discussion. Interestingly, a careful fit to the hadron masses \cite{Amorim:2021gat} seems to favor the intermediate choice {\bf{7a}} over the soft {\bf{5b}} and stiff {\bf{8b}} choices.

The Sakai-Sugimoto model has also been applied to the compact star context and been shown to lead to realistic outcomes \cite{Kovensky:2021kzl}. Similarly to the V-QCD construction discussed above, nuclear matter is treated in the homogeneous approximation, as a pointlike approximation for the baryons \cite{Bergman:2007wp} does not lead to a sequence of stars which would simultaneously satisfy the LIGO/Virgo tidal deformability constraint and support massive enough neutron stars \cite{Zhang:2019tqd}. More precisely, the work presented in \cite{Kovensky:2021kzl} uses the revisited treatment of \cite{Kovensky:2021ddl} with nonzero isospin chemical potential, which includes leptons in order to maintain charge neutrality in the presence of a disparity between neutrons and protons. The flavor action is treated in the Yang-Mills approximation. 

A novel component in \cite{Kovensky:2021kzl} is that the entire composition of the neutron star is modeled using holographic matter. In particular, the outer crust is constructed using a mixed phase description, where nuclear matter is spatially separated from the lepton gas. The ability to construct an entire star using holography is impressive: this has the advantage that the entire star has a single microscopic description. There are a few parameters left free, namely the 't Hooft coupling, the Kaluza-Klein scale $M_{KK}$, and a surface tension $\Sigma$, but choosing them appropriately one is able to produce EoSs inside the previously discussed striped band in the regime where matter is expected to be strongly coupled. Moreover, the mass-radius curves are also realistic. In fig.~\ref{fig:SShybrid}, we have shown a sample of EoSs as well as mass-radius curves. It should be noted that for all these parameter sets, the $1.4M_\odot$ stars conform with the LIGO/Virgo tidal deformability constraint and that the EoSs support two-solar-mass stars. 

%%%%%%%%%%%%%%%%%%%%%%%%%%%%%%%%%%%%%%%%%%%%%%%%%%%%%%%%%%%%%%%%%%%%%%%%
\begin{figure}[t!]
\begin{center}
\begin{tabular}{cc}
\includegraphics[width=0.45\textwidth]{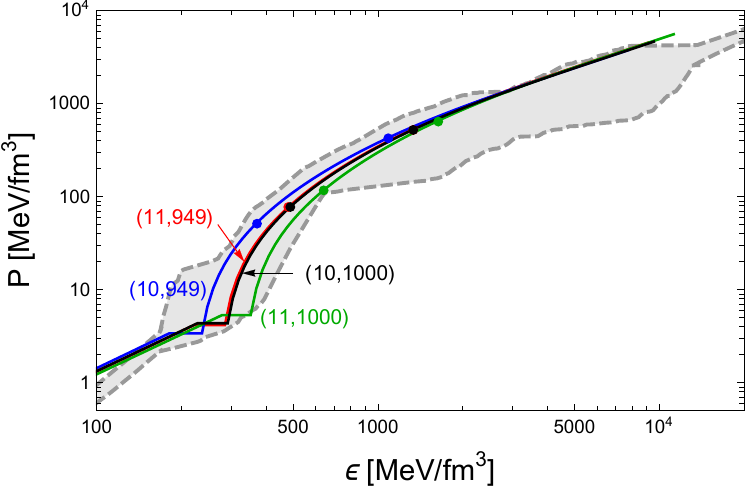} $\;\;$\includegraphics[width=0.45\textwidth]{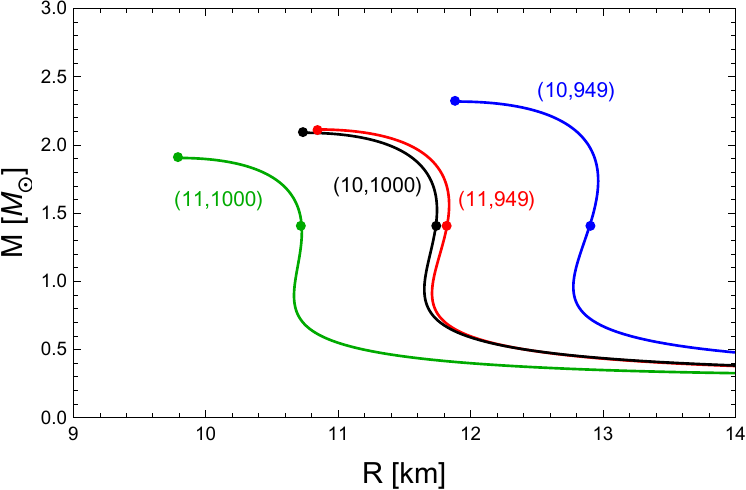} 
\end{tabular}
\end{center}
\caption{Left: The EoSs from the Sakai-Sugimoto model. Right: The corresponding mass-radius curves. The four curves correspond to four sets of parameters of the 't Hooft coupling and the Kaluza-Klein mass scale $(\laYM,M_{KK})$. Here, the surface tension is fixed to $\Sigma=1$MeV/fm$^2$. The dots represent the core densities for stars of 1.4 solar masses as well as the most massive stable ones. Figures adapted from \cite{Kovensky:2021kzl}.}\label{fig:SShybrid}
\end{figure}
%%%%%%%%%%%%%%%%%%%%%%%%%%%%%%%%%%%%%%%%%%%%%%%%%%%%%%%%%%%%%%%%%%%%%%%%

As we have seen above, both V-QCD and the Sakai-Sugimoto model lead to realistic neutron stars and have predictive power. Let us next discuss in more detail the latter aspect, i.e.~precisely what predictions we can make. In both cases, the nuclear matter phase is quite stiff which means that one generically finds larger radii than with approaches where CET results are extrapolated \cite{Capano:2019eae} to larger densities. In particular, for $M=1.4M_\odot$ stars, the radii $R(1.4M_\odot)\equiv R_{1.4}$ are
\bea
 \text{V-QCD \ :  } \qquad\qquad &  12.8\text{km} \geq R_{1.4} \geq 10.9\text{km} & \\
 \text{V-QCD with NICER constraints \ :  } \qquad\qquad &  12.8\text{km} \geq R_{1.4} \geq 12.0\text{km} & \\
 \text{Sakai-Sugimoto \ :  } \qquad\qquad &  R_{1.4} \approx (12.9, 11.8, 11.7, 10.7)\text{km} & \label{eq:SSradii} \ .
\eea
In comparison, recent results based on CET calculations yield $R_{1.4}=11.0_{-0.6}^{+0.9}$km \cite{Capano:2019eae}.
We note that for the V-QCD model the radii are constrained using the most recent {NICER} measurements for the radius of the millisecond pulsar PSR J0740+6620. We expect that implementing these constraints in the Sakai-Sugimoto model would similarly tighten the allowed window for $R_{1.4}$ especially on the lower end; the numbers quoted in (\ref{eq:SSradii}) correspond to the parameter sets $(\laYM,M_{KK})$ with fixed $\Sigma=1$MeV/fm$^2$ as used in fig.~\ref{fig:SShybrid}.

Next, let us recall that the LIGO/Virgo constraint on the tidal deformability is $\Lambda_{1.4}=190_{-120}^{+390}$. The variation of the model parameters both for the V-QCD and Sakai-Sugimoto model yield a window for tidal deformabilities $\Lambda_{1.4}$ and predict as lower bounds the somewhat higher values
\bea
 \text{V-QCD \ :  } \qquad\qquad &  \Lambda_{1.4} \geq 230 & \\
 \text{V-QCD  with NICER constraints\ :  } \qquad\qquad &  \Lambda_{1.4} \geq 435 & \\
 \text{Sakai-Sugimoto \ :  } \qquad\qquad &  \Lambda_{1.4} \geq 230 \label{eq:SStidal} & \ .
\eea
Similarly to the radius predictions, the V-QCD result is tightened using the {NICER} measurements; the number in the middle line results from the most stringent Miller et al. \cite{Miller:2021qha} constraint.  Interestingly, the V-QCD value would be equal to the Sakai-Sugimoto value \cite{Jokela:2020piw} if no input from the radius measurements is assumed. It is expected that implementing radius constraints for the Sakai-Sugimoto case would similarly elevate the lower value. If one further considers constraining the maximum mass of neutron stars to be $<2.19M_\odot$ following the results of \cite{Annala:2021gom}, obtained using the assumption that the GW170817 merger event was not a prompt collapse to a black hole but proceeded via an intermediate hypermassive neutron star, one finds that the minimal tidal deformability for the V-QCD model is further increased: $\Lambda_{1.4}\geq 480$ \cite{Jokela:2021vwy}. The most conservative estimate from \cite{Annala:2021gom} is on the other hand $<2.58M_\odot$, obtained by assuming gravitational collapse from a supramassive neutron star (see also \cite{Margalit:2017dij,Shibata:2017xdx,Rezzolla:2017aly,Ruiz:2017due,Shibata:2019ctb} for similar results). 

Finally, we can also spell out the key characteristics for the quark-hadron deconfinement-confinement phase transition in the V-QCD model. The transition is strongly first order and takes place at the densities $6.8\geq n_b/n_s \geq 4.3$ with a latent heat $\Delta\epsilon$ in the excess of $750$MeV/fm$^3$ if the NICER constraint is taken into account. Without considering this input the transition density does not happen below $4n_s$. In the Sakai-Sugimoto model, the transition to the quark-matter phase has not yet been studied but the expectation is that such a transition would take place at very high densities \cite{Kovensky:2021kzl}.

The prediction from the V-QCD model is that all compact stars are fully hadronic and there are no stable quark-matter cores. The results for the Sakai-Sugimoto model mildly support this too as one has seen indications that deconfinement transition could only happen at very high densities, though a systematic study of the quark-matter phase is still lacking. We can now contrast this result with the assertion on exactly the opposite \cite{Annala:2019puf}, {i.e.}, that it is likely that quiescent massive stars have sizable stable quark-matter cores. The key observation is that one of the main assumptions used in \cite{Annala:2019puf} is that when the adiabatic index $\gamma = \partial\log p/\partial\log\epsilon$ falls below a somewhat arbitrarily chosen value 1.75, the phase of the matter is classified as quark matter. We note that holographic nuclear matter behaves drastically differently than any model considered in \cite{Annala:2019puf}, which lead to the selection of 1.75 as the critical value for this parameter.  Note that the behavior of the adiabatic index can be compared with the behavior of the speed of sound as the quantities are related by a multiplication of $p/\epsilon$. Holographic nuclear matter is again seen to be stiffer than those following from CET, so a possible non-detection of quark-matter phases in neutron stars could be considered a positive sign for the holographic approach to modeling strongly coupled matter.

Finally, we note that low values of $\gamma$ can be also obtained for dense nuclear matter in models based on skyrmions \cite{Paeng:2017qvp,Ma:2020hno}. In the holographic context, however, while baryons resemble skyrmions, one does not find low values for the $\gamma$ neither in V-QCD  \cite{Ishii:2019gta} nor in the Sakai-Sugimoto models \cite{Zhang:2019tqd,Kovensky:2020xif,Kovensky:2021kzl} if one considers pointlike solitons. Thus far, only the homogeneous approximation for the holographic nuclear matter phase has been shown to lead to realistic neutron stars; see, however, a promising line of work in the six-dimensional AdS soliton setup which contains a superconducting  phase \cite{Ghoroku:2013gja,Ghoroku:2021fos}.

%%%%%%%%%%%%%%%%%%%%%%%%%%%%%%%%%%%%%%%
%%%%%%%%%%%%%%%%%%%%%%%%%%%%%%%%%%%%%%%
\section{Transport in compact stars}\label{sec:transport}
%%%%%%%%%%%%%%%%%%%%%%%%%%%%%%%%%%%%%%%
%%%%%%%%%%%%%%%%%%%%%%%%%%%%%%%%%%%%%%%

To a large extent, the structure of neutron stars is determined by the EoS and other bulk properties of matter in thermal equilibrium. However, there are many questions pertaining dynamical stability, thermal evolution and other processes such as accretion or phase transitions that require a detailed knowledge of out-of-equilibrium properties of the matter comprising the star. Transport properties describe how conserved quantities, such as energy, momentum or different charges are transferred from one region to another when the system is not in equilibrium. If the system is close to local thermal equilibrium and spatial gradients are not too large, then the transport properties and the evolution of the star should be well described by hydrodynamics. 

Transport coefficients can be defined via the response of conserved currents to an external source. Electric fields $E_i$, temperature gradients $\nabla_i T$ and perturbations of the spatial metric $g_{ij}=\delta_{ij}+h_{ij}(t)$ induce changes in the electric current $J^i$, heat current $Q^i$ and stress tensor $T_{ij}$ according to
\begin{equation}
\begin{split}
&\delta J^i=\sigma^{ij}E_j-\alpha^{ij}\nabla_j T,\ \ \delta Q^i=\kappa^{ij}\nabla_j T-T\alpha^{ij}E_j \\
&\delta T_{ij}=\eta \partial_t h_{ij}+\frac{1}{2}\left(\zeta-\frac{2}{3}\eta\right)\delta_{ij} \partial_t h^k_k \ ,
\end{split}
\end{equation}
where $\sigma^{ij}$, $\alpha^{ij}$, and $\kappa^{ij}$ are the electric, thermoelectric, and heat conductivities, respectively, and $\eta$ and $\zeta$ are the shear and bulk viscosities. In an isotropic fluid without parity and/or time reversal breaking, the conductivities are diagonal and independent of the spatial direction, i.e.~$\sigma^{ij}=\sigma \delta^{ij}$, $\kappa^{ij}=\kappa \delta^{ij}$, and $\alpha^{ij}=\alpha\delta^{ij}$. Furthermore, in a relativistic fluid the hydrodynamic constitutive relations imply that thermoelectric and thermal conductivities are determined by the electrical conductivity $\alpha=(\mu/T) \sigma$ and $\kappa=(\mu^2/T) \sigma$, with $\mu$ the chemical potential, so there are only three independent coefficients $\eta$, $\zeta$, and $\sigma$ at this order in the gradient expansion. Deviations from the hydrodynamic or relativistic regimes would in principle spoil this relation, so in general an independent calculation of each coefficient is necessary.

In the following, we first discuss what is known about these transport coefficients in dense matter from perturbative calculations, then discuss their holographic determination, and finally make more concrete predictions for neutron-star physics.

%%%%%%%%%%%%%%%%%%%%%%%%%%%%%%%%%%%%%%%%%%
\subsection{Relevant transport properties and perturbative estimates}
%%%%%%%%%%%%%%%%%%%%%%%%%%%%%%%%%%%%%%%%%%

A thorough review article on transport in neutron stars can be found from \cite{Schmitt:2017efp}, where one can also locate further references presenting some aspects in more detail. This is a rich and complicated topic as transport properties can vary wildly from the thin atmosphere made of atoms to the exotic matter of the core at ultrahigh densities several times larger than that of atomic nuclei. Each layer of the star requires in principle an appropriate and careful treatment, which may additionally vary with the phases of matter inside the star. Here, we will summarize results concerning quark matter at supranuclear densities, as transport properties in holographic models have been derived mostly for this type of matter. Although the scenario is still hypothetical, quark matter has been argued to be likely present at the cores of at least the most massive neutron stars \cite{Annala:2019puf}.

Transport properties typically depend crucially of the phase of matter. At asymptotically large densities where the differences in quark masses can be neglected and perturbation theory is reliable, quark matter is expected to be in a Color-Flavor-Locked (CFL) phase \cite{Alford:1997zt,Alford:1998mk}. This is a baryon superfluid phase with a Higgsing of the color group by the pairing of quarks. At lower densities, there is a plethora of possible phases with different symmetry breaking patterns, but our knowledge of the phase diagram is at best an educated guess from extrapolations of phenomenological models; see sec.~\ref{sec:intro} for more discussion. The simplest scenario is that quark matter in the inner cores of neutron stars is at densities large enough for deconfinement, but not for Cooper pair formation. Some amount of unpaired quark matter is expected in most phases, except in CFL, so the calculation of transport properties in the unpaired phase serves as a first approximation even if the precise phase in neutron stars has not been identified.

At the baryon densities met in neutron-star cores, interactions between the quarks are very strong, so kinetic theory calculations based on a quasiparticle description and perturbative expansion are not expected to be particularly good. Nevertheless, transport properties have been computed with this approach, so we briefly review them here, as it is interesting to compare them to the results from holographic models. It should be noted that some transport properties are dominated by electroweak processes, and are not as dependent on the characteristics of the strong interaction. These are not extracted directly from the holographic model\footnote{The holographic model could provide values for the QCD-dependent quantities that enter the electroweak calculations, but this requires a more refined study with different quark masses that what has been done so far; see sec.~\ref{sec:openquestions} for further discussion on this point.}, so we restrict our discussion to  those transport coefficients that are determined by the strong interaction alone. The values for the shear viscosity $\eta$, thermal conductivity $\kappa$\footnote{The thermal conductivity enters the heat current in the same way as the heat conductivity, but it is evaluated at a vanishing electric current $J^i=0$.} and electrical conductivity $\sigma$ (in units of $e^2/(\hbar c)$) in unpaired quark matter, at temperature $T$ and quark chemical potential $\mu$, read \cite{Heiselberg:1993cr}
\begin{eqnarray}\label{eq:perttrans}
\eta_{\text{per}} & \approx & 4.4\times 10^{-3} \frac{\mu^4 m_{_D}^{2/3}}{\alpha_s^2 T^{5/3}} \\
\kappa_{\text{per}} & \approx & 0.5 \frac{m_{_D}^2}{\alpha_s^2} \\
\sigma_{\text{per}} & \approx & 0.01 \frac{\mu^2 m_{_D}^{2/3}}{\alpha_s T^{5/3}} \ .
\end{eqnarray}
The contribution to the bulk viscosity of unpaired quark matter is considered to be negligible in a perturbative approximation, as the quark masses are much smaller than the typical energy of quarks at the Fermi surface \cite{Arnold:2006fz}.

At small but non-negligible temperatures, the one-loop Debye mass for $\Nf$ quark flavors and $\Nc$ colors is \cite{Vuorinen:2003fs}
\begin{equation}\label{eq:mD}
m_{_D}^2=2\frac{\alpha_s^2}{\pi}\left( \Nf\, \mu^2+(2 \Nc+\Nf)\frac{\pi^2}{3} T^2\right) \ .
\end{equation}
There is some dependence on the renormalization scheme that enters through the calculation of the strong coupling $\alpha_s=g_{_{QCD}}^2/(4\pi)$. The two-loop formula reads \cite{Baikov:2016tgj}
\begin{equation}
\begin{split}
\alpha_s^{-1}= & \frac{11 C_A-4 T_F}{6\pi^3} \log\frac{m_{ren}}{\Lambda_{_{pQCD}}} \\
 & +\frac{C_A(17 C_A-10 T_F)-6 C_F T_F}{2\pi^3(11 C_A-4 T_F) } \log\left( 2 \log\frac{m_{ren}}{\Lambda_{_{pQCD}}} \right) \ ,
\end{split}
\end{equation}
where
\begin{equation}
C_A=\Nc\ ,\ C_F=\frac{\Nc^2-1}{2\Nc}\ , \ T_F=\frac{\Nf}{2} \ .
\end{equation}
Following common conventions \cite{Kurkela:2009gj,Kurkela:2016was,Ghiglieri:2020dpq} the renormalization scale is set to $m_{ren}=x\sqrt{(2\pi T)^2+(2\mu)^2}$, with $x\sim 1/2-2$ a dimensionless number that parametrizes the scheme dependence. The perturbative scale is fixed to $\Lambda_{_{pQCD}}= 323.7~\text{MeV}$ by the condition $\alpha_s(m_{ren}=2~\text{GeV})=0.2994$ \cite{ParticleDataGroup:2008zun}.\footnote{If we take $\alpha_s$ rather than $\alpha_s^{-1}$ expanded to second order, the value is $\Lambda_{_{pQCD}}=377.9~\text{MeV}$.}

%%%%%%%%%%%%%%%%%%%%%%%%%%%%%%%%%%%%%%%%%%%%%%%
\subsection{Transport coefficients in holographic models}
%%%%%%%%%%%%%%%%%%%%%%%%%%%%%%%%%%%%%%%%%%%%%%%

Transport coefficients can be computed from correlators of conserved currents through the Green-Kubo formulas. Viscosities and thermal conductivities are obtained from correlators of the energy-momentum tensor, conductivities are  from correlators of electric or baryon currents, and thermoelectric conductivities from mixed correlators. All these can be extracted from the gravity dual by considering small (classical) perturbations of the metric and gauge fields around a black brane geometry \cite{Son:2002sd}. Compared to the kinetic and perturbation theory (quantum) derivation, the holographic calculation is consequently extremely simple. Furthermore, in many cases the value of DC transport coefficients is determined by the horizon geometry of an unperturbed classical black brane solution, i.e.~by thermodynamic quantities. This can be seen as a realization of the membrane paradigm first proposed in the 1980s \cite{Thorne:1986iy} as a way to understand black holes in GR by means of a fictional thin surface hovering just above the horizon. The fictitious membrane is described by a viscous fluid, and provides a natural framework to determine the low energy dynamics of the strongly coupled field theories described within the holographic correspondence \cite{Kovtun:2003wp}. This type of approach to transport in gravity duals has led to a powerful framework dubbed the fluid/gravity correspondence \cite{Bhattacharyya:2008jc}, which derives hydrodynamic equations from the gravity dual in systems which are close to thermal equilibrium. These ideas have been extensively extended to determine formulae for transport coefficients in terms of the near-horizon part of the solution to the equations of motion \cite{Iqbal:2008by,Eling:2011ms,Donos:2014yya,Gursoy:2014boa,Gouteraux:2018wfe}, thus avoiding the need to calculate fluctuations in very generic models. 

A paradigmatic example of the above is the Kovtun-Son-Starinets (KSS) relation \cite{Kovtun:2004de} between the shear viscosity and entropy density $s$, reading
\begin{equation}
 \frac{\eta}{s}=\frac{1}{4\pi}
\end{equation}
in natural units. The KSS relation indicates that strongly coupled systems with a gravitational dual favor diffusion of energy via thermal processes rather than mechanical ones. To date, strongly coupled plasmas with holographic duals are the most efficient systems known with this property. The KSS relation has been proven to hold in isotropic backgrounds for any gravity dual consisting of Einstein gravity coupled to matter \cite{Iqbal:2008by}. Anisotropies and deviations from the strong coupling limit in the form of higher curvature corrections in the gravity dual may modify this relation \cite{Rebhan:2011vd,Buchel:2004di,Mamo:2012sy}. For a review on the shear viscosity to entropy density ratio in holography, see \cite{Cremonini:2011iq}.

Another example of a transport coefficient that can be determined by evaluating the background at the horizon corresponds to the formula for the bulk viscosity, $\zeta$, given in \cite{Eling:2011ms,Hoyos:2013cba}.\footnote{A caveat exists about whether the horizon formula coincides in all cases with the one obtained through Kubo formulas \cite{Gubser:2008sz}, but so far there are no known counterexamples \cite{Buchel:2011wx}.} In contrast to the shear viscosity formula, it does not have a simple expression in terms of quantities defined on the field theory side, but it is determined by the value of the gravity fields at the horizon. Quite generically, the breaking of conformal invariance necessary to have a non-vanishing bulk viscosity is produced through relevant couplings to scalar operators. In the gravity dual this is realized by non-trivial profiles for a set of scalar fields $\phi_i$ that take values $\phi_i^H$ at the horizon. These values depend on the entropy and charge densities $s,\,\rho^a$, in such a way that the bulk viscosity is given by
\begin{equation}
 \frac{\zeta}{\eta}=\sum_{i} \left(s\frac{\partial \phi_i^H}{\partial s}+\rho^a\frac{\partial \phi_i^H}{\partial \rho^a} \right)^2 \ .
\end{equation} 
This formula can be generalized to other sources of breaking of conformal invariance, like vector operators inducing Lifshitz scaling \cite{Hoyos:2013cba}. Among the assumptions going into this formula one is that there are no charges outside the black brane horizon in the gravity dual, so in principle it does not apply to states with spontaneous symmetry breaking described by holographic superconductors \cite{Hartnoll:2008vx,Hartnoll:2008kx}. Deviations from strong coupling in the form of higher curvature corrections to the gravity dual affect to the value of both the shear and bulk viscosity \cite{Buchel:2018ttd}, so they may also modify the formula above.

Conductivities associated to currents of global symmetries in the field theory can also be extracted from the gravity solutions at the horizon, even for inhomogeneous solutions \cite{Donos:2014yya,Donos:2014cya,Donos:2018kkm,Gouteraux:2018wfe}. In a translationally invariant fluid an applied electric field or a forced temperature gradient will accelerate the charges and neutral components of the fluid in such a way that the electric and heat currents will grow with time. This implies that the DC conductivities are formally infinite, while the AC conductivity has a pole at zero frequency with a coefficient proportional to the charge and/or entropy density of the fluid. In a completely neutral fluid, the electric and thermoelectric DC conductivities remain finite and are determined by the coefficient $\sigma$ in the constitutive relations for a relativistic fluid. This finite contribution has been dubbed ``incoherent'' DC conductivity \cite{Davison:2015bea,Davison:2015taa}. There is an incoherent contribution to all the conductivities, and it is present in both charged and neutral fluids. 

For a fluid in equilibrium or in a steady state, the forces acting on it have to be compensated in such a way that the net acceleration of its components vanishes. In this case the conductivity consists solely of the incoherent contribution, which produces transport through purely diffusive processes. Denoting the temperature gradient as $\zeta_i=-\nabla_i T/T$ the no-force condition on a relativistic isotropic fluid reads
\begin{equation}
 \rho E_i+ Ts \zeta_i=0 \ .
\end{equation}
Since the electric and thermal gradients are not independent, it would be redundant to discuss thermoelectric conductivities. The electric and heat currents are proportional to the electric field and temperature gradient
\begin{equation}
 J^i=\sigma^{ij}E_j\ , \ Q^i=\kappa^{ij}\zeta_j \ ,
\end{equation}
with electric and heat conductivities that in an isotropic relativistic fluid take the forms
\begin{equation}\label{eq:finitecond}
 \sigma^{ij}=\frac{Ts}{\varepsilon+p}\sigma \delta^{ij}, \ \ \kappa^{ij}=\frac{\mu s}{\rho} \sigma^{ij} \ .
\end{equation}

The values of the electric and heat currents can be read of from fluxes that remain constant along the holographic radial direction. Assuming the electric field and temperature gradient both point along the $x$ direction, the fluxes associated to each current are
\begin{equation}
 {\cal J}=\sqrt{-g} Z F^{rx} \ ,  \ {\cal Q}=2 \sqrt{-g} G^{rx} \ ,
\end{equation}
where $F_{MN}$ are the components of the field strength of the gauge field $A_M$ dual to the current and $G_{MN}\sim \nabla_M k_N$ are the components of a two-form constructed from the Killing vector generating time translations, $k=\partial_t$ \cite{Donos:2014cya,Gouteraux:2018wfe}. The explicit forms of the coefficient $Z$ and the two-form $G$ depend on the action of fields in the gravity dual. In order to extract the conductivities, one introduces a perturbation of the gauge field and metric,
\begin{equation}
 \delta g_{tx}  \sim  \zeta_x\, t \ , \ \delta A_x  \sim  - \left(E_x - A_t\, \zeta_x \right) t   \ ,
\end{equation}
and evaluates the fluxes ${\cal J}$ and ${\cal Q}$ at the horizon. The conductivities can be read from the coefficients of $E_x$ and $\zeta_x$ in the resulting expression for the fluxes. Unfortunately, it seems that no simple expressions exist for them.

In addition to this general method, when the action for the gauge fields is of the DBI type (involving a square root), there is a method for probe flavor branes developed in \cite{Karch:2007pd}. In this method, the value of the conductivity is determined by demanding that the action remains real when an electric field is introduced at a fixed value of charge. In this case, the fields are not evaluated at the black brane horizon, but at an effective horizon that appears in the induced metric of the brane when the electric field is turned on.

%%%%%%%%%%%%%%%%%%%%%%%%%%%%%%%%%%%%%%%%%%%%%
\subsection{Applications of holographic transport results to compact stars}
%%%%%%%%%%%%%%%%%%%%%%%%%%%%%%%%%%%%%%%%%%%%%%

\begin{figure}[t!]
\begin{center}
\begin{tabular}{cc}
\includegraphics[width=0.42\textwidth]{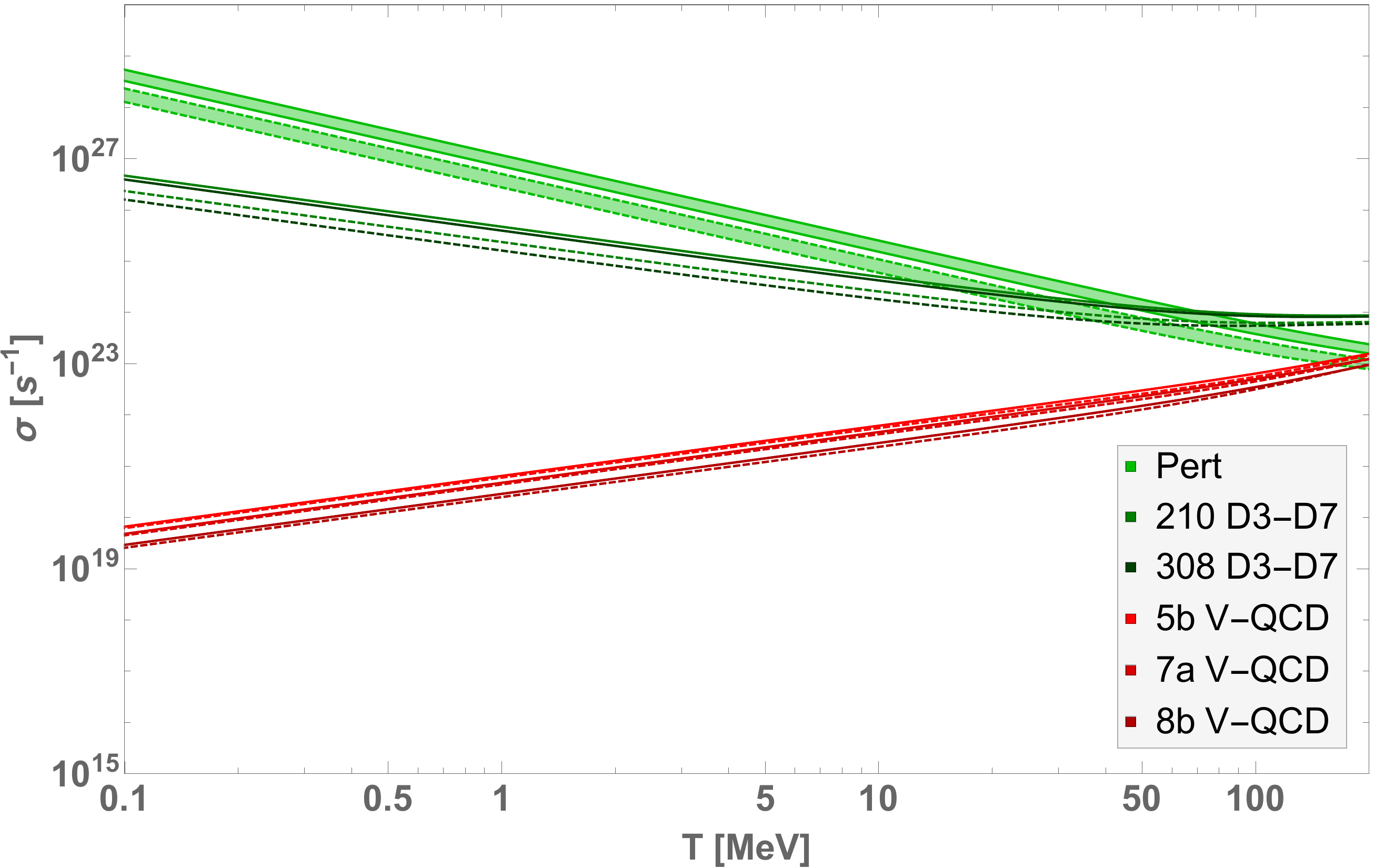}  $\;\;\;\;$ \includegraphics[width=0.42\textwidth]{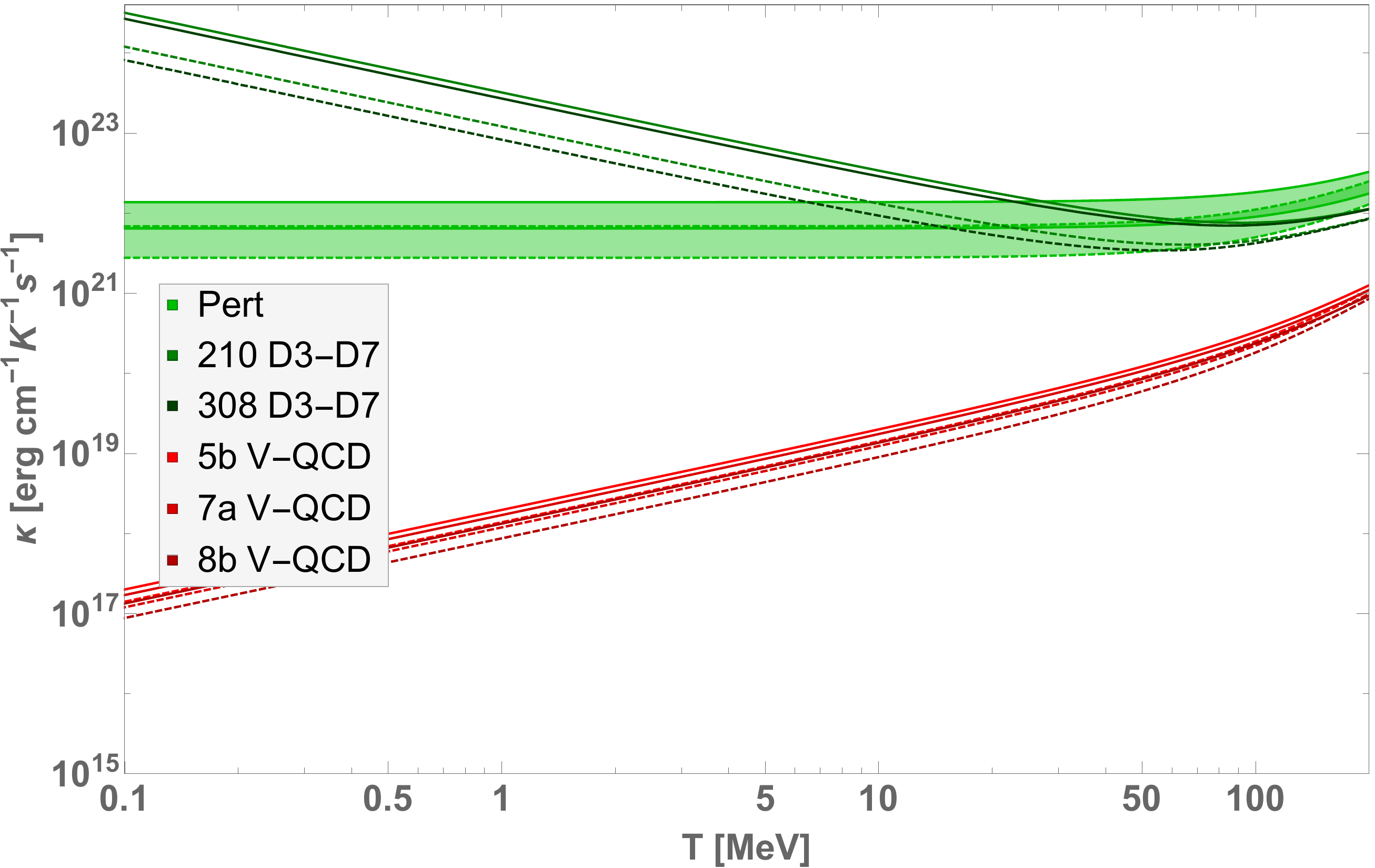}
\end{tabular}
\end{center}
\caption{The electric (left) and heat (right) conductivities as functions of temperature. The light green shaded areas correspond to the perturbative estimates, with the bands signifying uncertainties in the choice of the perturbative renormalization scale. The dark green lines correspond to the D3-D7 model for different values of the constituent quark masses indicated in MeV, and the reddish lines to the V-QCD model for different fits to lattice data, explained in \cite{Hoyos:2020hmq,Hoyos:2021njg}.The dashed lines correspond to a quark chemical potential $\mu=450$ MeV and the solid ones to $\mu=600$ MeV.
}\label{fig:cond}
\end{figure}

The holographic study of the transport properties of quark matter in neutron stars was initiated in \cite{Hoyos:2020hmq,Hoyos:2021njg} for the D3-D7 and V-QCD models. Numerical results comparing the perturbative and holographic values to each other are shown in fig.~\ref{fig:cond}  for the conductivities and in fig.~\ref{fig:visc} for the viscosities. They display remarkably different behaviors, even qualitatively speaking.  An analytic understanding of this difference can be grasped from the low-temperature scaling of transport coefficients. As the temperature decreases, the perturbative Debye mass of eq.~\eqref{eq:mD} saturates at a value fixed by the chemical potential $m_D\sim \mu$. From the expressions obtained for the perturbative transport coefficients \eqref{eq:pertrans}, both the shear viscosity and conductivity increase at even lower temperatures as an inverse power of $T$, while the thermal conductivity approaches a constant,
\begin{equation}\label{eq:pertrans}
 \eta_{\text{per}}\sim T^{ -5/3}\ ,  \ \sigma_{\text{per}}\sim T^{-5/3} \ ,  \ \ \kappa_{\text{per}} \sim T^0 \ .
\end{equation}
In the perturbative calculation, quarks are responsible for the transport of momentum and charge. As the temperature decreases, the mean free path increases, explaining qualitatively why the shear viscosity and conductivity are expected to grow. On the other hand, there will be less thermal energy available to transfer. In the free fermion gas picture, a constant thermal conductivity would be obtained for a relaxation time $\tau\sim 1/T$.

\begin{figure}[t!]
\begin{center}
\begin{tabular}{cc}
\includegraphics[width=0.42\textwidth]{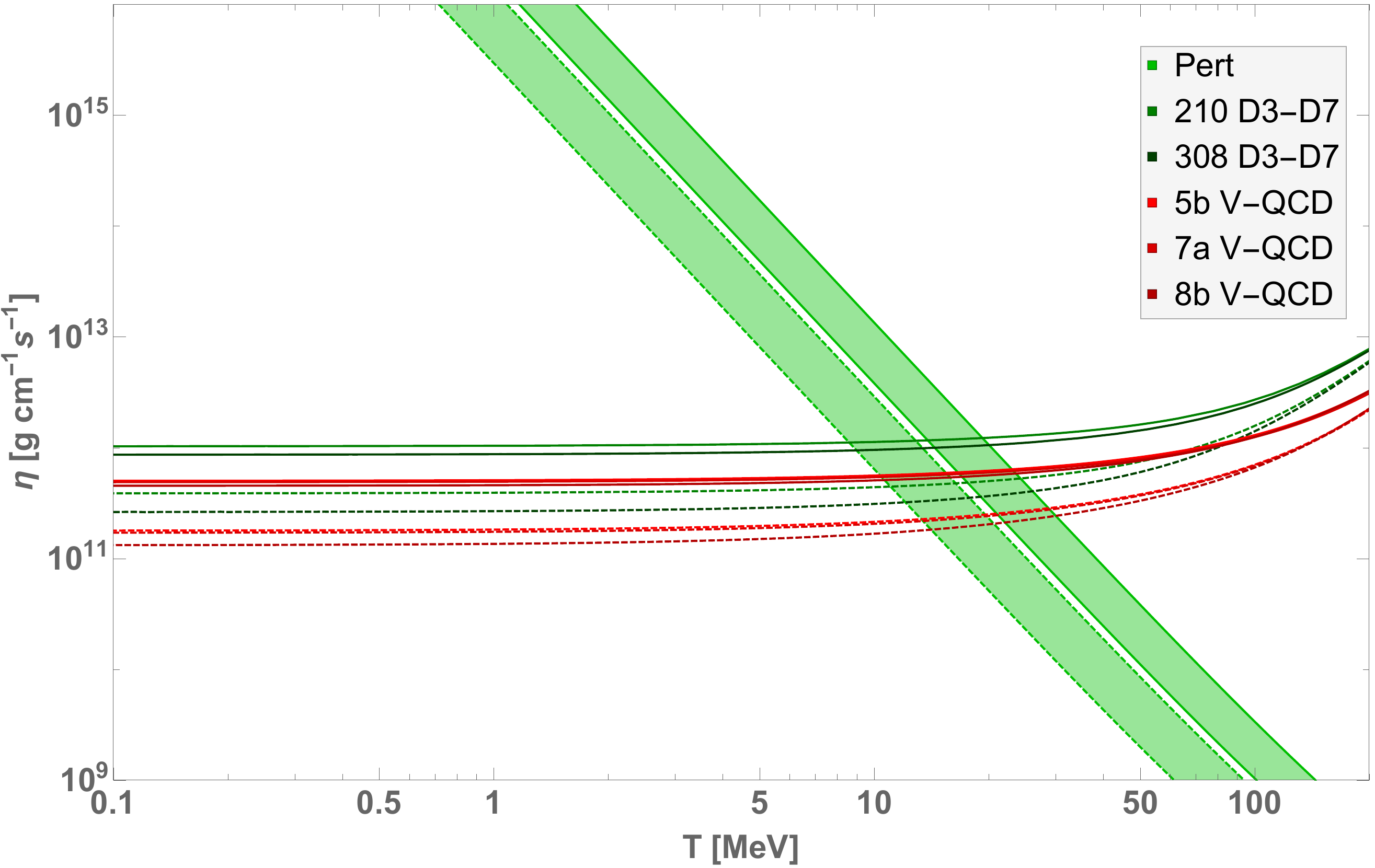}  $\;\;\;\;$ \includegraphics[width=0.42\textwidth]{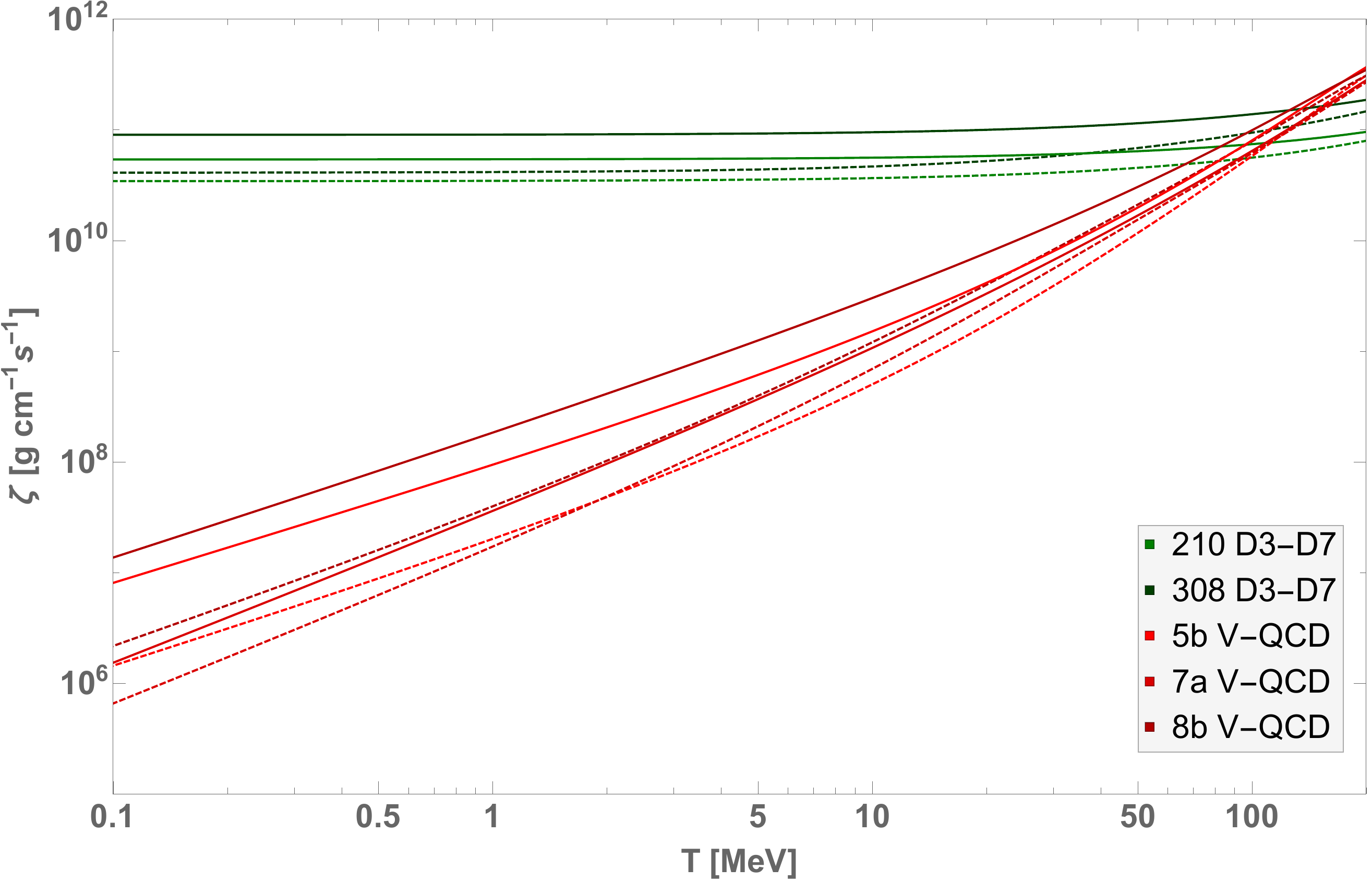}
\end{tabular}
\end{center}
\caption{The shear (left) and bulk (right) viscosities as functions of temperature. The plots follow the same conventions as those seen in fig.~\ref{fig:cond}. In the bulk viscosity plot the perturbative result is absent because it has been estimated to be negligible at large baryon density.}\label{fig:visc}
\end{figure}

The temperature dependence of the transport coefficients in pQCD is in contrast with the low-temperature behavior of the shear viscosity in holographic models. At very low temperatures, the shear viscosity reaches a fixed value determined by the chemical potential
\begin{equation}
 \eta_{_{D3D7}}\sim \eta_{_{VQCD}}\sim T^0 \ .
\end{equation}
For the D3-D7 model, it is furthermore possible to find an analytic expression for the leading contribution at low temperatures,
\begin{equation}\label{eq:etalowT}
    \eta_{_{D3D7}}  = \frac{\Nf\, \Nc }{8\pi\gamma^3 \sqrt{\laYM}}\, \mu(\mu^2-m^2) \ ,
\end{equation}
where $m$ is the constituent mass of the quarks and $\mu$ the quark chemical potential. In both of the holographic models considered here this is directly related to having nonzero entropy density even in the zero-temperature limit. 

Although quite general, the above behavior is not universal in all holographic models. From the gravity dual perspective, the zero-temperature state in the V-QCD model is an extremal black brane. Depending on the field content in the gravitational theory, the true zero-temperature state could be a geometry without a horizon, in which case the entropy would vanish. In this case, the shear viscosity is expected to decrease with the temperature, a behavior that would deviate even further from the perturbative result. The situation is slightly different in the D3-D7 model. In this case, the charge density on the brane can be understood as originating from strings extended between the D7-brane and the horizon \cite{Kobayashi:2006sb,Mateos:2007vc}. The density of strings will depend on the chemical potential as $\sim \mu^3$, while the energy of each string brings a contribution proportional to the thermal correction to the quark mass, $\Delta m\sim -T$ \cite{Herzog:2006gh}. Overall, this produces a contribution to the free energy linear in the temperature that results in a finite entropy density. The finite value of the $T=0$ shear viscosity is thus not expected to be a universal feature of holographic models based on probe branes. Generically the temperature dependence will depend on the thermal correction to the quark mass computed through a string in the holographic dual, a quantity that can vary for different background geometries. However, the thermal correction to the quark mass should be decreasing with the temperature, so if the shear viscosity follows a power-law dependence, $\eta \sim T^\alpha$, we can set a bound $\alpha>-1$, which is above the value seen in the perturbative result of eq.~\eqref{eq:pertrans}.

Turning now to conductivities, in contrast to the shear viscosity we see quite different behaviors depending on the model. In the D3-D7 model, conductivities increase as the temperature is lowered according to
\begin{equation}
 \sigma_{_{D3D7}}^{xx}\sim T^{ -1}\ ,  \ \kappa_{_{D3D7}}^{xx}\sim T^{-1} \ ,
\end{equation}
with the precise analytic forms at leading order reading
\begin{equation}\label{eq:sigmaxxZT}
    \sigma_{_{D3D7}}^{xx}= \frac{ \Nf\, \Nc\, }{\pi\, \gamma^3\laYM}\, \frac{(\mu^2-m^2)}{T} \ , \
\kappa_{_{D3D7}}^{xx} = \frac{ \Nf\, \Nc\, }{2\, \pi\, \gamma^3\sqrt{\laYM}}\, \frac{\mu\, (\mu^2-m^2)}{T} \ .
\end{equation}
On the contrary, in the V-QCD model the conductivities decrease with the temperature according to
\begin{equation}
 \sigma_{_{VQCD}}^{xx}\sim T \ ,  \ \kappa_{_{VQCD}}^{xx}\sim T \ .
\end{equation}
The reason behind this difference is likely the different treatment of flavors in the two models. In the D3-D7 model, the effect of the flavor degrees of freedom is captured in the holographic dual via the probe D7-branes. The absence of backreaction on the geometry amounts to neglecting the effect of flavor on the glue on the field theory side. As the temperature is lowered, the drag produced by the glue on the flavors becomes smaller, which effectively favors a larger mean free path, although this picture is not quite precise, as in the strongly coupled theory there is no good quasiparticle description. In the V-QCD model, the linear dependence on temperature arises from the explicit factor in eq.~\eqref{eq:finitecond}. The rest of the factors, including $\sigma$, remain finite and are determined by the chemical potential as the temperature vanishes.

Finally, as opposed to the naive expectation based on perturbative calculations, the bulk viscosity is smaller but not negligible compared to the shear viscosity. Just as seen in the conductivities, it also exhibits a different behavior in the D3-D7 and V-QCD models,
\begin{equation}
 \zeta_{_{D3-D7}}\sim T^0\ , \ \zeta_{_{VQCD}}\sim T^\alpha, \ 1\leq \alpha < 2 \ ,
\end{equation}
with the leading low-$T$ contribution in the D3-D7 model being
\be\label{eq:zetalowT}
 \zeta_{_{D3D7}} =  \frac{\Nf\, \Nc}{2\pi\gamma^3 \sqrt{\laYM} }m^2 \frac{\mu(\mu^2 - m^2)^2}{(3\mu^2 - m^2)^2} \ .
\ee
The value of $\alpha$ in the V-QCD model depends on the details of the gravitational action. The different qualitative behavior between the two models might be related to the probe approximation in the D3-D7 model, but also to the different sources of breaking of conformal invariance in each model. In the D3-D7 model, the breaking is produced by quark masses. The bulk viscosity is found to be proportional to the quark mass and to the density of quarks. On the other hand, in the V-QCD model the quarks are massless while conformal invariance is broken by the non-trivial beta function of the gauge coupling. Changes in the gravitational action result in changes in the running of the coupling constant with scale, and these are also reflected in the dependence of the bulk viscosity on the temperature.

Summarizing, the values of the transport coefficients in the holographic models studied so far point to a largely different qualitative behavior compared to perturbative calculations. There are some differences among the holographic models, stemming from different approximations. Since the V-QCD model does not rely on a quenched approximation, unlike the D3-D7 model, and it is fitted to lattice QCD results, the values obtained for the transport coefficients are {\em a priori} more likely to capture the correct physical behavior of transport coefficients in real QCD.

%%%%%%%%%%%%%%%%%%%%%%%%%%%%%%%%%%%%%%%%%%%%%%%%%%%%%%%%%%%%%%%%%%%%%%%%%%%%%%%%%%%%%%
\begin{figure}[t!]
\begin{center}
\includegraphics[width=0.65\textwidth]{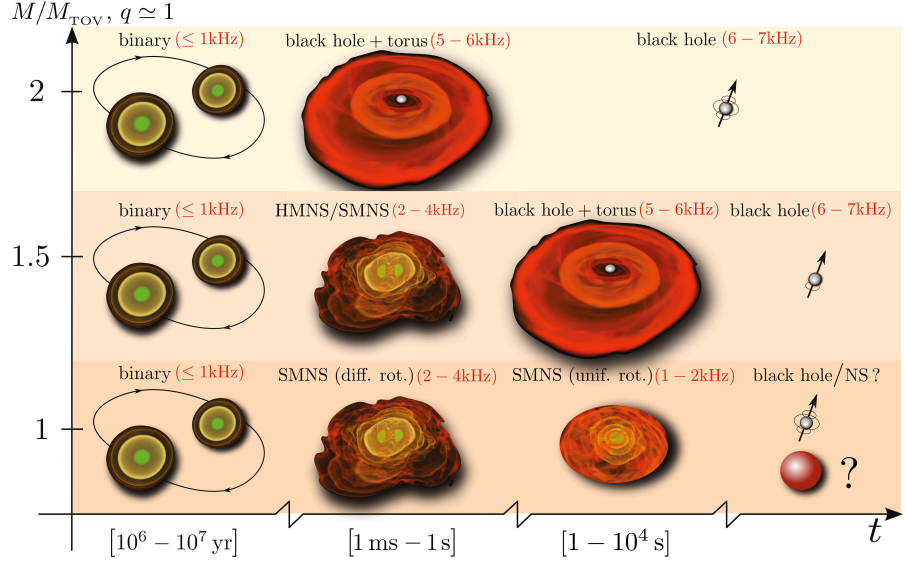} 
\end{center}
\caption{Possible stages and characteristic times in the evolution of a binary merger. After a long period of time, the amplitude and frequency of the GWs strongly increase in the final moments of the inspiral phase. After the merger, if the masses are large enough, there may be a prompt collapse to a black hole surrounded by an accretion disk, as shown in the upper row of the plot. For smaller masses there may be an intermediate hypermassive star that, held together by differential rotation, that eventually collapses to a black hole as in the middle row. Finally, for even smaller masses, the remnant may be a supramassive star supported by uniform rotation, that  after a longer period of time either collapses to a black hole or produces a stable compact star (bottom row). The plot was taken from \cite{Baiotti:2016qnr}.}\label{fig:evol}
\end{figure}
%%%%%%%%%%%%%%%%%%%%%%%%%%%%%%%%%%%%%%%%%%%%%%%%%%%%%%%%%%%%%%%%%%%%%%%%%%%%%%%%%%%%%%

%%%%%%%%%%%%%%%%%%%%%%%%%%%%%%%%%%%%%%%
%%%%%%%%%%%%%%%%%%%%%%%%%%%%%%%%%%%%%%%
\section{Binary mergers of compact objects}\label{sec:mergers}
%%%%%%%%%%%%%%%%%%%%%%%%%%%%%%%%%%%%%%%
%%%%%%%%%%%%%%%%%%%%%%%%%%%%%%%%%%%%%%

The first indirect evidence for the existence of GWs came from an observed decrease in the orbital period of the Hulse-Taylor binary pulsar, just as predicted by general relativity \cite{Taylor:1982zz,Weisberg:2010zz}. In such a binary system, two compact stars or black holes rotating around each other lose energy through the emission of GWs, going through an inspiral period and eventually colliding and merging. After the inspiral phase, the collision of two compact stars can produce either a new compact star or a black hole, depending on the masses of the binary components. The black hole may similarly be formed by prompt collapse or after an intermediate stage where a hypermassive object is formed. The main stages in the evolution of a merger of two compact objects are depicted in fig.~\ref{fig:evol}.

During the inspiral phase the GWs that are emitted  typically have frequencies below $1$ kHz, which are within the sensitivity window of the present GW detectors such as LIGO, Virgo, and KAGRA (see fig.~\ref{fig:sens}). During the collision and post-merger evolution, the frequency of GWs increases above the kHz range thus escaping the detectability range, although it may be within the reach of the next generation of observatories, such as advanced LIGO or the Einstein Telescope. In addition, from the observation of the emitted electromagnetic radiation and neutrinos one may extract some information about the post-merger evolution; see also our discussion in sec.~\ref{sec:observations}.

%%%%%%%%%%%%%%%%%%%%%%%%%%%%%%%%%%%%%%%%%%%%%%%%%%%%%%%%%%%%%%%%%%%%%%%%%%%%%%%%%%%%%%
\begin{figure}[t!]
\begin{center}
\includegraphics[width=0.65\textwidth]{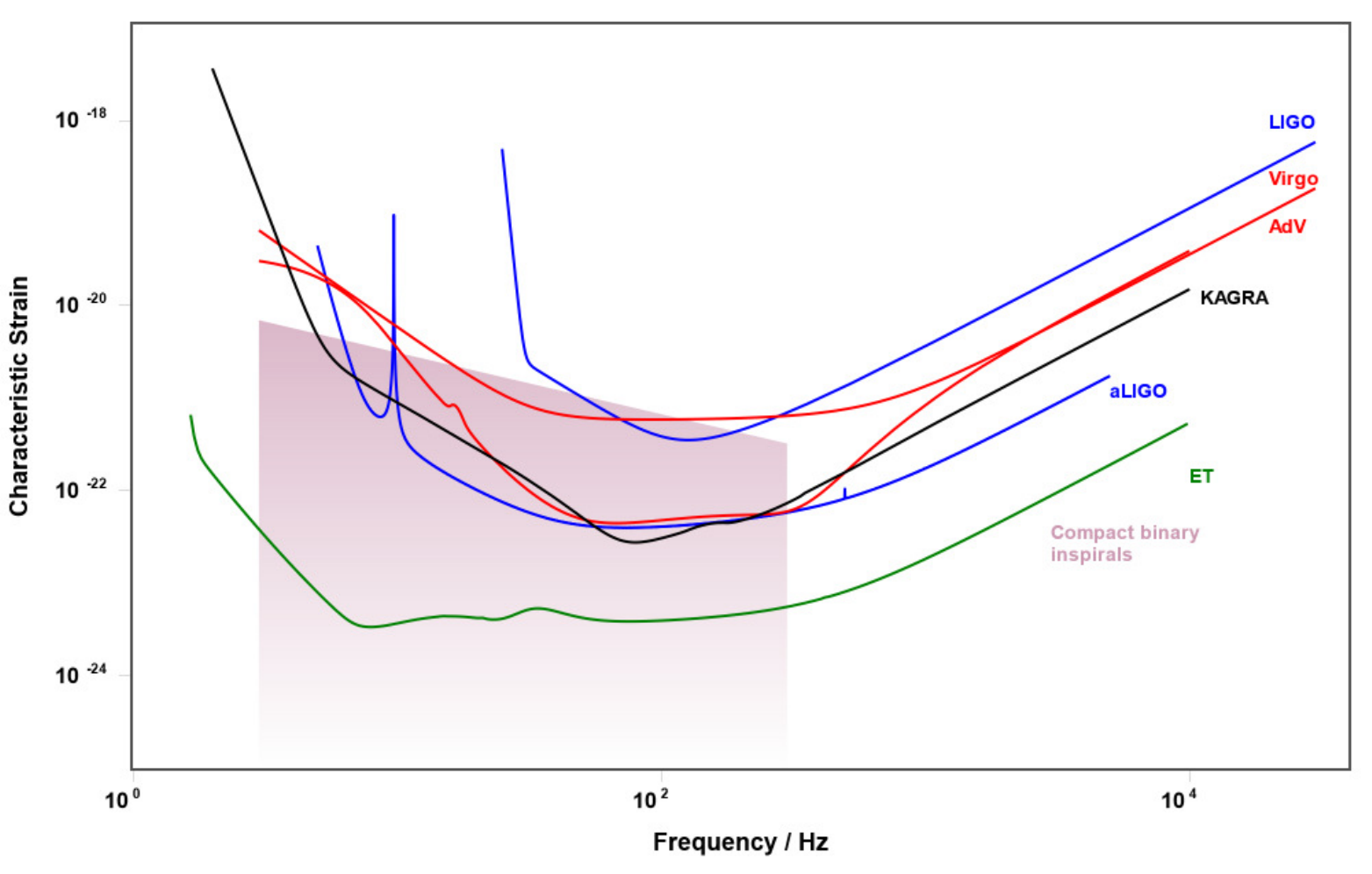} 
\end{center}
\caption{GW detector sensitivities as given by the strain amplitude as a function of frequency for LIGO and  advanced LIGO (blue), Virgo and  advanced Virgo (red), KAGRA (black) and the Einstein Telescope (green). The pink shadow corresponds to the expected signal from the inspiral phase of binary mergers. Image made with \url{http://gwplotter.com/}, based on \cite{Moore:2014lga}.}\label{fig:sens}
\end{figure}
%%%%%%%%%%%%%%%%%%%%%%%%%%%%%%%%%%%%%%%%%%%%%%%%%%%%%%%%%%%%%%%%%%%%%%%%%%%%%%%%%%%%%%

A GW signal from the final stages of the inspiral period, when the amplitude is large enough to be detectable by modern GW observatories on Earth, was first observed in 2015 by the LIGO collaboration and corresponded to the merger of two black holes \cite{LIGOScientific:2016aoc}. The observation of the first neutron-star merger was subsequently made in 2017 by the LIGO and Virgo collaborations \cite{TheLIGOScientific:2017qsa}, accompanied by a strong transient electromagnetic signal \cite{LIGOScientific:2017ync}, discussed in some length above in sec.~\ref{sec:observations}. Numerous binary merger observations have been made since then, with a particularly interesting neutron-star merger observed in 2019 \cite{LIGOScientific:2020aai}.

%%%%%%%%%%%%%%%%%%%%%%%%%%%%%%%%%%
\subsection{Tidal deformability}
%%%%%%%%%%%%%%%%%%%%%%%%%%%%%%%%%%

The GW signal produced during the inspiral is naturally affected by the properties of the components forming the binary (see e.g.~\cite{Baiotti:2016qnr,Dietrich:2020eud,Chatziioannou:2020pqz} for reviews on the topic). In particular, the shape of a compact star is deformed by the tidal forces produced by its companion, when the two objects come close together. This in turn modulates the GW signal, modifying its phase and frequency. The deformation in the shape of a star depends on the material properties of the matter inside it, so the GW signal provides indirect information about the properties of the EoS of matter in the interior of the star. The main quantity that can be extracted from the signal (besides the masses of the binary components) is the dimensionless tidal deformability
\begin{equation}
 \Lambda=\frac{2}{3C^5}k_2 \ , \ C=\frac{G M}{c^2R} \ .
\end{equation}
where $M$ and $R$ denote the mass and radius of the star, $C$ stands for its compactness, and $k_2$ its quadrupole Love number. To be more precise, the GW phase is  determined by the combined tidal deformability
\begin{equation}
 \tilde{\Lambda}=\frac{16}{13}\frac{(M_1+12 M_2)M_1^4\Lambda_1+(M_2+12 M_1)M_2^4x\Lambda_2}{(M_1+M_2)^5} \ ,
\end{equation}
where the subindex refers to each component of the binary system. In the simplified scenario of approximately identical stars $M=M_1\approx M_2$ and $\Lambda=\Lambda_1\approx \Lambda_2$, the combined tidal deformability coincides with the individual tidal deformability of each star, $\tilde{\Lambda}=\Lambda$.

On the other hand, black holes have zero tidal deformability \cite{Fang:2005qq,Damour:2009vw,Binnington:2009bb}.\footnote{There has been some discussion concerning  this result, but the conclusion is that it holds for non-dissipative tidal forces in four dimensions, see e.g.~\cite{Charalambous:2021mea} and references therein.} Then, in a merger of a  black hole and a neutron star, assuming that the black hole mass is larger than that of the neutron star, $M_{BH} > M_{NS}$, the combined tidal deformability is suppressed as 
\begin{equation}
 \tilde{\Lambda}\sim \left(\frac{M_{NS}}{M_{BH}}\right)^4\Lambda_2 \ .
\end{equation}
This makes the quantity hard to measure from the GW signal, and indeed no significant constraints have been found from such asymmetric mergers \cite{LIGOScientific:2021qlt}. As discussed already in sec.~\ref{sec:observations}, the constraints on the combined tidal deformability obtained from the two best neutron-star merger recordings are bounded by the 90\% credible intervals/upper limits \cite{LIGOScientific:2018hze,LIGOScientific:2018cki,LIGOScientific:2021qlt}
$$
\begin{array}{ccc}
\tilde{\Lambda} & \text{Low-spin prior} & \text{High-spin prior} \\ \hline
\text{GW}170817\  & \lesssim 720 & \lesssim 630\\
\text{GW}190425 & \lesssim 600 & \lesssim 1100
\end{array}
$$
Here, the low and high spin refer to the rotation of the stars. All the stars involved in these mergers have masses not too far from the typical value for pulsars, $M=1.4 M_\odot$. One can estimate the tidal deformability of a typical neutron star as $70\lesssim \Lambda_{1.4}\lesssim 580$ (low-spin) or $\Lambda_{1.4}\lesssim 1400$ (high-spin). Lower bounds on the tidal deformability have been estimated to be of the order $\tilde{\Lambda}\gtrsim 70-200$ \cite{LIGOScientific:2018hze,LIGOScientific:2018cki,De:2018uhw,LIGOScientific:2021qlt,Dietrich:2020eud}. A comparison of the tidal deformability derived from observations with the prediction from neutron-star modeling allows to impose further constraints on the EoS. As discussed in \ref{sec:interpolation}, a general study for piecewise-defined EoSs was performed in \cite{Annala:2017llu}, subsequently leading to indications for the existence of a quark-matter core in heavy neutron stars \cite{Annala:2019puf}. 

Starting with \cite{Annala:2017tqz}, there have been several works determining tidal deformabilities for stars built with a holographic component in the EoS with interesting results \cite{Jokela:2018ers,Zhang:2019tqd,Jokela:2020piw,BitaghsirFadafan:2020otb,Demircik:2020jkc}. Purely holographic EoSs based on the WSS model do not lead to compact stars compatible with both the tidal deformability constraints and other astrophysical observations \cite{Zhang:2019tqd}. The situation is better for hybrid EoSs, which are determined by some nuclear effective theory at low densities and by holographic models only  at higher densities. As discussed in sec.~\ref{sec:thermodynamics}, the phase transition to quark matter in holographic models is typically strongly first order, preventing the formation of stable stars with deconfined matter in their cores; this has been seen in both the D3-D7 \cite{Annala:2017tqz} and V-QCD models \cite{Jokela:2018ers}. The D3-D7 hybrid EoS allows for very exotic stars with quark matter at the surface or close to it \cite{Annala:2017tqz}, which turn out to be compatible with constraints from tidal deformability and mass measurements. The holographic EoS becomes more relevant in phenomenological deformations of the D3-D7 model that stiffen the quark-matter EoS \cite{BitaghsirFadafan:2019ofb}. In the deformed D3-D7 model, the hybrid EoS allows quark-matter cores compatible with the tidal deformability and other astrophysical bounds  \cite{BitaghsirFadafan:2020otb}.  Similarly, the hybrid EoS of the V-QCD model can be improved allowing the holographic model to describe the high density hadronic phase of matter close to the core. In this case, there are compact stars compatible with both the tidal deformability bounds and other astrophysical observations \cite{Jokela:2020piw,Demircik:2020jkc}.

%%%%%%%%%%%%%%%%%%%%%%%%%%%%%%%%%%
\subsection{Merger simulations}
%%%%%%%%%%%%%%%%%%%%%%%%%%%%%%%%%%

Determining the precise evolution of a binary merger  process and the emitted GW spectrum is a complicated  problem in numerical relativity. As discussed before, the main observable of relevance during the inspiral phase of a neutron-star merger is the tidal deformability. In the post-merger evolution, numerical simulations show that, if there is no prompt collapse to a black hole, the GW spectrum is peaked around three characteristic frequencies in the few kHz range, $f_1$, $f_2$, $f_3$, with $f_1<f_2\approx (f_1+f_3)/2<f_3$. A possible mechanism producing the peak frequencies $f_1$ and $f_3$ is through changes in the moment of inertia of the star due to the radial oscillation of the core, in which case $f_2$ would be related to the speed of rotation after the oscillation has been damped by dissipative effects \cite{Takami:2014tva}.

In the simulations of mergers, the energy-momentum tensor of matter is taken to be that of a fluid, with an EoS that initially is that of a cold quiescent star $p_c(\rho)$, with $p_c$ the pressure and $\rho$ the mass density. Heating produced during the collision is typically modeled by adding a thermal component to the pressure and the energy density according to $p=p_c+p_{th}$, $\varepsilon=\varepsilon_c+\varepsilon_{th}$ where
\begin{equation}
 p_{th}=\Gamma_{th}\rho (\varepsilon-\varepsilon_c)   \ , \ \varepsilon_{th}=\varepsilon-\varepsilon_c\ .  
\end{equation}
Here, $\Gamma_{th}\approx 1.5-2$ is the so-called thermal index that is usually taken  to stay constant for all densities, following an ideal gas approximation. From numerical simulations, the GW signal can be extracted from the Newman-Penrose or Weyl scalar $\psi_4$ \cite{Bishop:2016lgv}, proportional to some components of the Weyl tensor \cite{Newman:1961qr}. At the linearized level around flat space, $g_{\mu\nu}=\eta_{\mu\nu}+h_{\mu\nu}$, the Newman-Penrose scalar encodes the amplitude of plus $(h_+)$ and cross $(h_\times)$ polarizations of GWs in the transverse traceless gauge
\begin{eqnarray}
 \psi_4=\partial_t^2(h_+-ih_\times) \ .
\end{eqnarray}

The GW signal produced by the hybrid holographic EoSs of the V-QCD model was studied in \cite{Ecker:2019xrw} using a thermal index $\Gamma_{t}=1.75$ (see also \cite{Jarvinen:2021jbd} for a review with additional plots). Considering equal mass stars, the three possible types of evolution depicted in fig.~\ref{fig:evol} were observed for a range of masses between $1.3 M_\odot - 1.5 M_\odot$, with the lower masses leading to an apparently stable remnant, higher masses to prompt black-hole collapse, and intermediate masses $M\sim 1.4 M_\odot$ first to a hypermassive remnant that after a few milliseconds collapses to a black hole.

\begin{figure}[t!]
\begin{center}
\begin{tabular}{cc}
\includegraphics[width=0.48\textwidth]{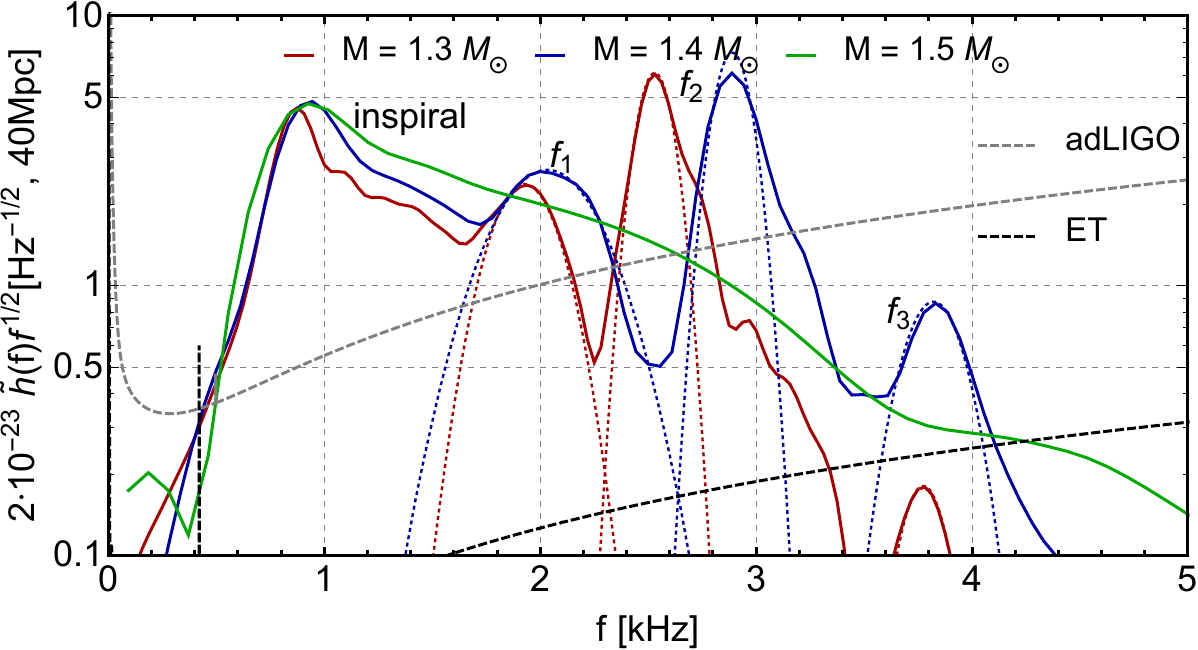}  \;\; \includegraphics[width=0.37\textwidth]{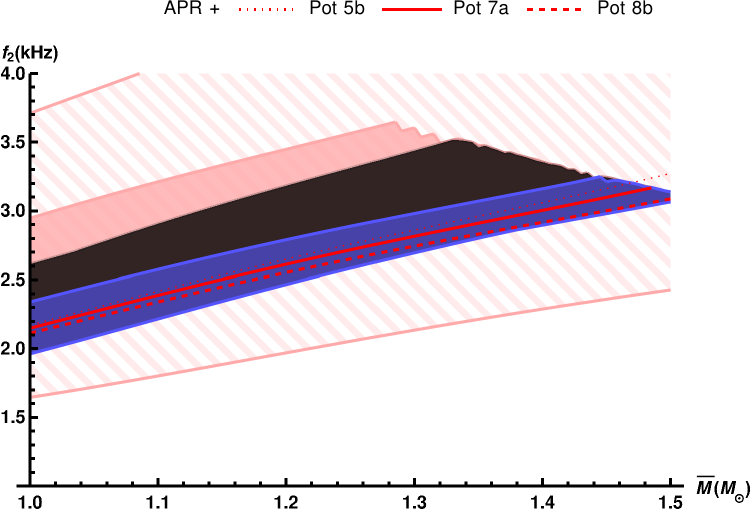}
\end{tabular}
\end{center}
\caption{Left: The GW spectrum obtained from numerical simulations with  a hybrid holographic EoS, for three masses representative of the three possible evolutions of the merger. Note the absence of peaks for the largest mass, for which there is prompt collapse to a black hole. The gray lines show the sensitivities of advanced LIGO and the Einstein Telescope  (plot from \cite{Ecker:2019xrw}). Right: The estimated $f_2$ frequency in equal-mass binaries obtained for hybrid holographic EoSs using universal relations \cite{Breschi:2019srl} (see also \cite{Jokela:2020piw,Jokela:2021bxn}); figure taken from~\cite{Jokela:2021vwy}. The dashed band covers the full set of hybrid EoSs, while the solid bands correspond to EoSs allowed by constraints from astrophysical observations, and the blue and black bands include constraints on the star radius from different analyses of NICER data \cite{Miller:2021qha} (blue) and \cite{Riley:2021pdl} (black). The cut indicates the values, for which one finds prompt collapse to a black hole.}\label{fig:gwmerger}
\end{figure}

In the merger simulations of \cite{Ecker:2019xrw}, the transition to quark matter as described by the holographic models was seen to lead to an immediate collapse to a black hole, since the holographic EoS for quark matter is much softer than the hadronic ones. As a consequence, no measurable amount of quark matter was produced. Interestingly, it turns out that the phase transition to quark matter during a merger might be detectable by comparing the tidal deformability before coalescence with the peak frequency after it \cite{Bauswein:2020ggy}. The peak frequencies for the hybrid EoSs were observed to be somewhat smaller than those of models without a holographic component,\footnote{The value of $f_2$ was estimated previously in \cite{Jokela:2020piw}, updated with recent NICER constraints in~\cite{Jokela:2021vwy} using universal relations from \cite{Breschi:2019srl}.} but still in the few kHz range. As seen in fig.~\ref{fig:gwmerger}, the peaks at the $f_1$ and $f_2$ frequencies for the hybrid EoSs have been shown to be observable with advanced LIGO and Einstein Telescope sensitivities at distances up to  $40$ Mpc from the merger.

%%%%%%%%%%%%%%%%%%%%%%%%%%%%%%%%%%%%%%%
%%%%%%%%%%%%%%%%%%%%%%%%%%%%%%%%%%%%%%%
\section{Open questions and future directions} \label{sec:openquestions}
%%%%%%%%%%%%%%%%%%%%%%%%%%%%%%%%%%%%%%%
%%%%%%%%%%%%%%%%%%%%%%%%%%%%%%%%%%%%%%%

Quantum Chromodynamics poses a formidable challenge in the limit of high baryon densities and small or moderate temperatures: the theory remains strongly coupled in all phenomenologically relevant settings, yet no nonperturbative first principles methods applicable for this part of its phase diagram exist \cite{Brambilla:2014jmp,deForcrand:2010ys}. To remedy the current situation for the systems of most pronounced observational activity --- neutron-star cores --- a number of novel approaches have been suggested. These include, e.g., the use of phenomenological models (see, e.g., \cite{Dexheimer:2009hi} and references therein), renormalization-group-inspired techniques \cite{Braun:2020bhy}, extrapolations of nuclear theory or perturbative QCD results outside their respective regions of validity \cite{Kurkela:2009gj}, or more recently, extrapolations and interpolations of physical quantities over the problematic density regime \cite{Hebeler:2013nza,Annala:2017tqz}. In the review article at hand, we have reviewed the foundations and practical results of one such method with a particularly solid theoretical background: the gauge/gravity duality, or holography. 

As discussed at length above, in its original formulation, the duality links a very nontrivial strongly coupled large-$N$ gauge theory to classical supergravity in a higher-dimensional spacetime, whereby very difficult field theory problems become mapped to (typically) tractable General Relativity calculations \cite{Maldacena:1997re,Gubser:1998bc}. Deformations of the original duality, featuring broken supersymmetry and conformal invariance \cite{Karch:2002sh,Sakai:2004cn}, have been introduced together with so-called bottom-up models, which are somewhat less rigorous but allow the description of quantum field theories closer to QCD \cite{Gursoy:2007cb,Gursoy:2007er}. Together, these approaches have solved a number of difficult open problems in particular in heavy-ion physics \cite{Casalderrey-Solana:2011dxg}, so it is only natural to ask whether they are applicable for the description of the high-density QCD matter found inside neutron-star cores.

The past ca.~5 years have witnessed increased activity in the study of holographic models describing strongly interacting fundamental-representation matter at high density, and in the subsequent application of the results to neutron-star physics. While the earliest such works concentrated on deformations of the original AdS/CFT conjecture by studying the unbackreacted D3-D7 model \cite{Hoyos:2016cob,Hoyos:2016zke}, the results have recently been generalized to both top-down (e.g.~the Sakai-Sugimoto) \cite{Kovensky:2021kzl} and bottom-up (e.g.~V-QCD) models \cite{Jokela:2018ers}. Similarly, while early activity concentrated almost solely on the EoS of quark matter, holographic calculations have recently addressed also the properties of strongly coupled nuclear matter \cite{Jokela:2020piw, Demircik:2020jkc} as well as transport in a dense and cold medium \cite{Hoyos:2020hmq,Ghoroku:2021fos}. Important challenges remain, however, and in this final section of our review, we want to discuss a number of directions, where we believe significant progress can be made in future years. The topics concern both technical improvements in previous calculations and novel physical phenomena that are yet to be addressed with holographic methods. As both past successes in other fields of physics and many of these examples highlight, the most important virtue of holography is that it allows addressing complicated physics questions that would be extremely challenging or downright intractable with traditional field theory tools.

\paragraph{Nonzero temperatures and magnetic fields}

After a short cooling period, the temperature of a neutron star becomes small enough so that a barotropic approximation to the EoS becomes valid, i.e.~the temperature can be altogether neglected \cite{Glendenning:1997wn}. The situation is, however, completely different in supernova explosions and neutron-star mergers, where it is imperative to account for the spatially and temporally varying temperatures \cite{Baiotti:2016qnr}. There are a few available nuclear matter EoSs that include $T$-dependence (see~\cite{Burgio:2021vgk} for a recent review), but most merger simulations currently use an approximation, where the so-called thermal index 
\begin{eqnarray}
\Gamma_\text{th}(T,n_B)&=&1+\frac{p(T,n_B)-p(0,n_B)}{\epsilon(T,n_B)-\epsilon(0,n_B)}\, 
\end{eqnarray}
is assumed to remain constant at all densities. Given that the properties of the GW signal of the merger are expected to be sensitive to the specific $T$-dependence of the EoS, and that many EoSs based on microscopic calculations do not agree with the constant-thermal-index approximation (see e.g.~\cite{Bauswein:2010dn} and references therein), it is clear that all holographic EoSs should be generalized to nonzero $T$. In the deconfined phase of the theory, it is typically straightforward to incorporate $T$-dependence by considering a black-hole geometry, which has indeed been done, e.g., in \cite{Chesler:2019osn}. In the confined phase, accounting for a nonzero temperature however requires going beyond the large-$N$ approximation, which has not been systematically implemented so far.

Another important generalization of current holographic calculations has to do with magnetic fields, which can also play an important role in the properties of individual neutron stars and their mergers. The strength of the magnetic fields generated in magnetars can reach values as high as $10^{10}-10^{12}$ T (see~\cite{Kaspi:2017fwg} for a review), which are the strongest found in nature except for the transient fields produced in heavy-ion collisions \cite{Kharzeev:2007jp}. Adding an external magnetic field in a holographic model is typically a relatively simple task, as this only requires turning on some of the spatial components of the gauge field dual to the baryon current e.g.~\cite{Erdmenger:2007bn,DHoker:2009mmn,DHoker:2009ixq,Jensen:2010vd,Evans:2010hi,DHoker:2010zpp,Kim:2010pu,Hoyos:2011us,Ammon:2012qs,Bergman:2012na,Gursoy:2018ydr,Ammon:2020rvg}. A generalization of holographic EoSs of dense matter to including magnetic fields should thus be relatively straightforward, but this has not been  explored so far. 

\paragraph{Differing quark masses}

So far, all holographic calculations aimed at describing nuclear or quark matter have made the same simplifying assumption: all dynamical quarks share the same mass. Recalling that the strange quark mass is neither negligibly small nor very large at the densities realized in neutron-star cores, the validity of this approximation can, however, clearly be questioned. A particularly striking example of a calculation where differing quark masses would play an important role is the determination of the bulk viscosity, the main contribution to which is expected to arise from chemical re-equilibration through electroweak processes (see, e.g.,  \cite{Kovensky:2021kzl} for recent discussion of the beta equilibration in the holographic context), such as
\begin{equation}
u+d \longleftrightarrow u+s
\end{equation}
for quark matter composed of the three flavors $u$, $d$, and $s$.

For a perturbation that changes the local baryon density with a characteristic frequency $\omega$ (related to the rotation frequency of the star), the effective bulk viscosity of three-flavor quark matter can be shown to take the approximate value \cite{Schmitt:2017efp,Madsen:1993xx}
\begin{eqnarray}
\zeta\approx \frac{\Gamma_1 B^2}{\Gamma_1 C^2+\omega^2} \ . 
\end{eqnarray}
Here, $\Gamma_1$ is the production rate of $d$ quarks for a small chemical imbalance between $d$ and $s$ quarks, $\delta\mu_d-\delta\mu_s$, and the two other constants are defined as $B=n_d\chi_d-n_s\chi_s$ and $C=\chi_d+\chi_s$, with $\chi_{d,s}=\frac{\partial \mu_{d,s}}{\partial n_{d,s}}$ standing for inverse susceptibilities. From the form of this result, it becomes clear that any setup that treats the $d$ and $s$ quarks in an identical manner is bound to miss this leading contribution to the bulk viscosity.

It turns out that the above bulk viscosity is maximized for $\Gamma_1 C^2\sim \omega^2$. Such a large viscosity dampens fluctuations that would otherwise spin down a rotating star, producing a ``stability window'' that allows pulsars to rotate with a  millisecond period \cite{Alford:2013pma}. Even though the physics related to this enhancement of the bulk viscosity is of  electroweak origin, the susceptibilities depend on QCD dynamics, in particular on the EoS. To this end, a determination of the high-density EoS in the limit of differing quark masses is imperative for a reliable computation of the bulk viscosity, and it is important to generalize existing holographic models in this direction.

\paragraph{EoS for an anisotropic system}

Most existing studies of neutron-star matter --- both holographic and others --- make the simplifying assumption that matter in neutron-star cores is isotropic, i.e.~the pressure is the same in the radial  and angular directions. Anisotropic EoSs are, however, not ruled out by observations, and furthermore produce stars with interesting differences relative to those built with isotropic EoS. In particular, anisotropic EoSs can evade Buchdahl's bound \cite{Buchdahl:1959zz} and reach compactnesses arbitrarily close to that of black holes. Studying the properties of compact stars in the black-hole limit has in fact already led to interesting insights \cite{Yagi:2015upa,Yagi:2016ejg,Alexander:2018wxr}. 

Strongly coupled anisotropic phases have been studied via holography in a variety of setups, including axionic/dilatonic sources \cite{Mateos:2011ix,Mateos:2011tv,Koga:2014hwa,Jain:2014vka,Banks:2015aca,Roychowdhury:2015cva,Roychowdhury:2015fxf,Misobuchi:2015ioa,Banks:2016fab,Avila:2016mno,Donos:2016zpf,Giataganas:2017koz,Itsios:2018hff,Arefeva:2018hyo,Liu:2019npm,Inkof:2019gmh}, electric \cite{Karch:2007pd,Albash:2007bk,Albash:2007bq} and magnetic fields \cite{Erdmenger:2007bn,DHoker:2009mmn,DHoker:2009ixq,Jensen:2010vd,Evans:2010hi,DHoker:2010zpp,Kim:2010pu,Hoyos:2011us,Ammon:2012qs,Gursoy:2016ofp,Gursoy:2018ydr,Gursoy:2020kjd,Ballon-Bayona:2020xtf}  or both \cite{Evans:2011tk,Kharzeev:2011rw,Donos:2011pn,Jokela:2015aha,Itsios:2016ffv,Arefeva:2018cli,Bohra:2019ebj,Bohra:2020qom,Arefeva:2020vae}, and in $p$-wave superfluids \cite{Gubser:2008wv,Ammon:2008fc,Basu:2008bh,Iizuka:2012iv,Donos:2012gg,Iizuka:2012wt}. Recently, smeared brane configurations in holographic models have also been used to discover phases with spontaneous anisotropy at zero density and temperature \cite{Hoyos:2020zeg}. 

When solving Einstein's equations, the regularity of solutions requires that the EoS becomes isotropic at the center of the star. This implies that the anisotropy corresponds to a spontaneous breaking of rotational invariance as opposed to an explicit one. In most holographic models, with the exception of smeared brane configurations and $p$-wave superfluids, this would require turning off an external source, such as electromagnetic fields, as one proceeds towards the center of the star. For a realistic application to compact stars, it would thus be interesting to generalize the smeared brane configurations to nonzero density and temperature and to derive the corresponding EoSs.

\paragraph{Quark pairing}

As discussed before, at asymptotically large densities, where QCD becomes perturbative and quark masses negligible, the ground state of three-flavor QCD is that of color-flavor-locked, or CFL, color-superconducting quark matter \cite{Alford:2000ze,Rajagopal:2006ig,Mannarelli:2006fy}. For the densities reached in the cores of physical neutron stars, the ground state is unknown, as it is unknown whether deconfined matter exists even inside the most massive stable neutron stars. In the case of quark matter, possibilities range from phases with totally and partially Higgsed color groups to ones featuring a spontaneous breaking of flavor symmetries, such as the 2SC \cite{Bailin:1983bm} and the  Larkin-Ovchinnikov-Fulde-Ferrell (LOFF) \cite{Larkin:1964wok,Fulde:1964zz} phases. Although paired phases have been studied in holographic models \cite{Chen:2009kx,Basu:2011yg,Rozali:2012ry,BitaghsirFadafan:2018iqr,Faedo:2018fjw,Henriksson:2019zph,Henriksson:2019ifu,Faedo:2019jlp,Ghoroku:2019trx,BitaghsirFadafan:2020otb,Ghoroku:2021fos}, a systematic study of their influence on the EoS and transport properties of matter in the cores of compact stars is still lacking.

\paragraph{Inhomogeneous and mixed phases}

A further possibility among phases potentially realized in neutron-star cores is that the ground state of QCD is inhomogeneous, such as in the case of the LOFF phase of paired quark matter. In holographic models, inhomogeneous phases have been considered both for deconfined \cite{Domokos:2007kt,Nakamura:2009tf,Ooguri:2010kt, Ooguri:2010xs, Bayona:2011ab, Bergman:2011rf, Jokela:2012vn, BallonBayona:2012wx, Donos:2012wi,Bu:2012mq, Jokela:2012se, Rozali:2012es, Withers:2013loa, Withers:2013kva, Rozali:2013ama,Donos:2013gda, Ling:2014saa, Jokela:2014dba,Jokela:2016xuy,Ammon:2016szz,Amoretti:2016bxs,Cremonini:2017usb,Jokela:2017fwa,Liu:2016hqb,Cremonini:2017qwq,Jokela:2017ltu,Ammon:2017ded,Amoretti:2017frz,Amoretti:2017axe,Donos:2018kkm,Gouteraux:2018wfe,Li:2018vrz,Ammon:2020rvg} and nuclear matter \cite{Kaplunovsky:2012gb,Kaplunovsky:2013iza,Kaplunovsky:2015zsa,Jarvinen:2020xjh}, but so far they have not been used to model compact stars. In many cases, the appearance of inhomogeneous phases has been linked to the axial anomaly, which can also produce interesting effects associated with chiral transport \cite{Ohnishi:2014uea,Kaminski:2014jda,Dvornikov:2014uza,Dvornikov:2015lea,Sigl:2015xva,Yamamoto:2015gzz}.

Finally, should the deconfinement transition be of first order and the surface tension between the two phases sufficiently small, there exists the possibility of having a mixture of nuclear and quark matter present in neutron-star cores \cite{Glendenning:1992vb}. It is extremely challenging to determine the surface tension parameter with traditional quantum field theory tools, but in a very recent holographic study, it was successfully computed in a strongly coupled field theory at high temperatures \cite{Ares:2021ntv}. A generalization of this result to the context of high-density QCD matter would clearly be very well-motivated.

\section*{Acknowledgements}
We would like to thank M.~J\"arvinen, L.~Rezzolla, and A.~Schmitt for discussions and correspondence during the writing of the review. C.H.~has been partially supported by the Spanish Ministerio de Ciencia, Innovaci\'on y Universidades through the grant PGC2018-096894-B-100; N.J. by the Academy of Finland grant no.~1322307; and A.V.~by the Academy of Finland grant no.~1322507 as well as by the European Research Council, grant no.~725369.

\bibliographystyle{elsarticle-num} 
\bibliography{biblio}

\end{document}